\documentclass[12pt]{article}

\usepackage{macros}
\numberwithin{equation}{section}
\usepackage{subcaption}
\usepackage{graphicx}
\usepackage{tikz}

\begin{document}

\thispagestyle{empty}
\vspace*{.5cm}

\begin{center}

{{\Large \textbf{The moments of the spectral form factor in SYK}}}\\

\begin{center}

\vspace{1cm}

\textbf{Andrea Legramandi$^{1,2}$, Neil Talwar$^{3,4}$}

\vspace{.8cm}

\footnotesize{$^1$Pitaevskii BEC Center, CNR-INO and Dipartimento di Fisica, Universit\'a di Trento, I38123 Trento, Italy\\ \vspace{.15cm}
$^2$INFN-TIFPA, Trento Institute for Fundamental Physics and Applications, Trento, Italy \\ \vspace{.15cm}
$^3$Department of Physics, Swansea University, Swansea, SA2 8PP, U.K.\\ \vspace{.15cm}
$^4$SISSA, Via Bonomea 265, 34136 Trieste, Italy}

\vspace{.8cm}

\texttt{andrea.legramandi@unitn.it, ntalwar@sissa.it}

\end{center}

\vspace{1cm}

{\bf Abstract}
\end{center}
\begin{quotation}
In chaotic quantum systems the spectral form factor exhibits a universal linear ramp and plateau structure with superimposed erratic oscillations. The mean signal and the statistics of the noise can be probed by the moments of the spectral form factor, also known as higher-point spectral form factors. We identify saddle points in the SYK model that describe the moments during the ramp region. Perturbative corrections around the saddle point indicate that SYK mimics random matrix statistics for the low order moments, while large deviations for the high order moments arise from fluctuations near the edge of the spectrum. The leading correction scales inversely with the number of random parameters in the SYK Hamiltonian and is amplified in a sparsified version of the SYK model, which we study numerically, even in regimes where a linear ramp persists. Finally, we study the $q=2$ SYK model, whose spectral form factor exhibits an exponential ramp with increased noise. These findings reveal how deviations from random matrix universality arise in disordered systems and motivate their interpretation from a bulk gravitational perspective.
\end{quotation}

\setcounter{page}{0}
\setcounter{tocdepth}{2}
\setcounter{footnote}{0}
\newpage

\parskip 0.1in

\setcounter{page}{2}
\begingroup
\hypersetup{linkcolor=black}
\tableofcontents
\endgroup

\pagebreak

\section{Introduction}

Quantum chaos plays a key role in characterising physical systems as diverse as quantum many-body systems and black holes. Various diagnostics for quantum chaos exist, such as out-of-time-order correlation functions \cite{larkin:1969,Maldacena:2015waa,Hosur:2015ylk,Chen_2016} which characterise chaos at early times and the distribution of nearby level spacings which relates to chaos at small energy differences or, equivalently, at late times \cite{Bohigas:1983er}. While diagnostics of classical chaos are firmly established, a unified notion of quantum chaos remains elusive. A significant idea in this context is \textit{random matrix universality}, which proposes that correlations between nearby energy levels in chaotic quantum systems are statistically equivalent to those of a Hamiltonian drawn from a random matrix ensemble of the appropriate symmetry class \cite{Dyson:1962eeu,Altland:1997zz}.

\begin{figure}
\centering
\begin{subfigure}{0.48\textwidth}
\raggedright
\begin{tikzpicture}
\node at (0,0) {\includegraphics[width=0.95\linewidth]{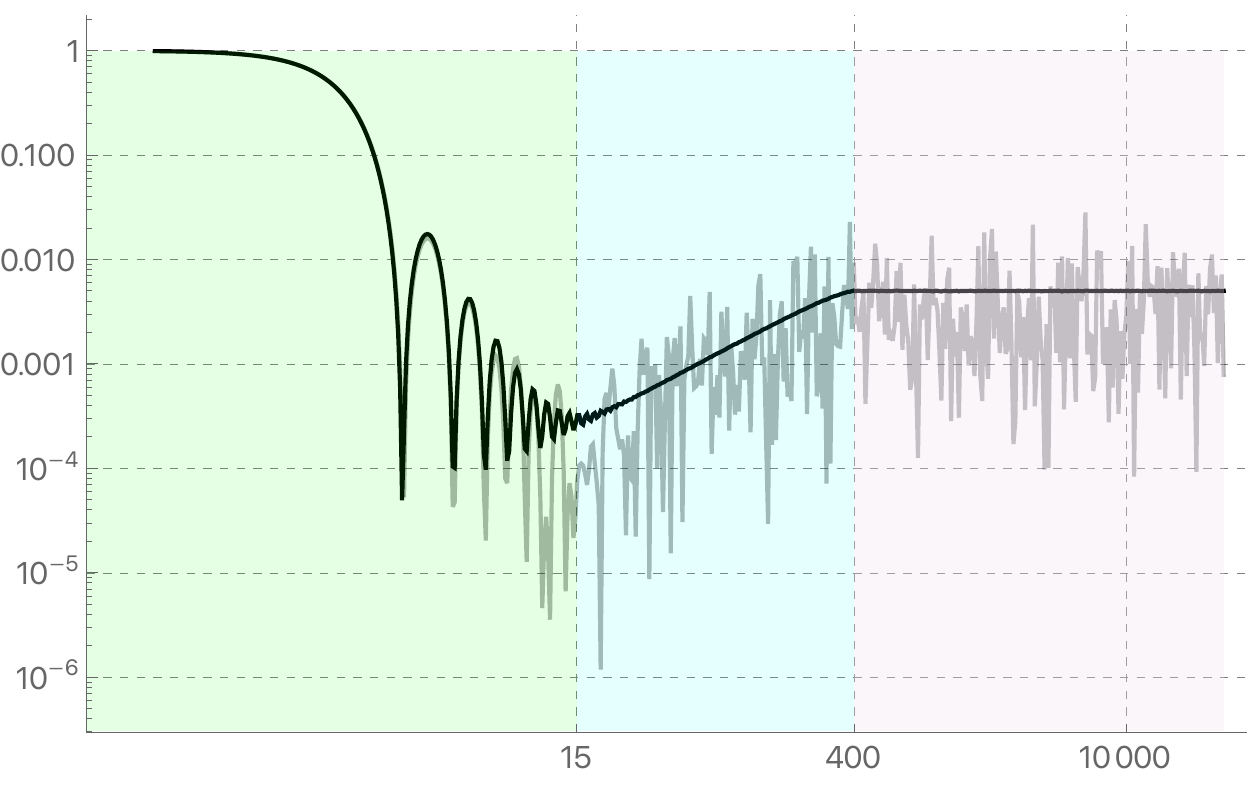}};
\node at (current bounding box.south) {{\scriptsize $T$}};
\node at ([xshift=-0.2em]current bounding box.west) {\scriptsize \rotatebox{90}{$\langle |Z(\ii T)|^2 \rangle$}};
\end{tikzpicture}
\end{subfigure}
% $\quad$
\begin{subfigure}{0.48\textwidth}
\raggedleft
\begin{tikzpicture}
\node at (0,0) {\includegraphics[width=0.95\linewidth]{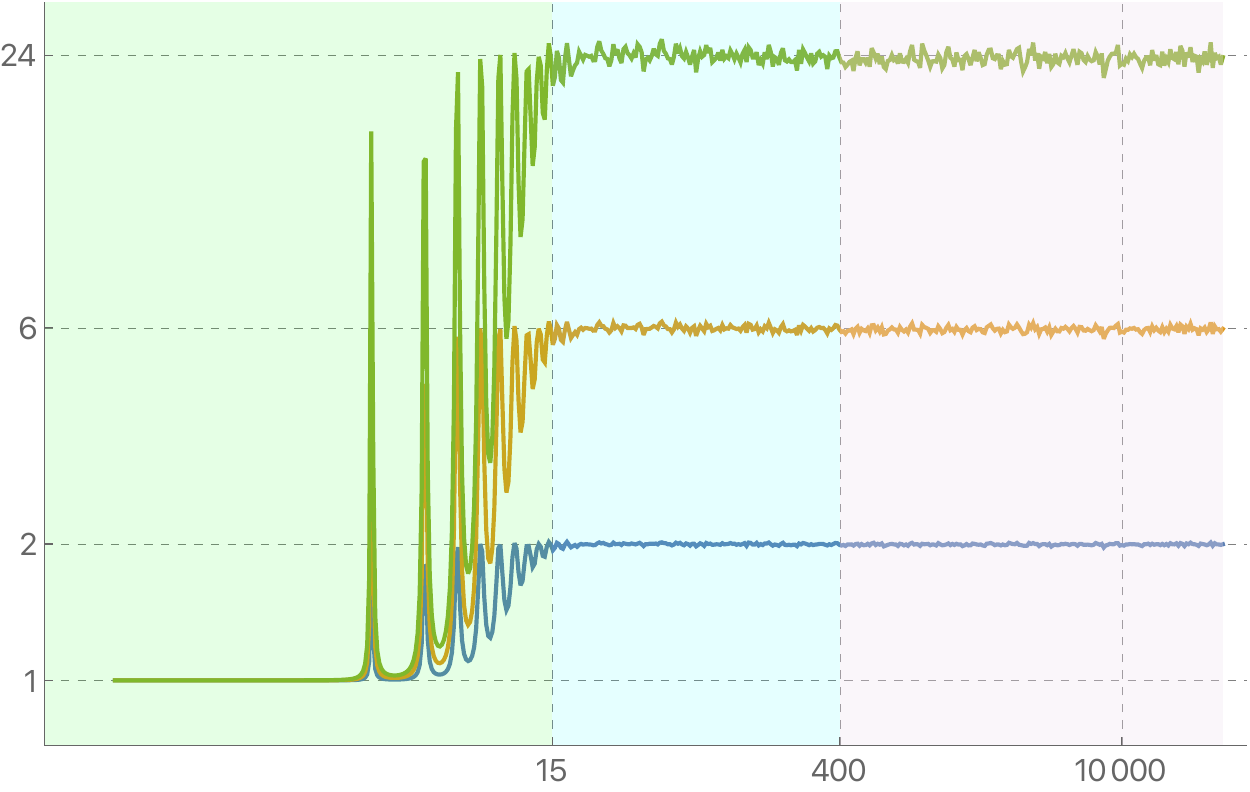}};
\node at (current bounding box.south) {{\scriptsize $T$}};
\node at (current bounding box.west) {\scriptsize \rotatebox{90}{$\langle |Z(\ii T)|^{2k} \rangle / \langle |Z(\ii T)|^2 \rangle^k$}};
\end{tikzpicture}
\end{subfigure}
\caption{Left panel: the spectral form factor (divided by $L^2$) for the Gausian unitary ensemble with $L=200$ for a single realisation (grey) and averaged over $50$K realisations (black). Right panel: the moments of the spectral form factor for $k=2$ (blue), $k=3$ (orange), and $k=4$ (green). At early times $|Z(\ii T)|^2$ is self-averaging whereas at late times $\langle |Z(\ii T)|^{2k} \rangle = k! \langle |Z(\ii T)|^{2} \rangle^k$ \eqref{eq:RMT_exp_intro}, reflecting the noise around the mean signal.}
\label{fig:momentsRMT}
\end{figure}
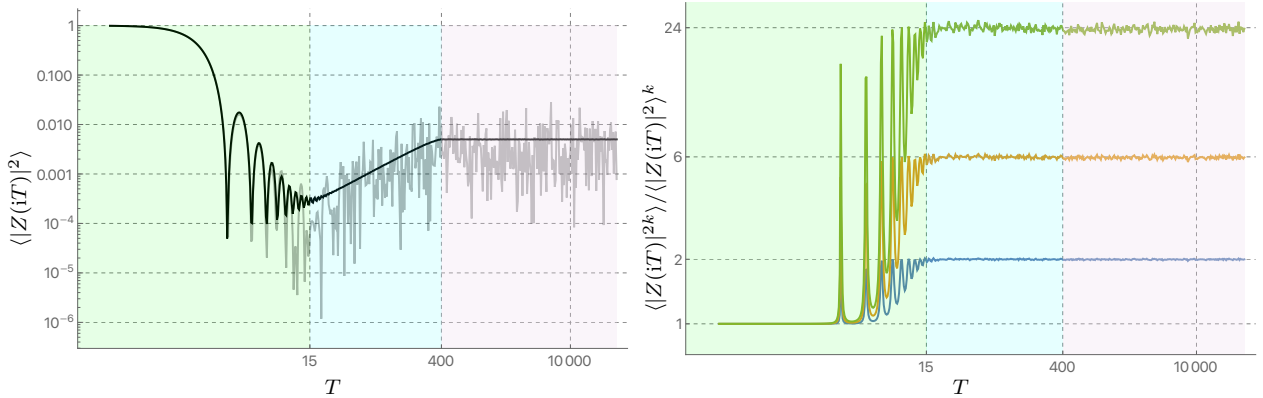

A useful diagnostic of late time quantum chaos is the spectral form factor\footnote{We define $Z(x) \equiv \Tr[e^{-x H}]$.} \cite{berry_1985,mehta2004random,brezin_1997}
\be
|Z(\ii T)|^2 = \Tr [ e^{-\ii T H}] \, \Tr [ e^{\ii T H}].
\ee
In random matrix theory (RMT), the spectral form factor exhibits a universal form: after decaying for a while, it begins to exhibit erratic oscillations, with the mean signal following a linear ramp before reaching a plateau, as in figure \ref{fig:momentsRMT}.\footnote{In all figures, we plot the rescaled spectral form factor $L^{-2} \langle |Z(\ii T)|^2 \rangle$, normalised so that it equals 1 at $T=0$. For simplicity, we refer to it as $\langle |Z(\ii T)|^2 \rangle$ in the plots.} The ramp and plateau structure only emerges after a suitable averaging, such as over Hamiltonians for random matrix ensembles, couplings in disordered systems, or time for individual systems. The noise is of the same order as the mean signal, so it is not a small effect, and results from the sensitivity of the spectral form factor to the specific details of the energy spectrum. While averaging largely removes this noise, its statistics can be examined by the variance and higher moments of the spectral form factor. Crucially, random matrix universality concerns not only the mean signal of the spectral form factor but also the statistics of the noise, as captured by its moments. This is especially relevant because, while often considered a hallmark of quantum chaos, a linear ramp---a consequence of long-range spectral rigidity---can be mimicked in systems whose nearby energy level statistics are not governed by a random matrix ensemble \cite{Das:2023yfj}.

The present paper focuses on studying the moments of the spectral form factor\footnote{The moments of the spectral form factor are sometimes referred to as high-point spectral form factors. See \cite{Cotler:2017jue,Liu:2018hlr} for related work in the context of RMT.}
\be \label{momentsintro}
\langle |Z(\ii T)|^{2k} \rangle, \qquad k \in \N,
\ee 
as a more comprehensive diagnostic of quantum chaos in the Sachdev--Ye--Kitaev (SYK) model \cite{Kitaev2015,Sachdev:2015efa,Polchinski:2016xgd}---a quantum mechanical model of $N$ Majorana fermions with random all-to-all $q$-fermion interactions. The angled brackets in \eqref{momentsintro} indicate an average over the random couplings. The SYK model is a rare example of a chaotic system which allows for an analytical understanding of the late time behaviour of the spectral form factor. Originally introduced to study non-Fermi liquids \cite{Sachdev1993}, the SYK model sparked renewed interest after it was shown that at low temperature its dynamics are dominated by the Schwarzian mode \cite{Maldacena:2016hyu,Jensen:2016pah,Kitaev:2017awl} that also describes Jackiw--Teitelboim (JT) gravity \cite{Maldacena:2016upp}, a theory of dilaton gravity in two-dimensional Anti de Sitter space \cite{Jackiw1984,Teitelboim1983}. 

The SYK model admits a reformulation in terms of \textit{collective field variables} which become semiclassical in the large $N$ limit, even away from low temperature. In this framework, quantities like $\langle \Tr [ e^{-\ii T H}] \, \Tr [ e^{\ii T H}] \rangle$ can be studied by introducing two \textit{replicas}, usually labelled $L$ and $R$, corresponding to the two traces in the spectral form factor. The collective field variables in this formulation become matrix-valued in replica space, with off-diagonal components representing correlations between the replicas. In this setting, Saad, Shenker, and Stanford \cite{Saad:2018bqo} identified a continuous family of saddle points in the SYK model that describe the linear ramp in the spectral form factor $\langle |Z(\ii T)|^2 \rangle$. The ramp arises from a compact zero mode, resulting from the saddle point spontaneously breaking the $\mathrm{U}(1)_L \times \mathrm{U}(1)_R$ time translation symmetry to the diagonal $\mathrm{U}(1)$.

The very late time behaviour of the moments of the spectral form factor can be understood by considering a long time average. Assuming the spectrum has no degeneracies, the moments saturate on a plateau:\footnote{For systems with an energy spectrum that is symmetric about zero, the $k!$ is instead replaced with $(2k-1)!!$ in equations \eqref{eq:long_time_average_SYK_RMT} and \eqref{eq:RMT_exp_intro}, see \cite{Winer:2020gdp}. This symmetry doesn't appear in the Wigner-Dyson ensembles \cite{Dyson:1962eeu} but can occur in other symmetry classes that appear in the 10-fold way of Altland and Zirnbauer \cite{Altland:1997zz}.  We review the relevant symmetry classes for SYK, discussed in \cite{Behrends_2019}, in \ref{app:rmt syk}.}
\be
\label{eq:long_time_average_SYK_RMT}
\lim_{T \to \infty} \frac{1}{T} \int_0^T \dd t \, |Z(\ii t)|^{2k} = \lim_{T \to \infty} \frac{1}{T} \int_0^T \dd t \, \bigg( \sum_{m,n=1}^L e^{-\ii t (E_m-E_n)}\bigg)^k = k! L^k,
\ee
where $E_n$ are the eigenvalues of the Hamiltonian $H$, and $L$ is the dimension of the Hilbert space. The result follows by only considering the terms where the phases cancel out. Up to a reshuffling of the $k$ copies of $Z(\ii t)$---responsible for the combinatorial factor $k!$---this happens when we set $E_m=E_n$ in each of the $k$ sums. This implies that for very late times $|Z(\ii T)|^2$ behaves like the modulus square of a complex Gaussian variable with mean zero and variance $L$. The behaviour of the moments of the spectral form factor for late times, but before the plateau time, can be understood as a perturbative effect in RMT \cite{Cotler_2017, AlSh86}, as we review in appendix \ref{app:rmt}. This leads to the result
\be
\langle |Z(\ii T)|^{2k} \rangle = k! (\upgamma \, T)^k,
\label{eq:RMT_exp_intro}
\ee
where $\upgamma$ is a constant which depends on symmetry class of the ensemble. This means the moments also exhibit a ramp, although a power law one rather than a linear one. In a similar way, this result implies that $|Z(\ii T)|^2$ behaves like the modulus square of a complex Gaussian variable with mean zero and variance $\upgamma \, T$. This is shown for the Gaussian unitary ensemble in figure \ref{fig:momentsRMT}.

In the SYK model, the moments of the spectral form factor $\langle |Z(\ii T)|^{2k} \rangle$ can be studied using collective fields by introducing $2k$ replicas. In this paper we describe a family of saddle points which describe the power law ramp in the moments. In addition to spontaneously breaking $k$ time translation symmetries, producing a factor $(\upgamma \, T)^k$, these saddle points spontaneously break the discrete $S_k \times S_k$ replica symmetry to the diagonal $S_k$, yielding the combinatorial factor $k!$.\footnote{This result applies to the SYK model with $q= 0 \mod 4$. For $q= 2 \mod 4$, the replica symmetry is enhanced to $S_{2k}$ and spontaneously broken to $S_k \times S_2^k$ by the saddle point, leading instead to the combinatorial factor $(2k-1)!!$. This is related to the energy spectrum of the $q= 2 \mod 4$ model being symmetric about zero.} The same result was obtained for a spin glass model in \cite{Winer:2022ciz}.
 
In the large $N$ limit, the moments of the spectral form factor in the SYK model align with expectations from random matrix universality. However, for any large but finite $N$, we expect to observe non-universal features, that is, deviations from RMT like behaviour. This aligns with the idea that SYK and RMT belong to different universality classes of chaotic systems, the \say{sparse} and \say{dense} classes, respectively \cite{Altland:2024ubs}. The key distinction lies in the number of independent random parameters in the Hamiltonian: for SYK it is approximately $N^q/q!$, scaling logarithmically with the Hilbert space dimension $L=2^{N/2}$, while for RMT it is approximately $L^2$, scaling polynomially. These differences are visible in the regime of large but finite $N$ in the perturbative corrections to the moments of the spectral form factor:
\be \label{eq:per_corr_intro}
\frac{\langle |Z(\ii T)|^{2k} \rangle}{\langle |Z(\ii T)|^2 \rangle^k} = k! \left[ 1 + \frac{k(k-1)}{4} \frac{q!}{N^{q}} T^2 |\overline{\Delta}_E(T)|^2 + \dots \right].
\ee
The correction arises from fluctuations near the edge of the spectrum, as discussed in section \ref{sec:syk}, which aligns with the expectation that differences between the sparse and dense classes of chaotic systems unfolds in physics near the spectral edge \cite{Altland:2024ubs}. While the time dependence of the correction appears challenging to compute analytically, we provide numerical evidence that it is approximately constant.

Although the correction to the moments is small for the low order moments, it becomes significant for the high order moments, specifically when $k$ approaches a fixed fraction of $N^{q/2-1}$. This scaling with $N$ follows because the quantity $\overline{\Delta}_E$ is extensive in $N$ so the leading correction ultimately scales as $N^{q-2}$. This marks a departure from RMT expectations where we expect the corrections to the moments to become significant only when $k$ approaches a fixed fraction of $L^\#$, which is exponential in $N$ in SYK. In this sense, SYK mimics RMT behaviour only for the low order moments of the spectral form factor.

Notably, the leading correction in \eqref{eq:per_corr_intro} is inversely proportional to the number of independent random parameters in the SYK Hamiltonian, approximately $N^q/q!$ for large $N$, suggesting that for more general disordered systems, the correction may scale similarly.\footnote{Similar corrections have appeared in different contexts \cite{Kitaev:2017awl,Garcia-Garcia:2018fns,Cotler:2017jue,Jia:2019orl,Berkooz:2020fvm}.}
We explore this possibility numerically in the sparse SYK model \cite{Xu:2020shn}, a sparsified version of the SYK model in which interaction terms are deleted with probability $1-p$. This model is particularly interesting because it has been argued to be holographic even in the highly sparsified regime, where the resources needed for its simulation are significantly reduced, allowing us to test the analogue of \eqref{eq:per_corr_intro} for sparse SYK with high accuracy. The observation that the leading correction is inversely proportional to the sparsification parameter $p$---supported by our numerical analysis of the sparse SYK model---indicates that the moments in the sparse SYK deviate from RMT expectations significantly earlier than in regular unsparsified SYK. This occurs even in the regimes where the onset of the linear ramp remains unchanged \cite{Orman:2024mpw}, raising questions about the conditions under which the sparse model continues to exhibit chaotic features.

The scaling of the leading correction as $1/N^{q-2}$ in equation \eqref{eq:per_corr_intro} also hints that the $q=2$ model is qualitatively different from the $q>2$ models. This is to be expected since the $q=2$ model is free and therefore isn't expected to exhibit features of many-body chaos. This results in drastic changes to the spectral form factor, which, in particular, features an exponential ramp rather than a linear one. The authors of \cite{Winer:2020mdc} showed how this property could be traced to an infinite enhancement of the $\mathrm{U}(1)_L \times \mathrm{U}(1)_R$ time translation symmetry which is spontaneously broken by the saddle point in a time dependent pattern. In this paper, we explain how this symmetry is further enhanced for the moments of the spectral form factor and demonstrate how its spontaneous breaking accounts for their behaviour. In stark contrast to the $q>2$ models, where the noise is the same order as the mean signal, the magnitude of the noise is exponential in $T \log N/T$. The large moments indicate distribution with heavy tails signalling that the spectral form factor has a higher probability of producing extreme values. The saddle point approximation breaks down as $T$ approaches $N$ and the exponential ramp gives rise to a plateau where the noise becomes exponential in $N$, in contrast to \eqref{eq:long_time_average_SYK_RMT}.

Finally, while SYK has an approximate low-energy description that includes JT gravity, the two systems are not equivalent. In contrast to the sparsely chaotic SYK model, pure JT gravity is dual to a double-scaled matrix integral \cite{Saad:2019lba} and so belongs to the dense class of chaotic systems. Observables like $\langle Z(\beta_1) \dots Z(\beta_n) \rangle$ in the matrix model are computed by the path integral of JT gravity with $n$ asymptotic boundaries with lengths $\beta_1,\dots,\beta_n$. These observables are related to the moments of the spectral form factor through analytic continuation and admit a genus expansion, where different topologies are weighted by $e^{S_0 \chi}$ with $\chi$ being the Euler characteristic. In a holographic interpretation, $S_0 \propto N$ represents the zero temperature entropy of the SYK model. In SYK these observables receive relatively large connected contributions, of order $N^{2-q}$, which are much larger than $e^{-S_0}$ contributions in JT gravity, arising from collective fluctuations in the spectrum. These fluctuations leave imprints on the moments of the spectral form factor in SYK at early and late times and mark a large departure from pure JT gravity. Understanding what additional bulk ingredients, such as matter fields \cite{Berkooz:2020fvm,Chen:2023hra}, are needed to capture this structure in SYK remains an interesting question.

The plan of this paper is as follows. In section \ref{sec:syk one time point} we compute the moments of the partition function in a zero-dimensional version of SYK known as SYK with one time point \cite{Saad:2021rcu}. This model is useful because it is simple enough to analyse in detail while still retaining some features of the regular SYK model. We compute the moments both exactly and in the saddle point approximation, including the leading $1/N$ contribution.
In section \ref{sec:syk} we review the saddle points that describe the linear ramp in the spectral form factor in SYK \cite{Saad:2018bqo} and explain how they may be used to construct saddle points that describe the power law ramp in the moments of the spectral form factor and compare the result with numerics based on exact diagonalisation of the SYK Hamiltonian. We also determine the leading in $1/N$ correction to the moments \eqref{eq:per_corr_intro} by considering perturbative fluctuations around the saddle point. In section \ref{syk:sec:q=2} we study the moments of the spectral form factor in the $q=2$ model, which requires a separate analysis due to the infinite symmetry enhancement of the model. In this model we also compute the behaviour of the very high moments with $k \gg N$. In section \ref{sec:sparse_SYK} we study the sparse SYK model numerically in order to understand the dependence of the leading in $1/N$ correction to the moments of the spectral form factor on $k$, $T$, and the sparsification parameter with an aim to understand the degree of deviation from RMT expectations. We conclude with a summary and discussion of future outlooks in Section \ref{sec:conclusion}.

In appendix \ref{app:rmt} we review how the power law ramp in the spectral form factor can be understood as a perturbative effective in RMT and review the relevant random matrix classification of SYK. In appendix \ref{app:SYK_perturbation} we provide details for the perturbative analysis in the SYK model. In appendix \ref{app:SYK_q=2} we provide more details on the $q=2$ model.

\section{SYK with one time point} \label{sec:syk one time point}

As a warm up to the SYK model, in this section we consider the following finite-dimensional integral over $N$ Grassmann variables $\psi_1, \dots, \psi_N$,\footnote{The measure is normalised such that $\int \dd \psi \, \psi = \ii^{1/2}$.}
\be
z = \int \dd^N \psi \exp \left\{ \ii^{q/2} \sum_{i_1 <\dots <i_q} J_{i_1 \dots i_q} \psi_{i_1}\dots \psi_{i_q}\right\},
\ee
where the $J_{i_1 \dots i_q}$ are drawn from a Gaussian distribution with zero mean and variance
\be
\langle J_{i_1 \dots i_q} J_{j_1 \dots j_q} \rangle = \frac{(q-1)!}{N^{q-1}} \delta_{i_1 j_1} \dots \delta_{i_q j_q}.
\ee
We assume that $q$ is an even integer and that $N$ is divisible by $q$, as $z$ vanishes otherwise. This model, introduced in \cite{Saad:2021rcu} as a very simple setup to study the factorization problem \cite{Maldacena:2004rf}, is known as SYK with one time point. This is because it can be viewed as a zero-dimensional version of the SYK model, where the time contour is shrunk to a single instant of time.

The goal of this section is to understand the statistics of $z$ for large $N$ by analysing its moments. We first show that $z$ follows Gaussian statistics in the large $N$ limit, addressing a suggestion made in \cite{Saad:2021rcu}, where the first four moments of $z$ were computed. We then show that perturbative $1/N$ corrections to this result indicate a deviation from Gaussian statistics for the very high moments of $z$ for any large but finite $N$. This model is useful because it is simple enough to analyse in detail while still retaining some features of the regular SYK model. Notably, the perturbative corrections to the moments in this model share a similar structure to those in regular SYK, but with fewer complications. The reader only interested in the results for regular SYK can skip ahead to section \ref{sec:syk} and jump back if needed.

\subsection{$\langle z^2 \rangle$}

Since the average of $z$ vanishes, the simplest nontrivial quantity is its variance, $\langle z^2 \rangle$. Doing the Gaussian integral over the $J_{i_1 \dots i_q}$ and anticommuting the Grassmann variables past each other we get
\be \label{<z^2> fermion}
\langle z^2 \rangle = \int \dd^{2N} \psi \exp \left\{ \frac{N}{q}  \left( \frac{1}{N} \sum_i \psi_i^L \psi_i^R \right)^q \right\},
\ee
where $L$ and $R$ label the two copies of Grassmann variables needed to represent the square of $z$. This integral can be represented as an integral over \textit{collective fields} $G$ and $\Sigma$,
\be \label{<z^2>}
\langle z^2 \rangle = \int_\R \dd G \int_{\ii \R} \frac{\dd \Sigma}{2\pi \ii/N} \exp\left\{N\left(\log(\Sigma)-\Sigma G + \frac{1}{q} G^q\right)\right\}.
\ee
This is achieved by inserting the following identity into \eqref{<z^2> fermion}:
\be \label{G Sigma identity}
1 = \int_\R \dd G \int_{\ii \R} \frac{\dd \Sigma}{2\pi \ii/N} \exp \bigg\{-N\Sigma \bigg(G - \frac{1}{N}\sum_i \psi_i^L \psi_i^R\bigg)\bigg\},
\ee
and then integrating out the Grassmann variables. Notice that in \eqref{G Sigma identity} the variable $\Sigma$ acts as a Lagrange multiplier enforcing the relation
\be
G = \frac{1}{N} \sum_{i=1}^N \psi_i^L \psi_i^R.
\ee
In an analogous computation in regular SYK one would require a matrix of collective fields $G_{LL}, G_{LR},$ and $G_{RR}$, and similarly for $\Sigma$. In this simple model, only the off-diagonal components $G \equiv G_{LR}$ and $\Sigma \equiv \Sigma_{LR}$ are needed. As we review in the following section, the integral \eqref{<z^2>} can be computed exactly \cite{Saad:2021rcu}. If $N$ is a multiple of $q$, the result is 
\be \label{<z^2> exact}
\langle z^2 \rangle = \frac{N! (N/q)^{N/q}}{N^N (N/q)!},
\ee
otherwise, if $N$ is not a multiple of $q$ it vanishes. For large $N$ the integral \eqref{<z^2>} can also be computed using the saddle point approximation. The saddle point equation is
\be \label{<z^2> eom}
G=\frac{1}{\Sigma},\qquad \Sigma=G^{q-1}.
\ee
There are $q$ solutions, each corresponding to a $q$\ts{th} root of unity and contributing as
\be
\frac{1}{\sqrt{q}} e^{-\left(1-\frac{1}{q}\right)N} e^{2\pi \ii mN/q},\qquad m=0,\dots,q-1,
\ee
where $1/\sqrt{q}$ is the contribution from the one loop determinant. The $q$-fold degeneracy is an artifact of having a single time point and is lifted in a model with two time points \cite{Saad:2021rcu}. If $N$ is a multiple of $q$, summing over the $q$ saddle points gives
\be \label{large n <z^2>}
\langle z^2 \rangle \sim \sqrt{q} e^{-\left(1-\frac{1}{q}\right)N},
\ee
which reproduces the large $N$ asymptotic form of the exact result \eqref{<z^2> exact}, which follows from Stirling's approximation. Otherwise, if $N$ is not a multiple of $q$, the sum vanishes.

\subsection{$\langle z^{2k} \rangle$}  \label{1tp z^2k}

Repeating the same trick for $\langle z^{2k} \rangle$ we end up with
\be \label{<z^2k>}
\langle z^{2k} \rangle = \int_\R \dd G_{ab} \int_{\ii \R} \frac{\dd \Sigma_{ab}}{2\pi \ii/N} \, \Pf(\Sigma)^N \exp\left\{N \sum_{a<b} \left(-\Sigma_{ab} G_{ab} + \frac{1}{q} G_{ab}^q\right)\right\},
\ee
where Pf is the Pfaffian, and $G_{ab}$ and $\Sigma_{ab}$ are $2k \times 2k$ antisymmetric matrices. The odd moments of $z$ vanish as the Pfaffian of odd dimensional matrix vanishes. We now demonstrate how this integral can be evaluated exactly. First, using the integral representation of the delta function, the integral over $\Sigma_{ab}$ gives\footnote{The basic identity used here is $\int_{\ii \R} \frac{\dd \Sigma}{2\pi \ii/N} \Sigma^N e^{-N \Sigma G} = (-N)^{-N} \partial_G^N \delta(G)$.}
\be \label{sigma integral}
\int_{\ii \R} \frac{\dd \Sigma_{ab}}{2\pi \ii/N} \, \Pf(\Sigma)^N \exp\left\{-N \sum_{a<b} \Sigma_{ab} G_{ab}\right\} = N^{-kN} \Pf(\partial_G)^N \prod_{a<b}\delta(G_{ab}),
\ee
where on the RHS $\partial_G$ is viewed as a matrix with components $(\partial_G)_{ab} = \partial_{G_{ab}}$. This formula implies that the integral \eqref{<z^2k>} localises to a neighbourhood of $G_{ab}=0$:
\be \label{<z^2k> pfaffian}
\langle z^{2k} \rangle = N^{-kN} \Pf(\partial_G)^N \exp\left\{N \sum_{a<b} G_{ab}^q \right\}\Bigg\rvert_{G_{ab}=0}.
\ee
Next, using the property that the Pfaffian can be expressed as a sum over pairings we find
\be \label{<z^2k> pairings}
\langle z^{2k} \rangle = N^{-kN} \sum_{n_\pi} \frac{N!}{\underset{\pi \in P_{2k}}{\prod} n_\pi !} \prod_{a<b} \left( \frac{N}{q} \right)^{m_{ab}} \frac{(q m_{ab})!}{m_{ab}!},\qquad q m_{ab} = \sum_{\underset{\{a,b\} \in \pi}{\pi \in P_{2k}:}} n_\pi,
\ee
where $P_{2k}$ denotes the set of pairings of $2k$ elements. In this expression the first sum is over integers $n_\pi \geq 0$ such that both $\sum_{\pi \in P_{2k}} n_\pi=N$ and $m_{ab}$ is an integer. For $k=2$ this formula simplifies: there are 3 parameters $n_1,n_2,$ and $n_3$ which correspond to the 3 pairings (or Wick contractions),
\[
\begin{tikzpicture}[thick,font=\small, node distance=1cm and 5mm, baseline={([yshift=-.5ex]current bounding box.center)}]
\tikzset{dot/.style={circle,fill=#1,inner sep=0,minimum size=4pt}}

\node (1) 				    {$L$};
\node (2)	[right=of 1] 	{$R$};
\node (3) 	[right=of 2]	{$L'$};
\node (4)	[right=of 3] 	{$R'$};

\node [dot=black] at ($ (1) + (0,-.3) $) {};	
\node [dot=black] at ($ (2) + (0,-.3) $) {};	
\node [dot=black] at ($ (3) + (0,-.3) $) {};	
\node [dot=black] at ($ (4) + (0,-.3) $) {};

\draw (1)	-- ++(0,-.7)		-| (2);
\draw (3)	-- ++(0,-.7)		-| (4);	
\end{tikzpicture}, \qquad
\begin{tikzpicture}[thick,font=\small, node distance=1cm and 5mm, baseline={([yshift=-.5ex]current bounding box.center)}]
\tikzset{dot/.style={circle,fill=#1,inner sep=0,minimum size=4pt}}

\node (1) 				{$L$};
\node (2)	[right=of 1] 	{$R$};
\node (3) 	[right=of 2]	{$L'$};
\node (4)	[right=of 3] 	{$R'$};

\node [dot=black] at ($ (1) + (0,-.3) $) {};	
\node [dot=black] at ($ (2) + (0,-.3) $) {};	
\node [dot=black] at ($ (3) + (0,-.3) $) {};	
\node [dot=black] at ($ (4) + (0,-.3) $) {};

\draw (1)	-- ++(0,-.7)		-| (3);	
\draw (2)	-- ++(0,-1.1)		-| (4);	
\end{tikzpicture}, \qquad
\begin{tikzpicture}[thick,font=\small, node distance=1cm and 5mm, baseline={([yshift=-.5ex]current bounding box.center)}]
\tikzset{dot/.style={circle,fill=#1,inner sep=0,minimum size=4pt}}

\node (1) 				{$L$};
\node (2)	[right=of 1] 	{$R$};
\node (3) 	[right=of 2]	{$L'$};
\node (4)	[right=of 3] 	{$R'$};

\node [dot=black] at ($ (1) + (0,-.3) $) {};	
\node [dot=black] at ($ (2) + (0,-.3) $) {};	
\node [dot=black] at ($ (3) + (0,-.3) $) {};	
\node [dot=black] at ($ (4) + (0,-.3) $) {};

\draw (1)	-- ++(0,-1.1)		-| (4);	
\draw (2)	-- ++(0,-.7)		-| (3);	
\end{tikzpicture},
\]
and 6 parameters $m_{ab}$,
\be
qm_{LR}=qm_{L'R'}=n_1,\quad qm_{LL'}=qm_{RR'}=n_2,\qquad qm_{LR'}=qm_{RL'}=n_3.
\ee
Since the $n_i$ must be divisible by $q$ we can rescale $n_i \to q n_i$ so that the sum is instead over integers $n_i \geq 0$ such that $n_1+n_2+n_3=N/q$. This gives
\be
\langle z^4 \rangle = \frac{N!}{N^2} \left(\frac{N}{q}\right)^{\frac{2N}{q}} \sum_{\underset{n_i\geq0}{n_1+n_2+n_3=N/q:}} \frac{(qn_1)!(qn_2)!(qn_3)!}{(n_1)!^2(n_2)!^2(n_3)!^2}.
\ee
For large $N$ and when $q>2$
\be
\frac{\langle z^4 \rangle}{\langle z^2 \rangle^2} = 3 + 6 \times \frac{q!}{q^2} \frac{1}{N^{q-2}} + \dots
\ee
The leading term arises from terms in the sum where one of the $n_i$ are equal to $N/q$ and the other $n_i$ vanish. The first correction arises from terms in the sum where one of the $n_i$ are equal to $N/q-1$, another $n_i$ is equal to $1$, and the third $n_i$ vanishes. The factors $3$ and $6$ count the multiplicity of such terms.

For general $k$, large $N$ and when $q>2$,
\be \label{<z^2k> exact ...}
\frac{\langle z^{2k} \rangle}{\langle z^2 \rangle^k} = (2k-1)!! \left[1 + k(k-1) \times \frac{q!}{q^2} \frac{1}{N^{q-2}} + \dots \right].
\ee
The leading term is determined by maximising the number of pairs $\{a,b\}$ in \eqref{<z^2k> pairings} for which $m_{ab}$ takes its maximal value, which is $N/q$. The maximum number of such pairs is $k$. This occurs when one of the $n_i$ are equal to $N$ and the other $n_i$ vanish. Since there are $|P_{2k}|=(2k-1)!! = \frac{(2k)!}{2^kk!}$ such terms this leads to the contribution
\be \label{<z^2k> leading}
\langle z^{2k} \rangle \supset (2k-1)!! \left[\frac{N!(N/q)^N}{N^N(N/q)!}\right]^k.
\ee
The first correction arises from a subset of the terms in the sum where one of the $n_i$ are equal to $N-q$, another $n_i$ is equal to $q$, and the remaining $n_i$ vanish. Let $\ell$ denote the number of pairs in common between the pairing with $n_i$ equal to $N-q$ and the pairing with $n_i$ equal to $q$. Then $\ell$ of the parameters $m_{ab}$ are equal to $N/q$, $k-\ell$ of parameters $m_{ab}$ are equal to $N/q-1$, another $k-\ell$ of parameters $m_{ab}$ are equal to $1$, and the remaining parameters vanish. The dominant contribution comes from the terms with $\ell=k-2$,
\be \label{<z^2k> subleading}
\langle z^{2k} \rangle \supset (2k-1)!! \times k(k-1) \frac{(N/q)^{kN/q} N!^{k-1} (N-q)! q!}{N^{kN} (N/q)!^{k-2} (N/q-1)!^2 }.
\ee
The combinatorial factor $(2k-1)!! \times k(k-1)$ can be understood as follows. For a fixed pairing we need to sum over the number of ways to pick another pairing such that the two pairings have $k-2$ pairs in common. There are $k(k-1)$ ways to do this. Summing over the number of ways of picking the first pairing that we fixed gives the factor $|P_{2k}|=(2k-1)!!$.
After using Stirling's approximation, this contribution gives rise to the second term in \eqref{<z^2k> exact ...}.

We learn from \eqref{<z^2k> exact ...} that, in the strict large $N$ limit, $z$ approaches a real Gaussian variable with zero mean and variance given by \eqref{large n <z^2>}, since $(2k-1)!!$ are the moments of the Gaussian distribution. On the other hand, for any large but finite $N$, the structure of the $1/N$ corrections in \eqref{<z^2k> exact ...} indicate that the distribution of $z$ is no longer Gaussian. The deviation of the moments becomes significant for sufficiently large $k$, specifically when $k$ approaches a fixed fraction of $N^{q/2-1}$ \cite{Saad:2021rcu}. The analogous $1/N$ correction in regular SYK has a richer structure and is discussed in more detail in section \ref{sec:1/N}. The formula \eqref{<z^2k> exact ...} also hints that the $q=2$ model is qualitatively different since the first $1/N$ correction becomes of order one. This is indeed the case and the $q=2$ model needs to be treated separately. We will not discuss this model here as it has been discussed in detail in \cite{Saad:2021rcu}.

\subsection{$\langle z^{2k} \rangle$ from saddle points}

As before, the leading behaviour \eqref{<z^2k> leading} of $\langle z^{2k} \rangle$ for large $N$ can also be obtained from the saddle point approximation. There are $(2k-1)!! \times q^k$ particularly simple saddle points which represent pairings between the $2k$ \textit{replicas}. Concretely, to each pairing $\pi \in P_{2k}$ of the $2k$ replicas set $G_{ab}=\Sigma_{ab}=0$ if $\{a,b\} \notin \pi$. One can check that this is a consistent ansatz. The saddle point equations then reduce to $k$ decoupled equations
\be \label{k decoupled}
G_{ab}=\frac{1}{\Sigma_{ab}},\qquad \Sigma_{ab}=G_{ab}^{q-1},
\ee
for the remaining variables $G_{ab}$ and $\Sigma_{ab}$ with $\{a,b\} \in \pi$. This is nothing but $k$ copies of the saddle point equation \eqref{<z^2> eom} for $\langle z^2 \rangle$. For each pair $\{a,b\} \in \pi$ there are $q$ solutions. Therefore, for each pairing $\pi \in P_{2k}$ there are $q^k$ solutions. Since there are $(2k-1)!!$ possible pairings, the total number of saddle points is $(2k-1)!! \times q^k$. Including the one loop determinant---whose computation is described in detail in the next section---and summing over the saddle points reproduces the large $N$ asymptotic form \eqref{<z^2k> leading} of the exact result.

Another way to understand the factor $(2k-1)!!$ is the following. The collective field representation \eqref{<z^2k>} of $\langle z^{2k} \rangle$ has a discrete symmetry $S_{2k}$ which acts as $G_{ab} \to G_{\sigma(a) \sigma(b)}$ and $\Sigma_{ab} \to \Sigma_{\sigma(a) \sigma(b)}$, permuting the $2k$ replicas. The factor $(2k-1)!!$ then corresponds to the size of the orbit of a single saddle point, corresponding to a pairing $\pi$, under the action of $S_{2k}$. This space is the quotient $S_{2k}/(S_k\times S_2^k)$, where $S_k \times S_2^k$ is the subgroup of $S_{2k}$ which leaves the saddle point fixed. Here $S_k$ is the subgroup that permutes the pairs in the pairing $\pi$, and $S_2^k$ is the subgroup that exchanges the saddle point with its transpose. The latter action automatically leaves the saddle point fixed as $G$ and $\Sigma$ are constrained to be antisymmetric.

\subsection{$1/N$ corrections}
\label{subsec:0d_SYK_1/N}

We begin this section by explaining how the one loop determinant \say{factorises} over the pairs before showing how the first $1/N$ correction to $\langle z^{2k} \rangle/\langle z^2 \rangle^k$ \eqref{<z^2k> exact ...} can be computed by including perturbative fluctuations around the saddle points. After expanding in small fluctuations
\be
G_{ab} = \mathsf{G}_{ab} + \delta G_{ab},\qquad \Sigma_{ab} = \mathsf{\Sigma}_{ab} + \delta \Sigma_{ab},
\ee
around one of the $(2k-1)!! \times q^k$ saddle points configurations $\mathsf{G}_{ab}$ and $\mathsf{\Sigma}_{ab}$, the action decomposes as $I=I_\text{cl}+\delta I$, where $I_\text{cl}$ is the classical action, and
\be \label{1tp delta I}
\frac{1}{N} \delta I = \sum_{n \geq 2} \frac{(-1)^n}{2n} \Tr [(\mathsf{G} \delta \Sigma)^n] + \sum_{a<b} \left[ \delta \Sigma_{ab} \delta G_{ab} - \frac{1}{q} \sum_{n=2}^q \binom{q}{n} \mathsf{G}_{ab}^{q-n} \delta G_{ab}^n \right].
\ee
In obtaining this expression we used the saddle point equations \eqref{k decoupled}. First, we consider the one loop contribution, which amounts to keeping only the quadratic part of the action
\be \label{1tp quadratic}
\frac{1}{N} \delta I^{(2)} = \frac{1}{4} \Tr [(\mathsf{G} \delta \Sigma)^2] + \sum_{a<b} \left[ \delta \Sigma_{ab} \delta G_{ab} - \frac{q-1}{2} \mathsf{G}_{ab}^{q-2} \delta G_{ab}^2 \right].
\ee
Any $2k \times 2k$ antisymmetric matrix $M$ can be decomposed as $M = M_\parallel + M_\perp$, where $M_\parallel$ lies in the $k$-dimensional subspace spanned by $2k \times 2k$ antisymmetric matrices of the form
\be \label{1tp block}
\begin{blockarray}{cccccccc}
\textcolor{gray}{\text{\small{$L_1$}}} & \textcolor{gray}{\text{\small{$R_1$}}} & \textcolor{gray}{\text{\small{$L_2$}}} & \textcolor{gray}{\text{\small{$R_2$}}} & \dots & \textcolor{gray}{\text{\small{$L_k$}}} & \textcolor{gray}{\text{\small{$R_k$}}} & \\
\begin{block}{(ccccccc)c}
0 & * & & & & & & \textcolor{gray}{\text{\small{$L_1$}}}\vspace{0.4em}\\
* & 0 & & & & & & \textcolor{gray}{\text{\small{$R_1$}}}\vspace{0.4em}\\
& & 0 & * & & & & \textcolor{gray}{\text{\small{$L_2$}}}\vspace{0.4em}\\
& & * & 0 & & & & \textcolor{gray}{\text{\small{$R_2$}}}\vspace{0.4em}\\
& & & & \ddots & & & \vdots\vspace{0.4em}\\
& & & & & 0 & * & \textcolor{gray}{\text{\small{$L_k$}}}\vspace{0.4em}\\
& & & & & * & 0 & \textcolor{gray}{\text{\small{$R_k$}}}\vspace{0.4em}\\
\end{block}
\end{blockarray}
\ee
and $M_\perp$ lies in its orthocomplement. If $\mathsf{G}$ and $\mathsf{\Sigma}$ represent the saddle point which takes the block diagonal form \eqref{1tp block} then $\mathsf{G}=\mathsf{G}_\parallel$ and $\mathsf{\Sigma}=\mathsf{\Sigma}_\parallel$. Equipped with this notation, the quadratic part of the action \eqref{1tp quadratic} can be written as
\be
\frac{1}{N} \delta I^{(2)} = \frac{1}{4} \Tr [(\mathsf{G} \delta \Sigma)^2] + \sum_{a<b} \left[ (\delta \Sigma_\parallel)_{ab} (\delta G_\parallel)_{ab} + (\delta \Sigma_\perp)_{ab} (\delta G_\perp)_{ab} - \frac{q-1}{2} \mathsf{G}_{ab}^{q-2} (\delta G_\parallel)_{ab}^2 \right].
\label{eq:0d_ddeltaI2}
\ee
Notice that $\delta G_\perp$ appears linearly and so integrating it out gives a delta function which sets $\delta \Sigma_\perp=0$. In other words, the fluctuations $\delta G_\perp$ and $\delta \Sigma_\perp$ don't contribute to the one loop determinant. This leaves us with an integral over the remaining variables $\delta G_\parallel$ and $\delta \Sigma_\parallel$ with quadratic action
\be
\frac{N}{4} \Tr [(\mathsf{G} \delta \Sigma_\parallel)^2] + N \sum_{a<b} \left[ (\delta \Sigma_\parallel)_{ab} (\delta G_\parallel)_{ab} - \frac{q-1}{2} \mathsf{G}_{ab}^{q-2} (\delta G_\parallel)_{ab}^2 \right],
\ee
which factorises into a sum over the $k$ blocks in \eqref{1tp block}, since the variables $\delta G_\parallel$ and $\delta \Sigma_\parallel$ are block diagonal. This implies that the one loop contribution from expanding in fluctuations around one of the $(2k-1)!! \times q^k$ saddle points for $\langle z^{2k} \rangle$ is given by the $k$\textsuperscript{th} power of the one loop contribution for $\langle z^{2} \rangle$. That is, for large $N$,
\be
\frac{\langle z^{2k} \rangle}{\langle z^2 \rangle^k} \sim (2k-1)!!.
\ee
This formula admits a $1/N$ expansion which can be obtained by separately considering the perturbative expansions around each of the saddle points for $\langle z^{2k} \rangle$ and $\langle z^2 \rangle$. However, since we are interested in the ratio $\langle z^{2k} \rangle / \langle z^2 \rangle^k$, the only perturbative corrections which contribute arise from the fluctuations $\delta G_\perp$ and $\delta \Sigma_\perp$. These corrections can be computed using the Feynman rules for the action $\delta I$ \eqref{1tp delta I} together with the free propagators
\be \label{1tp free prop}
\ba
\langle \delta G_{ab} \delta G_{cd} \rangle &= \frac{1}{N} \left[ \mathsf{G}_{ac} \mathsf{G}_{bd} - \mathsf{G}_{ad} \mathsf{G}_{bc} -(1+1/q) \mathsf{G}_{ab}^2 \delta_{ac} \delta_{bd} \right],\\
\langle \delta G_{ab} \delta \Sigma_{cd} \rangle &= \frac{1}{N} \left[ \delta_{ac} \delta_{bd} - (1-1/q) \Pi_{ab,cd} \right],\\
\langle \delta \Sigma_{ab} \delta \Sigma_{cd} \rangle &= - \frac{1}{N} (1-1/q) \mathsf{G}_{ab}^{q-2} \delta_{ac} \delta_{bd},
\ea
\ee
obtained from the quadratic part of the action \eqref{1tp quadratic}. In this expression the angular brackets represent an average weighted by the quadratic action \eqref{1tp quadratic} (and normalised so that $\langle 1 \rangle = 1$) and $\Pi$ is the projector onto the subspace of $2k \times 2k$ antisymmetric matrices of the form \eqref{1tp block}. Crucially,
\be
\langle \delta \Sigma_\perp \delta \Sigma_\parallel \rangle = \langle \delta \Sigma_\perp \delta \Sigma_\perp \rangle = \langle \delta \Sigma_\perp \delta G_\parallel \rangle = 0,
\ee
which implies that the fluctuations arising from $\delta G_\perp$ and $\delta \Sigma_\perp$ arise from Feynman diagrams with at least one vertex corresponding to a $\delta G_\perp$ interaction term. By inspecting $\delta I$ \eqref{1tp delta I} we find that there is only one $\delta G_\perp$ interaction term which contributes to the action:
\be
\delta I \supset -\frac{N}{q} \sum_{a<b} (\delta G_{\perp})_{ab}^q.
\ee
The upshot is that the $1/N$ expansion of $\langle z^{2k} \rangle / \langle z^2 \rangle^k$ can be obtained by considering all Feynman diagrams involving at least one $\delta G_\perp^q$ vertex. Since each diagram contributes as $N^{V-P}$, where $V$ is the number of vertices and $P$ is the number of propagators, naively, the leading correction arises from a diagrams with a single $\delta G_\perp^q$ vertex. However, these contributions vanish,
\be
\frac{N}{q} \sum_{a<b} \langle (\delta G_\perp)_{ab}^q \rangle = \frac{q! N}{q^{q/2+1}} \sum_{a < b} (\mathsf{G}_\perp)_{ab}^q = 0.
\ee
This follows since \eqref{1tp free prop} implies that a Wick contraction between $(\delta G_\perp)_{ab}$ and itself is proportional to $(\mathsf{G}_\perp)_{ab}=0$. Another possibility is to have a diagram consisting of one $\delta G_\perp^q$ vertex and one $\delta \Sigma^r$ vertex. These diagrams contribute as
\be
(-1)^{r+1} \frac{N^2}{2qr}\sum_{a < b} \langle \Tr [ (\delta G_\perp)_{ab}^q (\mathsf{G} \delta \Sigma)^r] \rangle.
\ee
Since a Wick contraction between $(\delta G_\perp)_{ab}$ and itself vanishes, we need to contract each $(\delta G_\perp)_{ab}$ with a $\delta \Sigma$ to get a nonzero result, which requires $r \geq q$. Naively, then, the leading contribution has $r=q$, since these diagrams contribute at order $N^{2-(q+r)/2}$. However, using \eqref{1tp free prop}, we find that this contribution also vanishes,
\be
- \frac{q!}{2q^2 N^{q-2}}\sum_{a < b} (\mathsf{G}_\perp)_{ab}^q=0.
\ee
Finally, let's consider diagrams consisting of one $\delta G_\perp^q$ vertex and one $\delta G^r$ vertex. By a similar argument, to get a nonzero result requires $r \geq q$. The leading contribution has $r=q$, and contributes as
\be
\ba
\frac{N^2}{2q^2} \sum_{\substack{a<b \\ c<d}} \langle (\delta G_\perp)_{ab}^q \delta G_{cd}^q \rangle &= \frac{q!}{2q^2 N^{q-2}} \sum_{\substack{a<b \\ c<d}} \sum_{a'<b'} \Pi^\perp_{ab,a'b'} (\mathsf{G}_{a'c} \mathsf{G}_{b'd} - \mathsf{G}_{a'd} \mathsf{G}_{b'c})^q\\
&= \frac{q!}{2q^2 N^{q-2}} \Tr \, \Pi^\perp\\
&= k(k-1)\frac{q!}{q^2 N^{q-2}},
\label{eq:just_correction_0d}
\ea
\ee
where we used the fact that the nonzero entries of $\mathsf{G}_{ab}^q$ are equal to 1. Notice that this precisely reproduces the leading correction \eqref{<z^2k> exact ...} obtained from the exact formula \eqref{<z^2k> pairings}. In regular SYK, there is no analogue of the exact formula \eqref{<z^2k> pairings} for $\langle |Z(\ii T)|^{2k} \rangle / \langle |Z(\ii T)|^2 \rangle^k$, so a direct proof identifying the leading term in the perturbative expansion is needed. This is done in appendix \ref{app:perturbative_exp}. Although it is not required here, it is possible to adapt the argument given there to the case at hand.

\section{The SYK model} \label{sec:syk}

In this section we consider the moments of the spectral form factor in the SYK model. The SYK model is a quantum mechanical model of $N$ Majorana fermions $\psi_1,\dots,\psi_N$ with random couplings. The Hamiltonian is given by
\be \label{HSYK}
H = \ii^{q/2} \sum_{1 \leq i_1 < \dots < i_q \leq N} J_{i_1 \dots i_q} \psi_{i_1} \dots \psi_{i_q}, \qquad \left\{ \psi_i, \psi_j \right\} = \delta_{ij}.
\ee
The independent components of the antisymmetric tensor of couplings $J_{i_1 \dots i_q}$ are drawn from a Gaussian distribution with zero mean and variance
\be \label{J variance}
\langle J_{i_1 \dots i_q} J_{j_1 \dots j_q} \rangle = \frac{J^2 (q-1)!}{N^{q-1}} \delta_{i_1 j_1} \dots \delta_{i_q j_q}.
\ee
Throughout we assume both $N$ and $q$ are even. Since $N$ is even, the fermion algebra \eqref{HSYK} has an irreducible representation in a Hilbert space of dimension $L=2^{N/2}$.

We start by reviewing the collective field representation for the moments of the spectral form factor and its symmetries. Next, we review the result of Saad, Shenker, and Stanford (SSS) in \cite{Saad:2018bqo}, which traces the origin of the linear ramp in the spectral form factor to a compact zero mode in the collective field description. Drawing on this, we identify saddle points for the moments of the spectral form factor and show how this gives rise to a result consistent with expectations from RMT. The ensembles of random matrices relevant to SYK and the behaviour of the moments of the spectral form factor in each of these ensembles is reviewed in appendix \ref{app:rmt}. Finally, we end this section by computing the leading perturbative in $1/N$ correction to the moments of the spectral form factor which indicate a deviation from RMT expectations for the very high moments.

\subsection{Collective fields}

The spectral form factor may be represented as a path integral
\be
|Z(\ii T)|^2 = \int D\psi \exp\left\{\ii \int_0^T \dd t \left[ \frac{\ii}{2} \psi_i^a \partial_t \psi_i^a-J_{i_1\dots i_q}\left(\ii^{\frac{q}{2}}\psi_{i_1}^L \dots \psi_{i_q}^L-(-\ii)^{\frac{q}{2}}\psi_{i_1}^R \dots \psi_{i_q}^R\right)\right]\right\},
\ee
with antiperiodic boundary conditions for the fermions, $\psi^a_i(T) = - \psi^a_i(0)$, around the real time circle parameterised by $t \in [0,T]$. In this expression $a \in \{ L, R\}$ is a replica index and the integral over $\psi^L$ computes $\Tr[e^{-\ii T H}]$ while the integral over $\psi^R$ computes $\Tr[e^{\ii T H}]$. The kinetic term involves an implicit sum over both $i$ and $a$ while the interaction term involves an implicit sum over $i_1 < \dots < i_q$.

The $k$\ts{th} moment of the spectral form factor may be represented as a path integral over bilocal \textit{collective fields} $G$ and $\Sigma$,
\be \label{moments collective fields}
\ba
{}&\langle |Z(\ii T)|^{2k} \rangle = \int D G D \Sigma \, e^{-I[G, \Sigma]},\\
{}&\frac{1}{N} \, I[G, \Sigma] = - \frac{1}{2} \log \det(\partial_t - \Sigma) + \frac{1}{2} \int_0^T\int_0^T \dd t \dd t' \left[ \Sigma_{ab}(t,t') G_{ab}(t,t') - \frac{J^2}{q} s_{ab} G_{ab}(t,t')^q \right],\\
{}&s_{L_\#L_\#} = s_{R_\#R_\#} = - 1, \qquad s_{L_\#R_\#} = s_{R_\#L_\#} = \ii^q.
\ea
\ee
Here $\Sigma_{ab}(t,t')$ was originally introduced as a Lagrange multiplier enforcing the relation
\be
G_{ab}(t,t') = \frac{1}{N} \sum_{i=1}^N \psi_i^a(t) \psi_i^b(t').
\ee
The indices $a,b \in \{ L_1, R_1, \dots L_k, R_k \}$ label the $L$ and $R$ replicas, which were each introduced to represent the factors of $\Tr[e^{-\ii T H}]$ and $\Tr[e^{\ii T H}]$ in the spectral form factor. The collective fields $G, \Sigma$ are antiperiodic with period $T$ in each argument, a property inherited from the antiperiodic boundary conditions on the fermions. Also, not all components of $G$ and $\Sigma$ are independent since $G_{ab}(t,t')=-G_{ba}(t',t)$ and $\Sigma_{ab}(t,t')=-\Sigma_{ba}(t',t)$, which follows from the Grassmann nature of the fermions. Finally, in the action \eqref{moments collective fields}, the determinant represents the ordinary determinant in replica space as well as the functional determinant in time space, and the sum over $a$ and $b$ is left implicit.

The action \eqref{moments collective fields} has a $\mathrm{U}(1)^{2k}$ time translation symmetry which acts independently on each of the $L$ and $R$ systems. In addition, there is a discrete replica symmetry
\be \label{replica sym!}
G_{ab} \to G_{\sigma(a)\sigma(b)}, \qquad \Sigma_{ab} \to \Sigma_{\sigma(a)\sigma(b)}.
\ee
For $q=0$ mod 4, $\sigma \in S_k^L \times S_k^R$ permutes the $L$ and $R$ replicas separately, whereas, for $q=2$ mod 4, $\sigma \in S_{2k}$ can also permute the $L$ and $R$ replicas among each other. The dependence of the replica symmetry on $q$ mod 4 stems from the fact that exchanging an $L$ and an $R$ replica is only a symmetry if $Z(\ii T)$ and $Z(-\ii T)$ are the same variables, or, in other words, if the energy spectrum is symmetric about zero. This only occurs for the model with $q=2$ mod 4, where the Hamiltonian anticommutes with a time reversal operator, as briefly reviewed in appendix \ref{app:rmt syk}. The signature of this in the collective field description \eqref{moments collective fields} lies in the dependence of $s_{ab}$ on $q$ mod 4.

Assuming that each of the saddle point configurations only depend on the difference of times, the saddle point equations can be written as
\be \label{saddle all q}
\ba
G(\omega_n) &= - \left( \ii \omega_n + \Sigma(\omega_n) \right)^{-1},\\
\Sigma_{ab}(t) &= s_{ab} J^2 G_{ab}(t)^{q-1}.
\ea
\ee
Such configurations automatically preserve a diagonal $\mathrm{U}(1)^k \subset \mathrm{U}(1)^{2k}$ symmetry. In the first equation $G(\omega_n)$ and $\Sigma(\omega_n)$ are the Fourier modes of the collective fields and the frequencies are fermionic Matsubara frequencies due to the antiperiodic boundary conditions on the fermions.

%We can relate $G_{ab}(t,t')$ more explicitly to the original fermionic fields by writing, for example,
%\begin{align}
   % \int D G D \Sigma \, e^{-I[G, \Sigma]} G_{L_\# L_\#}(t,t') =& \frac{1}{N} \sum_{i=1}^N \langle \Tr \left[ e^{-\ii T H} \mathcal{T} \psi_i(t) \psi_i(t')  \right] \Tr[e^{-\ii T H}]^{k-1} \Tr[e^{\ii T H}]^{k}\rangle \, ,\nonumber \\
    %\int D G D \Sigma \, e^{-I[G, \Sigma]} G_{R_\# R_\#}(t,t') =& \frac{1}{N} \sum_{i=1}^N \langle %\Tr \left[ e^{\ii T H} \tilde{\mathcal{T}} \psi_i(t) \psi_i(t')  \right] \Tr[e^{-\ii T H}]^{k} \Tr[e^{\ii T H}]^{k-1}\rangle \, ,     \label{eq:G_integrated} \\
   % \int D G D \Sigma \, e^{-I[G, \Sigma]} G_{L_\# R_\#}(t,t') =& \frac{1}{N} \sum_{i=1}^N \langle \Tr \left[ e^{-\ii T H} \psi_i(t)\right]\Tr \left[ e^{\ii T H} \psi_i(t')  \right] |\Tr[e^{-\ii T H}]|^{2k-2}\rangle \, , \nonumber
%\end{align}
%where $\mathcal{T}$ ($\tilde{\mathcal{T}}$) is the (anti-)time ordering operator. The saddle point solutions of \eqref{saddle all q} does not necessarily have to share all the symmetries of the integrated version above, since some of the can be spontaneously broken. However, we will discuss in the following that they do share some of this symmetries. For example, we already discussed the time-translation invariance of \eqref{saddle all q}. Another relevant property of     \eqref{eq:G_integrated} that we will find in the saddle point solution is that
%\begin{equation}
   % G_{L_\# L_\#}(t) = G_{R_\# R_\#}(t)^* \, .
  %  \label{eq:G_LL_G_RR_relation}
%\end{equation}

\subsection{The spectral form factor} \label{sec:sff}

We now review the saddle points that describe the slope and the linear ramp in the average spectral form factor $\langle |Z(\ii T)|^2 \rangle$ in SYK. The saddle points describing the linear ramp were first identified by Saad, Shenker, and Stanford (SSS) \cite{Saad:2018bqo}.

Let us first consider the saddle point that describe the slope. This saddle points correspond to a disconnected contribution to the average spectral form factor and is identified by assuming a replica diagonal ansatz in which the off-diagonal components of the collective fields vanish, i.e. $G_{LR}=\Sigma_{LR}=0$. Under this ansatz, the saddle point equations \eqref{saddle all q} decouple into separate equations for the $L$ and $R$ systems. These equations simply compute the disconnected contribution
\be
\langle |Z(\ii T)|^2 \rangle \supset |\langle Z(\ii T) \rangle|^2,
\ee
where the inclusion symbol indicates that this is only one of the possible saddle points that can contribute. This contribution oscillates with a decaying envelope, tending to zero at late times, and describes the slope.

We now consider the continuous family of saddle points, identified by SSS, that describe the linear ramp and can be understood semi-analytically. These saddle points correspond to a connected contribution to the average spectral form factor and have nonzero off-diagonal components. An immediate consequence of this is that these configurations spontaneously break the relative time translation symmetry. Hence there is a continuous family of saddle points, generated by acting on the saddle point solution with the generator of the broken $\mathrm{U}(1)$ symmetry. Concretely, this symmetry acts on the saddle points by leaving $G_{LL}$ and $G_{RR}$ unchanged but shifting $G_{LR}(t) \to G_{LR}(t-\Delta)$. There is a similar action on $\Sigma$. Since the collective fields are antiperiodic with period $T$, the relative time shift parameter $\Delta$ is valued on a circle of circumference $2T$. The origin of the linear ramp is simply the volume of the orbit of the saddle point under the action of the broken symmetry which is $2T$. In a holographic interpretation, a nonzero correlation between the $L$ and $R$ replicas, signaled by a nonzero value for $G_{LR}$, corresponds to a wormhole configuration, in this case the double cone \cite{Saad:2018bqo}, connecting the $L$ and $R$ conformal boundaries.

As argued by SSS, there is another free parameter which labels these saddle points and can be understood by considering the following auxiliary problem. The idea is to think about computing the thermal partition function
\be
Z(\beta_\text{aux}) = \Tr [ e^{- \frac{\beta_\text{aux}}{2} H} e^{-\ii T H} e^{- \frac{\beta_\text{aux}}{2} H}  e^{\ii T H}],
\ee
by a path integral on an elaborate Schwinger-Keldysh contour. In this problem the field configurations on the two Lorentzian parts of the contour are joined to each other along the Euclidean parts of the contour. In contrast, the original problem of computing $\Tr[e^{-\ii T H}] \Tr[e^{\ii T H}]$, requires separately periodically identifying the two Lorentzian contours. However, for large $T$, the saddle point configurations for the Lorentzian parts of the contour of the auxiliary problem are approximately same as that for the original problem. This suggests we can construct an approximate solution for the original problem by taking fermion correlators in the thermofield double state,
\be
\label{eq:G_aux_def}
G_{ab}^{(\beta_\text{aux})}(t) = \frac{1}{N} \sum_{i=1}^N \bra{\Psi_\text{TFD}} \psi_i^a(t) \psi_i^b(0) \ket{\Psi_\text{TFD}},
\ee
and summing over images
\be \label{G_ab}
G_{ab}(t) = \sum_{n=-\infty}^\infty (-1)^n G^{(\beta_\text{aux})}_{ab}(t+n T).
\ee
More precisely, SSS showed that for large $T$ this solution lies close to an exact solution, in the sense that the saddle point equations can be solved using numerical iteration with \eqref{G_ab} as a starting point. Although also $G^{(\beta_\text{aux})}_{ab}$ cannot generally be described analytically, in the limit of large $\beta_\text{aux}$, the SYK model is dominated by a soft mode governed by the Schwarzian theory. In this limit \cite{Saad:2018bqo},
\be
\ba
G_{LL}^{(\beta_\text{aux})}(t) &= G_{RR}^{(\beta_\text{aux})}(t) = b \left[\frac{\beta_\text{aux}}{\pi} \sinh \left(  \frac{\pi t}{ \beta_\text{aux}} \right) \right]^{-\frac{2}{q}} \text{sign} (t), \\
G_{LR}^{(\beta_\text{aux})}(t) &= \ii b \left[\frac{\beta_\text{aux}}{\pi} \cosh \left(  \frac{\pi (t-\Delta)}{ \beta_\text{aux}} \right) \right]^{-\frac{2}{q}},
\ea
\ee
where $b$ is a constant determined by $J^2 b^q \pi \cot \frac{\pi}{q} = \frac{1}{2}-\frac{1}{q}$. To justify \eqref{G_ab}, it is crucial that $G_{ab}^{(\beta_\text{aux})}(t)$ decays exponentially as $t \to \infty$ \cite{Saad:2018bqo, Winer:2022ciz}. Since the conformal solution sets the scale for exponential decay to $\beta_\text{aux}$, \eqref{G_ab} is only accurate for $\beta_\text{aux} \ll T$. For $\beta_\text{aux} \gtrsim T$ the solution obtained by numerical iteration can differ significantly from \eqref{G_ab}.

The upshot of studying this auxiliary problem is that it makes it clear that there is another zero mode, namely $\beta_\text{aux}$. Also, since the classical action for the two Lorentzian parts of the contour in the auxiliary problem ought to cancel out, it suggests the action for these saddle points is zero. A more careful argument shows the action is zero up to exponentially small correction in $T$ \cite{Saad:2018bqo}. Hence, these saddle points contribute as
\be
\langle |Z(\ii T)|^2 \rangle \supset \int_0^\infty \dd \beta_\text{aux} \, \mu(\beta_\text{aux}) \int_0^{2T} \dd \Delta,
\ee
where $\mu(\beta_\text{aux})$ is the one loop determinant for the nonzero modes. For large $T$, it turns out that the measure $\mu(\beta_\text{aux}) \, \dd \beta_\text{aux}$ becomes a flat measure in terms of the energy. This gives
\be \label{ziT^2}
\langle |Z(\ii T)|^2 \rangle \supset 2T \int \frac{\dd E_\text{aux}}{2\pi} \times  \begin{cases}
2 & \text{for $q=0$ mod 4},\\
1 & \text{for $q=2$ mod 4}.
\end{cases}
\ee
The model with $q=0$ mod 4 has a time reversal symmetry which leads to an extra factor of two since we need to sum over the saddle point and its time reversal conjugate.

\subsection{General $k$} \label{sec:general k}

There is a simple class of saddle points that contribute to the moments of the spectral form factor and can be described in terms of pairings of the replicas. These saddle points were also described for a spin glass model in \cite{Winer:2022ciz}. Consider a configuration where $G$ and $\Sigma$ are block diagonal
\be \label{block matrix 2}
\begin{blockarray}{cccccccc}
\textcolor{gray}{\text{\small{$L_1$}}} & \textcolor{gray}{\text{\small{$R_1$}}} & \textcolor{gray}{\text{\small{$L_2$}}} & \textcolor{gray}{\text{\small{$R_2$}}} & \dots & \textcolor{gray}{\text{\small{$L_k$}}} & \textcolor{gray}{\text{\small{$R_k$}}} & \\
\begin{block}{(ccccccc)c}
* & * & & & & & & \textcolor{gray}{\text{\small{$L_1$}}}\vspace{0.4em}\\
* & * & & & & & & \textcolor{gray}{\text{\small{$R_1$}}}\vspace{0.4em}\\
& & * & * & & & & \textcolor{gray}{\text{\small{$L_2$}}}\vspace{0.4em}\\
& & * & * & & & & \textcolor{gray}{\text{\small{$R_2$}}}\vspace{0.4em}\\
& & & & \ddots & & & \vdots\vspace{0.4em}\\
& & & & & * & * & \textcolor{gray}{\text{\small{$L_k$}}}\vspace{0.4em}\\
& & & & & * & * & \textcolor{gray}{\text{\small{$R_k$}}}\vspace{0.4em}\\
\end{block}
\end{blockarray}
\ee
This corresponds to the pairing $\{ \{L_1,R_1\}, \dots \{L_k,R_k\} \}$. Under this ansatz, the saddle point equations \eqref{saddle all q} decouple into $k$ copies of the saddle point equations for $\langle |Z(\ii T)|^2 \rangle$. As seen in the previous section, there are two kinds of saddle points for each block: a \textit{disconnected} solution and a \textit{connected} solution. Here the disconnected solution refers to the saddle point where $G$ and $\Sigma$ are replica diagonal, while the connected solution refers to the saddle point \eqref{G_ab} which correlates the $L$ and $R$ replicas in pairs. For early times the dominant solution is given by taking all the blocks to be the disconnected solution and gives rise to the fully disconnected contribution
\be \label{disc!}
\langle |Z(\ii T)|^{2k} \rangle \supset |\langle Z(\ii T) \rangle|^{2k}.
\ee
For later times the dominant solution is given by taking all the blocks to be the connected solution. In addition to breaking $k$ of the relative time translation symmetries, which gives rise to a factor $T^k$, this solution breaks the replica symmetry from $S_k \times S_k$ to the diagonal $S_k$ for $q=0$ mod 4 or from $S_{2k}$ to $S_{k} \times S_2^k$ for $q=2$ mod 4. This gives rise to the contribution
\be \label{k! (2k-1)!!}
\ba
\langle |Z(\ii T)|^{2k} \rangle \supset \langle |Z(\ii T)|^2 \rangle^k \times
\begin{cases}
k! & \text{for $q=0$ mod 4},\\
(2k-1)!! & \text{for $q=2$ mod 4},
\end{cases}
\ea
\ee
with $\langle |Z(\ii T)|^2 \rangle$ given by \eqref{ziT^2}. The combinatorial factors compute the size of the orbit of a single pairing solution under the action of the discrete replica symmetry group. This simply counts the number of pairing solutions. For $q=0$ mod 4, these are pairings between the $L$ and $R$ replicas, while for $q=2$ mod 4, these can include pairings in which the $L$ and $R$ replicas are paired among themselves. In writing \eqref{k! (2k-1)!!} we have assumed that the one loop determinant for $\langle |Z(\ii T)|^{2k} \rangle$ \say{factorises} over the pairings, i.e. it is given by the $k$\ts{th} power of the one loop determinant for $\langle |Z(\ii T)|^2 \rangle$---a fact we verify in the next section.

At early times, \eqref{disc!} shows that the moments are self-averaging. In contrast, at later times \eqref{k! (2k-1)!!} shows that, although $|Z(\ii T)|^2$ is not self-averaging, its distribution is simple: for $q=0$ mod 4 it behaves as the modulus square of a complex Gaussian variable with zero mean and variance \eqref{ziT^2}, while for $q=2$ mod 4 it behaves as the square of real Gaussian variable with zero mean and variance \eqref{ziT^2}.

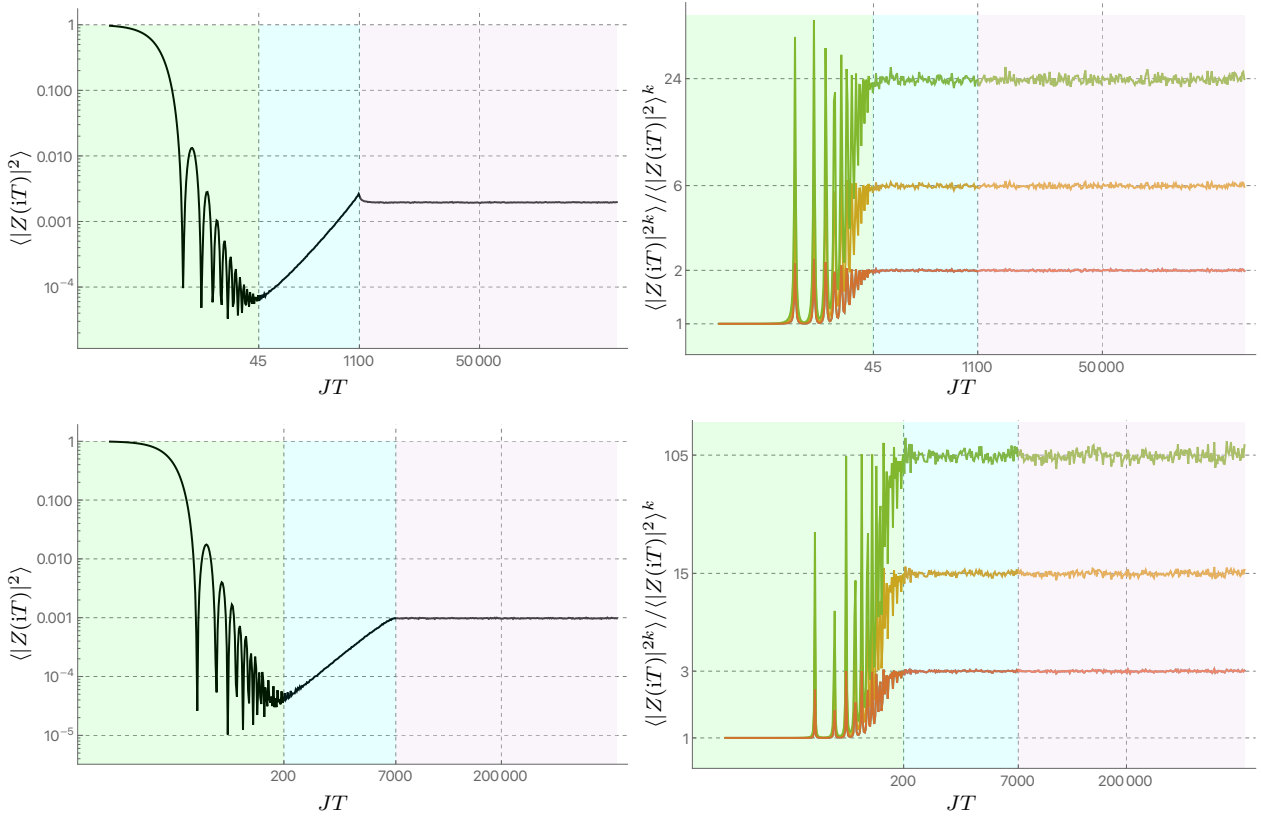
\begin{figure}
\centering
\begin{subfigure}{0.48\textwidth}
\raggedleft
\begin{tikzpicture}
\node at (0,0) {\includegraphics[width=0.95\linewidth]{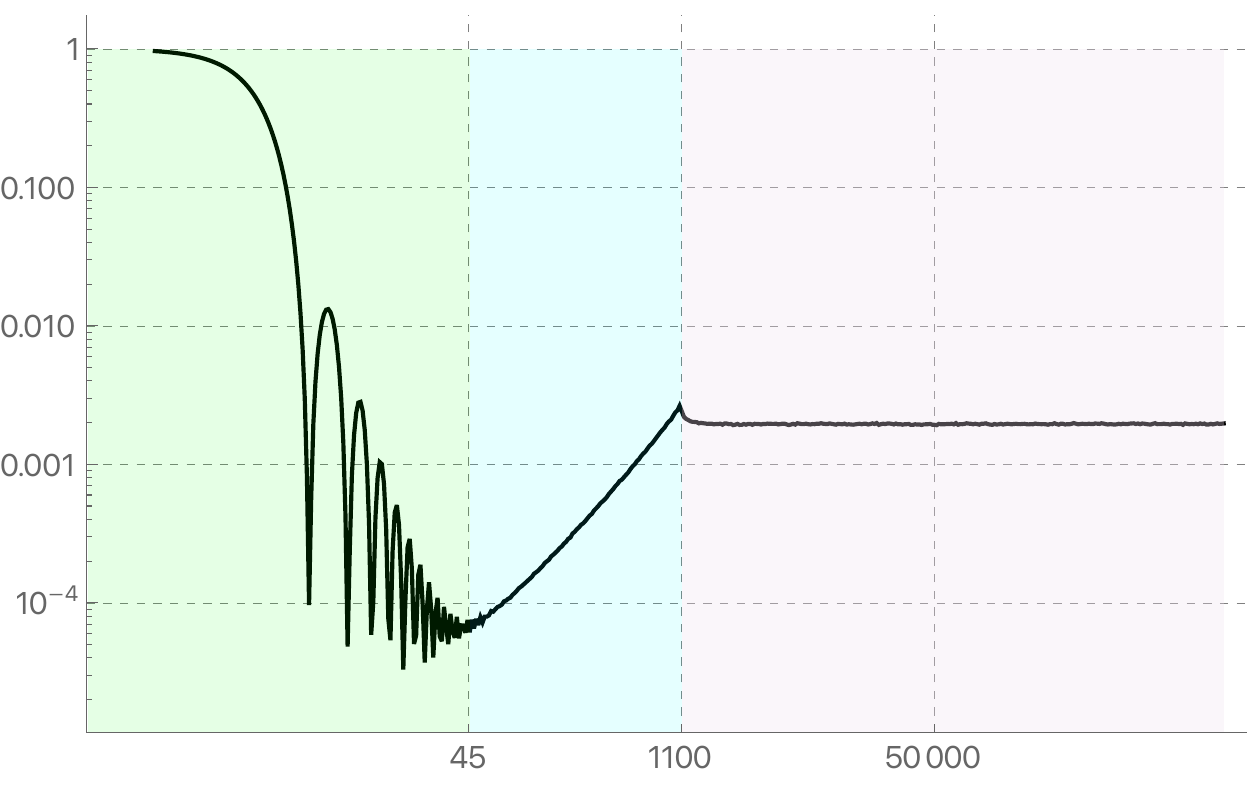}};
\node at (current bounding box.south) {{\scriptsize $JT$}};
\node at ([xshift=-0.2em]current bounding box.west) {\scriptsize \rotatebox{90}{$\langle |Z(\ii T)|^2 \rangle$}};
\end{tikzpicture}
\end{subfigure}
% $\quad$
\begin{subfigure}{0.48\textwidth}
\raggedright
\begin{tikzpicture}
\node at (0,0) {\includegraphics[width=0.95\linewidth]{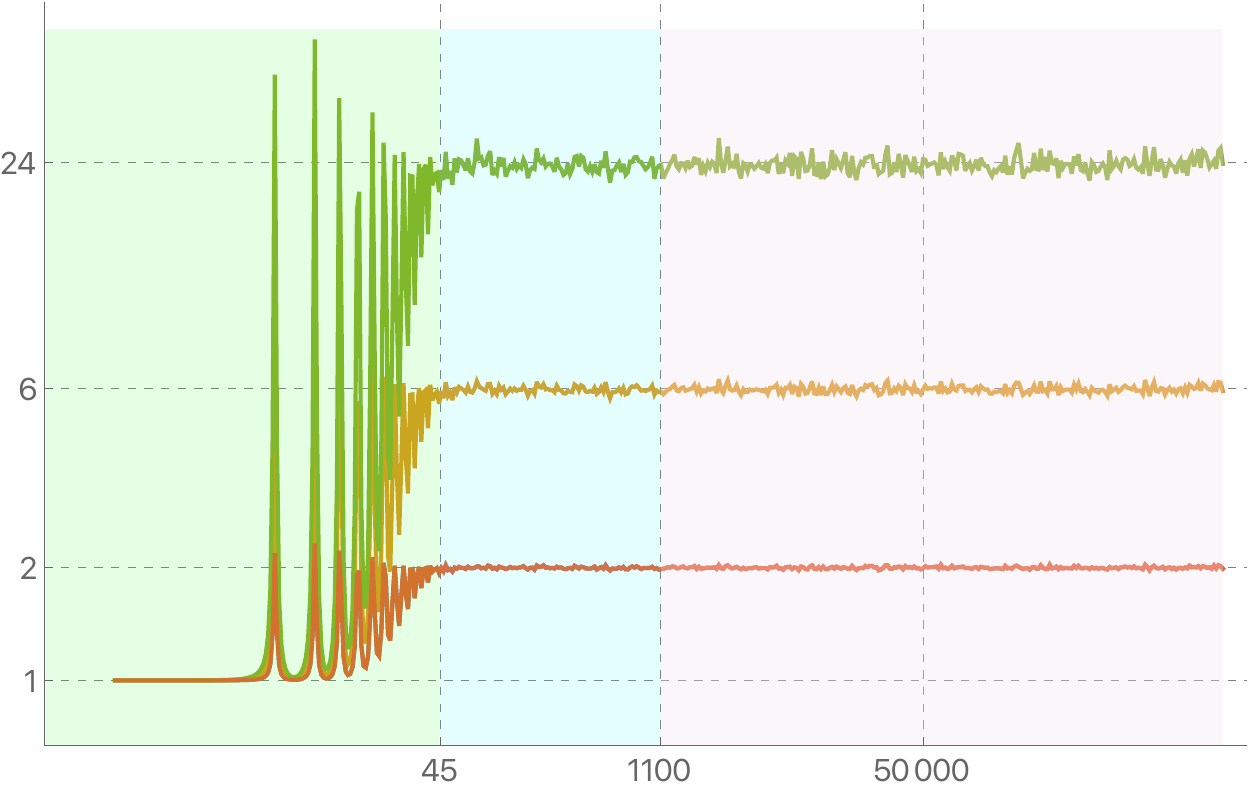}};
\node at (current bounding box.south) {{\scriptsize $JT$}};
\node at (current bounding box.west) {\scriptsize \rotatebox{90}{$\langle |Z(\ii T)|^{2k} \rangle / \langle |Z(\ii T)|^2 \rangle^k$}};
\end{tikzpicture}
\end{subfigure}
\begin{subfigure}{0.48\textwidth}
\raggedleft
\begin{tikzpicture}
\node at (0,0) {\includegraphics[width=0.95\linewidth]{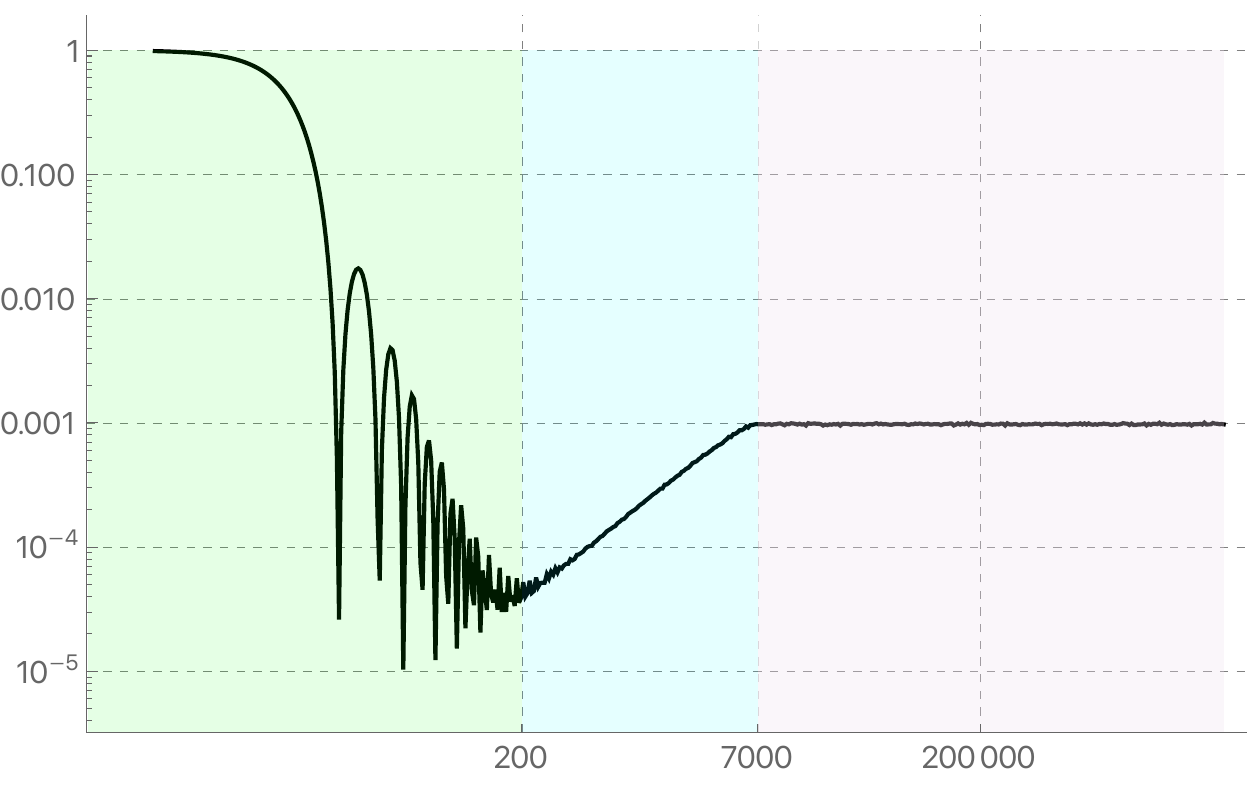}};
\node at (current bounding box.south) {{\scriptsize $JT$}};
\node at ([xshift=-0.2em]current bounding box.west) {\scriptsize \rotatebox{90}{$\langle |Z(\ii T)|^2 \rangle$}};
\end{tikzpicture}
\end{subfigure}
% $\quad$
\begin{subfigure}{0.48\textwidth}
\raggedright
\begin{tikzpicture}
\node at (0,0) {\includegraphics[width=0.95\linewidth]{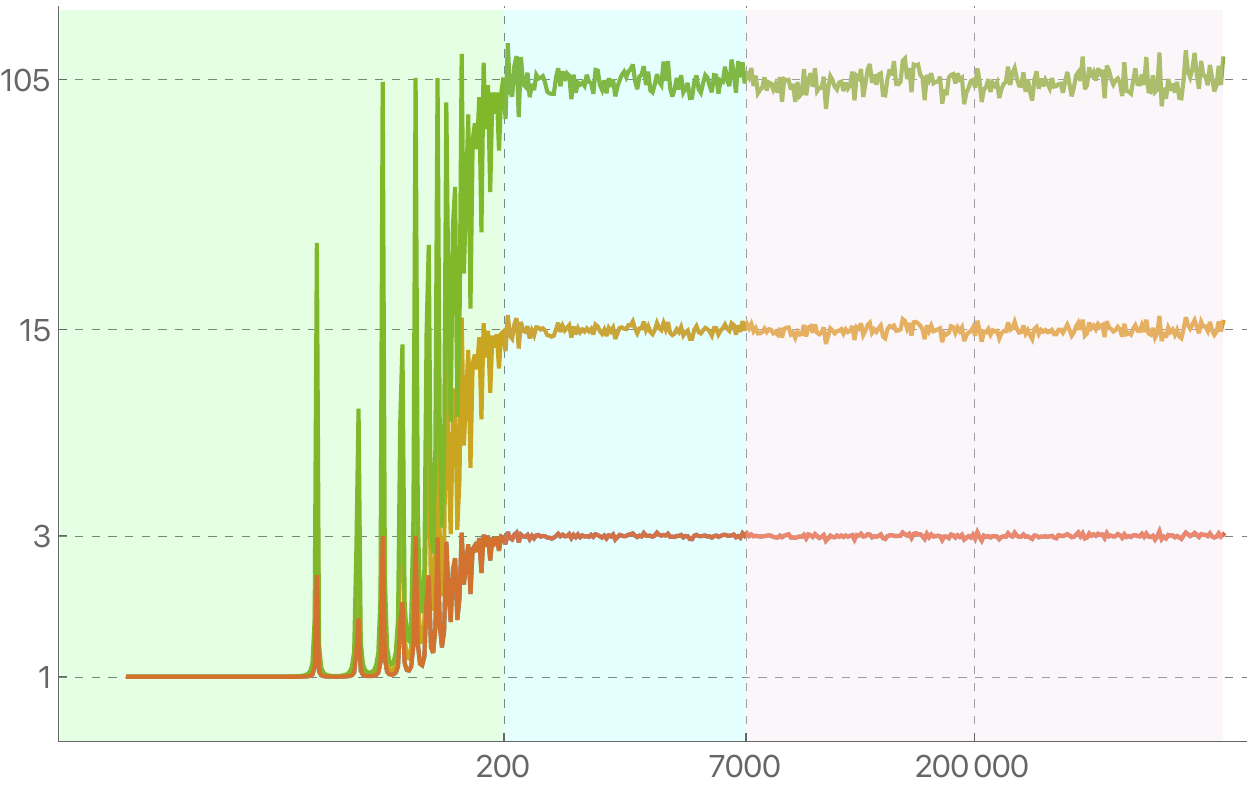}};
\node at (current bounding box.south) {{\scriptsize $JT$}};
\node at (current bounding box.west) {\scriptsize \rotatebox{90}{$\langle |Z(\ii T)|^{2k} \rangle / \langle |Z(\ii T)|^2 \rangle^k$}};
\end{tikzpicture}
\end{subfigure}
\caption{The spectral form factor (divided by $2^N$) and its moments for $q=4$ SYK (top) and $q=6$ SYK (bottom) with $N=20$ and averaged over $20$K realisations. The red, orange, and green curves are $k=2$, $k=3$, and $k=4$, respectively. Initially the spectral form factor is self-averaging. At later times, it fluctuates around the predicted values \eqref{k! (2k-1)!!}: $k!$ for $q=4$ and $(2k-1)!!$ for $q=6$. The oscillations in the moments near the dip arise because the slope contribution oscillates and so can temporarily exchange dominance with the ramp contribution.}
\label{fig:moments46}
\end{figure}

In a holographic interpretation, the saddle points discussed here correspond to wormhole pairings, that is, wormholes which only connect two conformal boundaries. In principle, there could be contributions from wormhole configurations which connect multiple boundaries. In figure \ref{fig:moments46} we compute $\langle |Z(\ii T)|^{2k} \rangle / \langle |Z(\ii T)|^2 \rangle^k$ through exact numerical diagonalisation in SYK for $q=4$ and 6. These results support the idea that other saddle points do not contribute in an important way, consistent with the analysis in SYK with one time point in section \ref{sec:syk one time point}.\footnote{Further numerical evidence is provided in section \ref{sec:sparse_SYK}, where it is shown, for the sparse SYK model, that the leading correction is dominated by fluctuations around the pairwise connected saddle points, discussed in the next section.}

There are also other saddle points where we take $k-\ell$ of the blocks to be the disconnected contribution and the remaining $\ell$ blocks to be the connected solution, with $\ell$ running from 0 to $k$. These kinds of saddle points also appear in RMT (see appendix \ref{app:rmt}) and contribute as
\be \label{more:)}
\ba
\langle |Z(\ii T)|^{2k} \rangle \sim \sum_{\ell=0}^k |\langle Z(\ii T) \rangle|^{2(k-\ell)} \langle |Z(\ii T)|^2 \rangle_c^{\ell} \times 
\begin{cases}
\ell! \binom{k}{\ell}^2 & \text{for $q=0$ mod 4},\\
(2\ell-1)!! \binom{2k}{2\ell} & \text{for $q=2$ mod 4},
\end{cases}
\ea
\ee
where the subscript $c$ stands for the connected contribution. These other saddle points never dominate over the \say{fully disconnected} solution ($\ell=0$) or the \say{pairswise connected} solution ($\ell=k$). This is true except near the dip time, where the disconnected contribution oscillates and can be small enough so that all these saddle points contribute equally. The formula \eqref{more:)} describes the spikes in the moments seen in figure \ref{fig:moments46} just before the dip time.

\subsection{$1/N$ corrections}
\label{sec:1/N}

We begin this section by explaining how the one loop determinant factorises over the pairs before showing how the first $1/N$ correction to the moments can be computed by including perturbative fluctuations around the saddle points. This section closely follows section \ref{subsec:0d_SYK_1/N}. We begin by expanding in small fluctuations
\be
G_{ab}=\mathsf{G}_{ab}+\delta G_{ab},\qquad \Sigma_{ab}=\mathsf{\Sigma}_{ab}+\delta \Sigma_{ab},
\ee
around one of the pairing saddle point configurations $\mathsf{G}_{ab}$ and $\mathsf{\Sigma}_{ab}$. The action decomposes as $I=I_\text{cl}+\delta I$, where $I_\text{cl}$ is the classical action and
\be \label{delta I}
\frac{1}{N} \, \delta I = \sum_{n \geq 2} \frac{1}{2n} \Tr[(\mathsf{G} \delta \Sigma)^n] + \frac{1}{2} \int_0^T \int_0^T \dd t \dd t' \left[ \delta \Sigma_{ab} \delta G_{ab} - \frac{J^2}{q} s_{ab} \sum_{n=2}^q \binom{q}{n} \mathsf{G}_{ab}^{q-n} \delta G_{ab}^n \right].
\ee
In obtaining this expression we used the saddle point equations \eqref{saddle all q}. First, we consider the one loop contribution, which amounts to keeping only the quadratic part of the action
\be \label{quadratic action}
\frac{1}{N} \, \delta I^{(2)} = \frac{1}{4} \Tr[(\mathsf{G} \delta \Sigma)^2] + \frac{1}{2} \int_0^T \int_0^T \dd t \dd t' \left[ \delta \Sigma_{ab} \delta G_{ab} - J^2 \frac{q-1}{2} s_{ab} \mathsf{G}_{ab}^{q-2} \delta G_{ab}^2 \right].
\ee
As in section \ref{subsec:0d_SYK_1/N}, it is useful to decompose the fluctuations as $\delta G = \delta G_\parallel + \delta G_\perp$, where $\delta G_\parallel$ lies in the subspace spanned by matrices of the form \eqref{block matrix 2} and $\delta G_\perp$ lies in its orthocomplement. There is a similar decomposition for $\delta \Sigma$. If $\mathsf{G}$ and $\mathsf{\Sigma}$ represent the saddle point which takes the block diagonal form \eqref{block matrix 2}, then $\mathsf{G}=\mathsf{G}_\parallel$ and $\mathsf{\Sigma}=\mathsf{\Sigma}_\parallel$. Equipped with this notation, the quadratic part of the action \eqref{quadratic action} reads
\be
\frac{1}{N} \delta I^{(2)}=\frac{1}{4} \Tr[(\mathsf{G} \delta \Sigma)^2] + \frac{1}{2} \int \! \dd t \dd t' \left[ (\delta \Sigma_\parallel)_{ab} (\delta G_\parallel)_{ab}\! +\! (\delta \Sigma_\perp)_{ab} (\delta G_\perp)_{ab} \! - \! J^2 \frac{q-1}{2} s_{ab} \mathsf{G}_{ab}^{q-2} (\delta G_\parallel)_{ab}^2 \right]\!.
\ee
Since $\delta G_\perp$ appears linearly, integrating it out gives a delta function which sets $\delta \Sigma_\perp=0$. In other words, the fluctuations $\delta G_\perp$ and $\delta \Sigma_\perp$ don't contribute to the one loop determinant. This leaves us with an integral over the remaining variables $\delta G_\parallel$ and $\delta \Sigma_\parallel$ with quadratic action
\be
\frac{1}{4} \Tr[(\mathsf{G} \delta \Sigma_\parallel)^2] + \frac{1}{2} \int_0^T\int_0^T \dd t \dd t' \left[ (\delta \Sigma_\parallel)_{ab} (\delta G_\parallel)_{ab} - J^2 \frac{q-1}{2} s_{ab} \mathsf{G}_{ab}^{q-2} (\delta G_\parallel)_{ab}^2 \right],
\ee
which factorises into a sum over the $k$ blocks in \eqref{block matrix 2}, since the variables $\delta G_\parallel$ and $\delta \Sigma_\parallel$ are block diagonal. This implies that the one loop contribution from expanding in fluctuations around one of the pairing saddle points for $\langle |Z(\ii T)|^{2k} \rangle$ is given by the $k$\textsuperscript{th} power of the one loop contribution for $\langle |Z(\ii T)|^{2} \rangle$. That is, for large $N$,
\be
\frac{\langle |Z(\ii T)|^{2k} \rangle}{\langle |Z(\ii T)|^2 \rangle^k} \sim \begin{cases}
k! & \text{for $q=0$ mod 4},\\
(2k-1)!! & \text{for $q=2$ mod 4}.
\end{cases}
\ee
This formula admits a $1/N$ expansion which can be obtained by separately considering the perturbative expansions around each of the saddle points for $\langle |Z(\ii T)|^{2k} \rangle$ and $\langle |Z(\ii T)|^2 \rangle$. However, since we are interested in the ratio $\langle |Z(\ii T)|^{2k} \rangle / \langle |Z(\ii T)|^2 \rangle^k$, the only perturbative corrections which contribute arise from the fluctuations $\delta G_\perp$ and $\delta \Sigma_\perp$. These corrections can be computed using the Feynman rules for the action $\delta I$ \eqref{delta I}. The relevant propagators are (repeated indices are not summed):
\begin{subequations}
\begin{align}
&\langle \delta G_\perp \delta \Sigma_\parallel \rangle = \langle \delta G_\parallel \delta \Sigma_\perp \rangle =\langle \delta G_\perp \delta G_\parallel \rangle = \langle \delta \Sigma_\perp \delta \Sigma_\parallel \rangle=0 ,\label{eq:2point_mixed} \\[2mm]
&\langle \delta \Sigma_\perp \delta \Sigma_\perp \rangle =0 , \label{eq:2point_SS}\\[2mm]
&\langle \delta \Sigma_\perp(t_1,t_2)_{ab} \delta G_\perp(t_3,t_3)_{cd} \rangle = \frac{\pi^\perp_{ab}}{N} \delta_{ac} \delta_{bd} \delta(t_1-t_3)\delta(t_2-t_4),\label{eq:2point_SG} \\[2mm]
&\langle \delta G_\perp(t_1,t_2)_{ab} \delta G_\perp(t_3,t_4)_{c d} \rangle = \frac{\pi^\perp_{ab}}{N}\left[\mathsf{G}_{ac}(t_1,t_3) \mathsf{G}_{bd}(t_2,t_4)-\mathsf{G}_{ad}(t_1,t_4) \mathsf{G}_{bc}(t_2,t_3)\right].\label{eq:2point_GG}
\end{align}
\end{subequations}
Here the angular brackets represent an average weighted by the quadratic action \eqref{quadratic action} (and normalised so that $\langle 1 \rangle = 1$) and $\pi^\perp$ is a matrix which is equal to zero on the diagonal blocks of \eqref{block matrix 2} and 1 elsewhere. Crucially, as in section \ref{subsec:0d_SYK_1/N},
\be
\langle \delta \Sigma_\perp \delta \Sigma_\parallel \rangle = \langle \delta \Sigma_\perp \delta \Sigma_\perp \rangle = \langle \delta \Sigma_\perp \delta G_\parallel \rangle = 0,
\ee
which implies that the fluctuations arising from $\delta G_\perp$ and $\delta \Sigma_\perp$ arise from Feynman diagrams with at least one vertex corresponding to a $\delta G_\perp$ interaction term. By inspecting $\delta I$ \eqref{delta I} we find that there is only one $\delta G_\perp$ interaction term which contributes to the action as
\be
\delta I \supset - N \, \frac{J^2}{2q} \int_0^T \int_0^T \dd t \dd t' \, s_{ab} (\delta G_\perp)_{ab}^q.
\label{eq:relevant_vertex_G^q} 
\ee
The upshot is that the $1/N$ expansion of $\langle |Z(\ii T)|^{2k} \rangle / \langle |Z(\ii T)|^2 \rangle^k$ can be obtained by considering all Feynman diagrams involving at least one $\delta G_\perp^q$ vertex. Since each diagram contributes as $N^{V-P}$, where $V$ is the number of vertices and $P$ is the number of propagators, naively, the leading correction arises from a diagrams with a single $\delta G_\perp^q$ vertex. However, these contributions vanish,
\be
\label{eq:1_term_expansion}
N \, \frac{J^2}{2q} s_{ab} \int \dd t \dd t' \, \langle \delta G_\perp(t,t')_{ab}^q \rangle=0.
\ee
This follows since $\langle (\delta G_\perp)_{ab} (\delta G_\perp)_{ab} \rangle \propto \pi^\perp_{ab}(\mathsf{G}_{ab} \mathsf{G}_{ab}-\mathsf{G}_{aa} \mathsf{G}_{bb} ) = 0$ as $\mathsf{G}=\mathsf{G}_\parallel$.
All other nonzero terms involve correspond to diagrams with at least two vertices. Identifying the leading term in the perturbative expansion is tedious exercise. The detailed analysis of all the potential contributions is presented in appendix \ref{app:perturbative_exp}. The leading term is given by terms with one $\delta G_\perp^q$ vertex and one $\delta G^q$ vertex, as was the case in section \ref{subsec:0d_SYK_1/N}:
\be \label{two G}
\frac{N^2}{2} \frac{J^4}{4q^2} s_{ab} s_{cd} \int \prod_{j=1}^4 \dd t_j \, \langle \delta G_\perp(t_1,t_2)_{ab}^q \delta G_{cd}(t_3,t_4)^q \rangle.
\ee
This contributes at order $N^{2-q}$ since it corresponds to a diagram with 2 vertices and $q$ propagators. Since there is a parameter $\beta_\text{aux}$ for each block (or pair) in \eqref{block matrix 2}, after a little algebra we find this term contributes as
\be \label{eq:B_interm}
\frac{q!}{N^{q-2}} \frac{J^4}{4q^2} \pi^\perp_{ab} s_{ab} s_{cd} \prod_{i=1}^k \int_0^\infty \dd \beta_\text{aux}^i \, \widetilde{\mu}(\beta_\text{aux}^i) \int \prod_{j=1}^4 \dd t_j \, \mathsf{G}_{ad}(t_1,t_4)^q \mathsf{G}_{bc}(t_2,t_3)^q,
\ee
where we have left the dependence of the saddle point configuration $\mathsf{G}$ on $\beta_\text{aux}^i$ implicit and where
\be
\widetilde{\mu}(\beta_\text{aux}) = \frac{\mu(\beta_\text{aux})}{\int_0^\infty \dd \beta_\text{aux} \mu(\beta_\text{aux})}.
\ee
In \eqref{two G} the $q!$ arises from the $q$ possible Wick contraction between $\delta G_{ab}$ and $\delta G_{cd}$. We also used
\begin{equation}
\pi^\perp_{ab}\left(\mathsf{G}_{ac} \mathsf{G}_{bd}-\mathsf{G}_{ad} \mathsf{G}_{bc}\right)^q = \pi^\perp_{ab} \left(\mathsf{G}_{ac}^q \mathsf{G}_{bd}^q + \mathsf{G}_{ad}^q \mathsf{G}_{bc}^q \right),
\end{equation}
which follows since $\mathsf{G}_{ac} \mathsf{G}_{bd}$ is only nonzero when $\mathsf{G}_{ad} \mathsf{G}_{bc}$ vanishes and vice versa. To further simplify \eqref{eq:B_interm} we use the identity $s_{ab} s_{cd} =  s_{ad} s_{bc}$ and write $\pi^\perp_{ab} = 1 - \pi^\parallel_{ab}$, so that
\begin{equation}
    \pi^\perp_{ab} s_{ab} s_{cd} \mathsf{G}_{ad}^q \mathsf{G}_{bc}^q = (s_{ab} \mathsf{G}_{ab}^q )^2 - \pi^\parallel_{ac} s_{ab} s_{cd} \mathsf{G}_{ab}^q \mathsf{G}_{cd}^q \, .
    \label{eq:intermediate:GG}
\end{equation}
The first term on the RHS gives rise to a contribution which is proportional to the square of
\begin{equation}
s_{ab} \int \dd t \dd t' \, \mathsf{G}_{ab}(t,t')^q = - \frac{T}{J^2} \partial_t \mathsf{G}_{aa}(t) \rvert_{t\to0^+} = \ii \frac{qT}{NJ^2} \sum_{i=1}^k \Delta_E(\beta_\text{aux}^i),
    \label{eq:identity_piece_B}
\end{equation}
where we used the equations of motion in the first step and we introduced the quantity
\be
\Delta_E(\beta_\text{aux}) = \ii \, \frac{N}{q} \partial_t \left( \mathsf{G}_{LL} + \mathsf{G}_{RR} \right) \rvert_{t\to0^+},
\label{eq:deltaE_def}
\ee
which is purely imaginary since the saddle point solution satisfies $\mathsf{G}_{LL}=\mathsf{G}_{RR}^*$. In the auxiliary problem $\Delta_E$ is the difference between energies of the $L$ and $R$ systems in the thermofield double state, which vanishes exactly. However, the solution to the auxiliary problem is only accurate for large $T$ and so $\Delta_E$ does not vanish exactly, although it is exponentially small in $T$ for $\beta_\text{aux} \ll T$ \cite{Saad:2018bqo}. We comment more on the behvaiour of $\Delta_E$ below and in the next subsection. From the second term on the RHS of \eqref{eq:intermediate:GG} we get
\begin{equation}
\pi^\parallel_{ab} s_{ab} s_{cd} \int \prod_{j=1}^4 \dd t_j \, \mathsf{G}_{ab}(t_1,t_2)^q \mathsf{G}_{cd}(t_3,t_4)^q = - \left(\frac{qT}{NJ^2}\right)^2 \sum_{i=1}^k \Delta_ E(\beta_\text{aux}^i)^2.
\end{equation}
Altogether \eqref{eq:B_interm} leads to the correction:
\be
\frac{\langle |Z(\ii T)|^{2k} \rangle}{\langle |Z(\ii T)|^2\rangle^k} = \left[ 1 + \frac{k(k-1)}{4} \frac{q!}{N^{q}} T^2 |\overline{\Delta}_E|^2 + \dots \right] \times \begin{cases}
k! & \text{for $q=0$ mod 4},\\
(2k-1)!! & \text{for $q=2$ mod 4},
\label{eq:final_result_correction}
\end{cases}
\ee
where
\be \label{bar delta E}
\overline{\Delta}_E= \int_0^\infty \dd \beta_\text{aux} \, \widetilde{\mu}(\beta_\text{aux}) \Delta_E(\beta_\text{aux}).
\ee
Since $\Delta_E(\beta_\text{aux})$ is exponentially small for large $T$, schematically $e^{-T/\beta_\text{aux}}$ \cite{Saad:2018bqo}, $\overline{\Delta}_E$ is dominated by the part of the integral where the temperature is very low, $\beta_\text{aux} \gtrsim T$, corresponding to the spectrum's edge.\footnote{Since large $\beta_\text{aux}$ corresponds to the Schwarzian limit where $\mu(\beta_\text{aux}) \propto 1/\beta_\text{aux}^3$ we can be more explicit: near the, say, lower edge $E-E_\text{min} \propto N/J \beta_\text{aux}^2$.} This means the correction can be understood as arising from fluctuations near the edge of the spectrum. While $\overline{\Delta}_E$ appears difficult to compute, we provide evidence in section \ref{sec:sparse_SYK} that the total correction is independent of time $T$.

The fact that the leading correction is only polynomially small in $N$ rather exponentially small marks a departure from RMT expectations. This is to be expected as SYK and RMT belong to different universality classes of chaotic systems \cite{Altland:2024ubs}. The key distinction lies in the number of independent random parameters in the Hamiltonian: for SYK it is approximately $N^q/q!$, scaling logarithmically with the Hilbert space dimension $L=2^{N/2}$, while for RMT it is approximately $L^2$, scaling polynomially. These differences unfold in physics near the edge of the spectrum in the regime of large but finite $L$. In sparse systems, which contain at most order $\log L$ random parameters, like SYK, the fluctuations from near the edge of the spectrum are more pronounced than in dense systems, which contain at most order $L$ random parameters, like RMT or JT gravity. Specifically, the corrections to the spectral density near the edge scale inversely with $\log L$ in sparse systems compared to $L$ in dense systems \cite{Altland:2024ubs}. Surprisingly, the first correction to the moments of the spectral form factor scales inversely with the number of independent random parameters $N^q/q!$ rather than simply $1/N$\cite{Kitaev:2017awl,Garcia-Garcia:2018fns,Cotler:2017jue,Jia:2019orl,Berkooz:2020fvm}. Notice that since $\Delta_E$ is extensive in $N$, the correction ultimately scales as $1/N^{q-2}$ . 

Since the correction in \eqref{eq:final_result_correction} scales inversely with $N^{q-2}$ it hints that the $q=2$ model is qualitatively different and may exhibit enhanced noise. This is consistent with the fact that the $q=2$ model is a free theory and hence isn't expected to display signatures of many-body chaos unlike its $q>2$ cousins. The enhanced symmetries of the $q=2$ model necessitate a separate treatment which is the focus of the next section.

Although the correction to the moments \eqref{eq:final_result_correction} from fluctuations near the edge of the spectrum is very small for the low order moments, it becomes significant in the very high order moments, when $k$ approaches a fixed fraction of $N^{q/2-1}$.
This signals a departure from RMT expectations where we anticipate the corrections to the moments to become significant only when $k$ approaches a fixed fraction of $L^\#$; however we are not aware of a proof which determines when this breakdown occurs.\footnote{An exception is for the very late time behaviour of spectral form factor in the circular unitary ensemble (CUE), the ensemble of unitary matrices with the Haar measure. The spectral form factor displays a linear ramp and plateau: $\langle |\Tr[U^\texttt{T}]|^2 \rangle_\text{CUE} = \min\{\texttt{T},L\}$, with the discrete variable $\texttt{T}$ playing the role time. On the plateau $\langle |\Tr[U^\texttt{T}]|^{2k} \rangle_\text{CUE} = k! L^k (1-\frac{k(k-1)}{4L}+ \dots)$, signalling large deviations from Gaussianity when $k \sim \sqrt{L}$. This result follows from expanding the exact formula \cite{pastur2004moments} (see theorem 2.1) for the moments for large $L$. Curiously, the correction comes with a minus sign. We will see a similar behaviour in sparse SYK in seciton \ref{sec:sparse_SYK} from numerics.} In this sense, SYK mimics a RMT for the low order moments of the spectral form factor, specifically for $k \ll N^{q/2-1}$. This structure is reminiscent of the \say{mock Gaussian} behaviour seen in the circular ensembles \cite{diaeva}, the ensembles of orthogonal, unitary, or symplectic matrices with the Haar measure. The term mock Gaussian refers to the fact that while first few moments are Gaussian, the overall distribution is not. For example, for the circular unitary ensemble (CUE)
\be
\langle \Tr[U]^k \Tr[U^*]^{k'} \rangle_\text{CUE} = k! \, \delta_{kk'},
\ee
provided $k \leq L$, where $L$ is the size of the matrix. A similar expression holds for the mixed moments; see \cite{Meckes_2019}. Really \eqref{eq:final_result_correction} indicates that an approximate version of mock Gaussian property holds SYK, since for any finite $N$ the moments are not exactly Gaussian although they are approximately so provided $k \ll N^{q/2-1}$.

The fact that the correction scales inversely with $N^q/q!$ also raises the possibility that, more generally, in sparse systems the leading correction scales inversely with the number of independent random parameters. To explore this idea, we study a sparsified version of SYK, where interaction terms are deleted with some probability, in section \ref{sec:sparse_SYK}. Numerical results indicate that the leading correction increases as the system becomes more sparsified and the number of random parameters decreases.  The larger magnitude of the correction enables its observation at relatively small moments, where it is feasible to stabilize numerically the noise. This behaviour suggests a faster departure of sparse SYK from the RMT expectation compared to the regular, unsparsified, SYK model and suggests generalization of equation \eqref{eq:final_result_correction} to other contexts.

A similar $1/N$ expansion applies to the moments of the spectral form factor at early times when the disconnected solution $\mathsf{G}_{ab}^\text{disc}$, which described the slope, dominates. This can be obtained by repeating the steps but redefining $M_\parallel$ to project a matrix $M$ onto its diagonal components. The final result is:
\be
\frac{\langle |Z(\ii T)|^{2k} \rangle}{|\langle Z(\ii T)\rangle|^{2k}} = 1 + \frac{q!}{N^{q}} \left( \frac{k(k-1)}{4} T^2 |\Delta_E|^2 + \frac{k}{2}  T^2 |F|^2 \right) + \dots,
\label{eq:final_result_correction_slope}
\ee
where 
\be
\Delta_E = \ii \, \frac{N}{q} \partial_t (\mathsf{G}_{LL}^{\text{disc}}+\mathsf{G}_{RR}^{\text{disc}}) \rvert_{t\to0^+} , \quad |F|^2 = \frac{N}{q} \partial_t \mathsf{G}_{LL}^{\text{disc}}\rvert_{t\to0^+} \times \frac{N}{q} \partial_t  \mathsf{G}_{RR}^{\text{disc}}  \rvert_{t\to0^+}.
\ee
For $k=1$ the latter term describes a connected contribution to $\langle |Z(\ii T)|^2 \rangle$ arising from a perturbative expansion around the disconnected saddle point and has already been obtained in \cite{Berkooz:2020fvm}. The term $T \Delta_E$ starts at zero, increases to a maximum, and then tends to zero again at late times. The vanishing of $\Delta_E = \ii \, \partial_T I_\text{cl}$ at late times occurs because the classical action $I_\text{cl}$ approaches a constant for late times.\footnote{We are grateful to Alex Windey for discussions on this point and for assistance with numerically solving the saddle point equations.} Large oscillations in the spectral form factor near the dip time make it difficult to identify these corrections numerically. This issue can be circumvented by studying a microcanonical version of the spectral form factor where the oscillations are tamed.

\subsection{Microcanonical spectral form factor} \label{sec:finite_beta}

While the spectral form factor is sensitive to the whole spectrum, we could also imagine studying a microcanonical version of it where the sum over states is weighted by a \say{filter function} $f$ which focuses on some part of the spectrum \cite{Saad:2018bqo}
\be \label{Yf}
|Y_f(T)|^2 = \Tr[f(H) e^{-\ii T H}] \Tr[f(H) e^{\ii T H}].
\ee
A standard choice for $f$ is a Gaussian
\be \label{gausshimself}
f(H) = \exp\left(-\frac{(H-E)^2}{2\Delta^2}\right),
\ee
which roughly restricts to energies within a window $\Delta$ of $E$. For the Gaussian filter function, the microcanonical spectral form factor can be computed as
\be
\langle |Y_f(T)|^2 \rangle \propto \int \dd \beta_L \dd \beta_R e^{(\beta_L + \beta_R)E+\frac{1}{2}(\beta_L^2+\beta_R^2)\Delta^2} \langle Z(\beta_L-\ii T)Z(\beta_R+\ii T) \rangle,
\ee
and similarly for its moments. For nonzero $\beta_L$ and $\beta_R$ the approximate zero mode $\beta_\text{aux}$ is lifted classically \cite{Saad:2018bqo} and so in the saddle point approximation we look for stationary points with respect to $\beta_L, \beta_R$, and $\beta_\text{aux}$. For the moments we would still have \cite{Winer:2022ciz}
\be
\frac{\langle |Y_f(T)|^{2k} \rangle}{\langle |Y_f(T)|^2 \rangle^k} \sim \begin{cases}
k! & \text{for $q=0$ mod 4},\\
(2k-1)!! & \text{for $q=2$ mod 4},
\end{cases}
\ee
with the equality for $q=2$ mod 4 holding provided $f$ is an even function, i.e. $E=0$ for the Gaussian filter function \eqref{gausshimself}. Focusing away from the edge of the spectrum, say by setting $E=0$ and a choosing a suitable $\Delta$, we expect the leading correction to this formula to still be given by \eqref{eq:final_result_correction}. However, the integral over $\beta_\text{aux}$ in \eqref{bar delta E} is now restricted to a range where $\Delta_E (\beta_\text{aux})$ becomes exponentially small in $T$. In terms of the auxiliary energy $E_\text{aux}$, the integral is roughly restricted to a window centered around zero of width $\Delta$ and away from the spectrum's edge. We examine this numerically in sparse SYK in section \ref{sec:sparse_SYK}, see figure \ref{fig:spars_moments_filtered}. This result aligns with the expectation that SYK behaves like a RMT away from the edge of the spectrum. We note that this argument does not rule out the possibility of further corrections in the $1/N$ expansion which are not exponentially supressed in $T$.

In contrast, at early times, when the disconnected saddle point dominates, the corrections to the moments persist. The only modification is that now the disconnected solution $\mathsf{G}_{ab}^\text{disc}$ depends on the saddle point values of the $\beta_{L,R}$ variables. Since these corrections are always present, even away from the edge, it suggests that they correspond to global fluctuations in the spectrum, as proposed in \cite{Berkooz:2020fvm,Altland:2017eao,Jia:2019orl}.

\section{$q=2$ SYK} \label{syk:sec:q=2}

The $q=2$ SYK model is a free theory of fermions with a random mass matrix and as such is not expected to display features of many-body chaos. However, it retains certain signatures of chaos in its single-particle energy levels which are given by the eigenvalues of a random antisymmetric hermitian matrix and exhibit eigenvalue repulsion. This feature makes the $q=2$ model valuable for studying the transition from single-body chaos to many-body chaos.

\begin{figure}
\centering
\begin{subfigure}{0.48\textwidth}
\raggedright
\begin{tikzpicture}
\node at (0,0) {\includegraphics[width=0.95\linewidth]{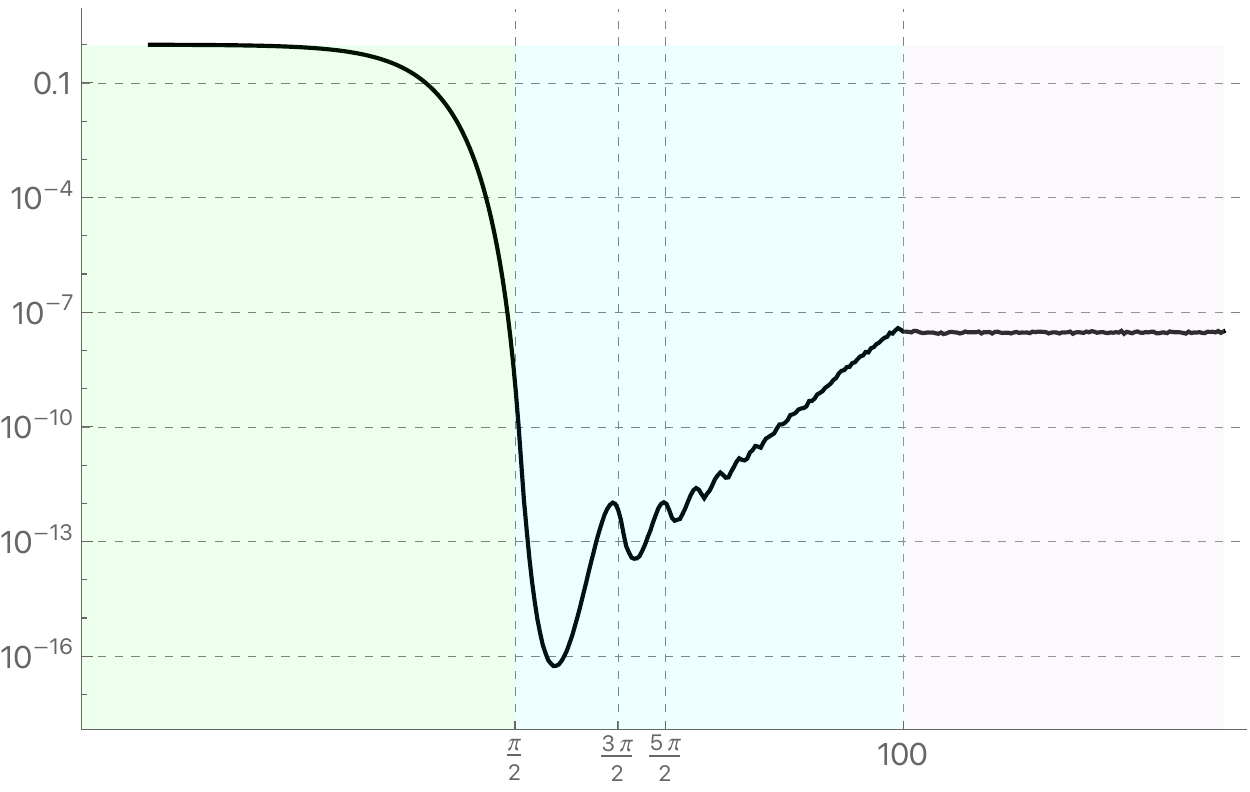}};
\node at (current bounding box.south) {{\scriptsize $JT$}};
\node at ([xshift=-0.2em]current bounding box.west) {\scriptsize \rotatebox{90}{$\langle |Z(\ii T)|^2 \rangle$}};
\end{tikzpicture}
\end{subfigure}
% $\quad$
\begin{subfigure}{0.48\textwidth}
\raggedleft
\begin{tikzpicture}
\node at (0,0) {\includegraphics[width=0.95\linewidth]{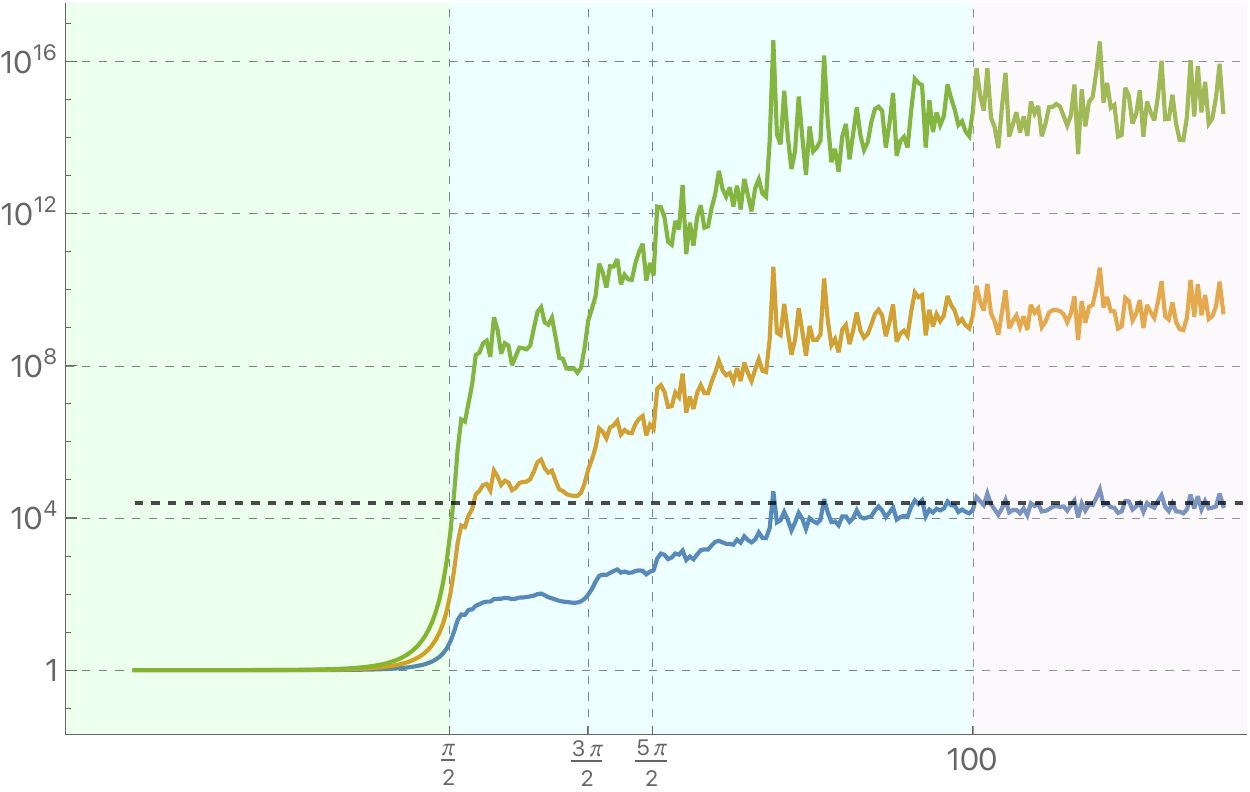}};
\node at (current bounding box.south) {{\scriptsize $JT$}};
\node at (current bounding box.west) {\scriptsize \rotatebox{90}{$\langle |Z(\ii T)|^{2k} \rangle / \langle |Z(\ii T)|^2 \rangle^k$}};
\end{tikzpicture}
\end{subfigure}
\caption{Left panel: the spectral form factor (divided by $L^2$) for the $q=2$ SYK model with $N=50$, averaged over $2 \times 10^6$ realisations. The plateau value is at $JT=2N$, as calculated in \cite{Liao:2020lac}. Right panel: the moments of the spectral form factor for $k=2$ (blue), $k=3$ (orange), and $k=4$ (green). Every time a Matsubara mode enters the symmetry broken regime $|\omega_n| < 2J$, the moments display a discrete jump due to the volume of the zero mode, as showed in \eqref{eq:volume_contribution_SYK2}. The black dashed line is the plateau prediction according to \eqref{eq:plateau_SYK2} for $k=2$.}
\label{fig:SFFSYK2}
\end{figure}

In contrast to the $q>2$ model, the spectral form factor of the $q=2$ model, shown in figure \ref{fig:SFFSYK2}, features an exponential ramp rather than a linear one. The authors of \cite{Winer:2020mdc} showed how this property could be traced to an infinite enhancement of the U(1)$_L \times$ U(1)$_R$ time translation symmetry of the problem which is spontaneously broken by the saddle point in a time dependent pattern.\footnote{See \cite{Liao:2020lac} for a different treatment for the complex $q=2$ model.} The exponential ramp eventually gives rise to a plateau at a time scale of order $N$, much before the plateau time in RMT. In this section we extend this calculation to the moments of the spectral form factor. We begin by showing how the $\mathrm{U}(1)^{2k} \times S_{2k}$ symmetry---the direct product of time translation and replica symmetry--- for $q>2$ is infinitely enhanced in the special case that $q=2$ to a U$(2k)$ replica symmetry for each Matsubara frequency $\omega_n$. A time dependent symmetry breaking pattern gives rise to a zero mode manifold whose volume is proportional to
\be
\prod_{|\omega_n| < 2J} N^{k^2} \text{vol} \left(\frac{\mathrm{U}(2k)}{\mathrm{U}(k)^2}\right), \label{eq:volume_contribution_SYK2}
\ee
which grows exponentially in time $T$, as the number of Matsubara frequencies which enter the regime $|\omega_n| < 2J$ increases. This is clearly seen in the numerical plot of the moments of the spectral form factor in figure \ref{fig:SFFSYK2}. At discrete time steps, a new Matsubara frequency enters the symmetry broken regime and causes a jump in the moments due to the volume of its zero mode. The enhanced zero mode manifold explains why the spectral form factor in the $q=2$ model exhibits much larger erratic oscillations. Finally, we discuss the large $k$ behaviour of the moments.

\subsection{Symmetries}

For the $q=2$ model the discrete replica symmetry $S_{2k}$ is infinitely enhanced. After integrating out $G$, which appears quadratically in the action \eqref{moments collective fields}, we can write the result as
\be \label{moments collective fields q=2}
\ba
{}&\langle |Z(\ii T)|^{2k} \rangle = \int D \Sigma \, e^{-I[\Sigma]}\\
{}&\frac{1}{N} \, I[\Sigma] = - \frac{1}{2} \log \det(\partial_t - \Sigma) + \frac{1}{4J^2} \Tr\Sigma^2.
\ea
\ee
In the infrared limit, the derivative operator $\partial_t$ can self-consistently be dropped from the action, resulting in an effective theory with equation of motion $\Sigma^2 = J^2$, or, more explicitly
\be
\int_0^T \dd s \, \Sigma_{ac}(t,s) \Sigma_{cb}(s,t') = J^2 \delta_{ab} \delta(t-t').
\ee
This equation is invariant under the transformation
\be
\Sigma(t,t') \to \Sigma^U(t,t') = \int_0^T\int_0^T \dd s \dd s' \, U(t,s)\Sigma(s,s') U(t',s')^\tr,
\ee
for a matrix valued distribution $U(t,t')$ which is antiperiodic with period $T$ in each argument and satisfies
\be
\int_0^T \dd s \, U(t,s) U(t',s)^\tr = \delta(t-t'),
\ee
where $U(t,t')^\tr$ is its transpose. It's useful to define $U^\tr$ to be the operator with matrix elements given by transposing both the replica indices and time labels of $U$, i.e. $(U^\tr)(t,t') = U(t',t)^\tr$ where on the RHS $\tr$ is the ordinary transpose in replica space. The transformation can then be concisely written as
\be
\Sigma \to \Sigma^U = U \Sigma U^\tr,\qquad UU^\tr=1.
\ee
This corresponds to an infinite enhancement of the emergent $\text{diff}(S^1)$ symmetry for each replica of the $q>2$ model.\footnote{The $\text{diff}(S^1)$ subgroup is parameterised by $U(t,t') = \phi'(t)^\Delta \delta(\phi(t)-t')$ with $\phi \in \text{diff}(S^1)$. The constraint $U^\tr U=1$ fixes $\Delta=\frac{1}{2}$.} The infrared symmetries of the single replica theory have been discussed in \cite{Anninos:2023lin}. Away from the strict infrared limit of the theory, this symmetry is explicitly broken by the presence of the derivative operator $\partial_t$ in the action. Indeed, the full equation of motion is
\be
\partial_t \Sigma - \Sigma^2 = - J^2,
\ee
and so for $U$ to be a symmetry of the full theory it must commute with the derivative operator
\be
U^\tr \partial_t U = \partial_t.
\ee
It is simple to check that this can be solved by a $U(t,t')$ which only depends on the difference of times, i.e. $U(t,t')=U(t-t')$, since
\be
\ba
\int_0^T \dd s \, U(s-t)^\tr \partial_s U(s-t') &= - \partial_{t'} \int_0^T \dd s \, U(s-t)^\tr U(s-t')\\
&= - \partial_{t'} \delta(t-t').
\ea
\ee
In going to the second line we used the constraint $U U^\tr = 1$. Since $U$ is antiperiodic it's Fourier series may be written as
\be
U(t) = \frac{1}{\sqrt{T}} \sum_{n \in \Z} U(\omega_n) e^{-\ii \omega_n t},\qquad \omega_n = \frac{2\pi(n+\frac{1}{2})}{T}.
\ee
In order to preserve the reality condition on $\Sigma$, we require $U$ to be real valued and so $U(\omega_n) = U(-\omega_n)^*$. It is then straightforward to verify that the constraint $U U^\tr=1$ implies that the Fourier modes $U_n=U(\omega_n)$ satisfy $U_n U_n^\dagger = 1$, i.e. $U_n \in \mathrm{U}(2k)$. The action of this symmetry on the Fourier modes of $\Sigma$ is given by
\be \label{SigmaU}
\Sigma_{mn} \to (\Sigma^U)_{mn} = U_m \Sigma_{mn} U_n^\tr.
\ee
To summarise, the $\mathrm{U}(1)^{2k} \times S_{2k}$ symmetry---the direct product of time translation and replica symmetry--- for $q>2$ such that $q=2$ mod 4 is enhanced in the special case that $q=2$ to the infinite dimensional group $\prod_{n \geq 0} \mathrm{U}(2k)$.

\subsection{Saddle points} \label{sec-sp-q2}

To simplify the saddle point equation
\be
\partial_t \Sigma - \Sigma^2 = - J^2,
\ee
we make an ansatz that $\Sigma$ only depends on the difference of times, i.e. $\Sigma(t,t')=\Sigma(t-t')$. With this ansatz, configurations which are not invariant under this diagonal time translation symmetry only contribute as fluctuations in the one loop determinant in the saddle point approximation. As such, it is useful to warm up to the full problem by studying a slightly simpler path integral defined similarly to \eqref{moments collective fields q=2} but where we only integrate over time translation invariant configurations. We will add back the contribution of the configurations which are non invariant under the diagonal time translation symmetry later. Writing the Fourier series of $\Sigma$ as
\be
\Sigma(t) = \sum_{n \in \Z} \Sigma(\omega_n) e^{-\ii \omega_n t},\qquad \omega_n = \frac{2\pi(n+\frac{1}{2})}{T},
\ee
the path integral decomposes into an infinite product of decoupled hermitian matrix integrals\footnote{This matrix integral has also been studied in a different context \cite{kamenev1999level}.}
\be \label{product matrix integrals}
\langle |Z(\ii T)|^{2k} \rangle = 2^{kN} \prod_{n=0}^\infty \left( \frac{N}{2\pi J^2} \right)^{2k^2} \int_{\R^{(2k)^2}} \dd \Sigma_n \exp \left\{ N \Tr \log \left(1 + \frac{\Sigma_n}{\ii \omega_n}\right) - \frac{N}{2J^2} \Tr \Sigma_n^2\right\},
\ee
where
\be
\dd \Sigma_n = \prod_a \dd (\Sigma_n)_{aa} \prod_{a<b} \sqrt{2} \, \dd \text{Re}\,(\Sigma_n)_{ab} \sqrt{2} \, \dd \text{Im}\,(\Sigma_n)_{ab},
\ee
is the standard measure on the space of $2k \times 2k$ hermitian matrices, which is a copy of $\R^{(2k)^2}$. That $\Sigma_n = \Sigma(\omega_n)$ is hermitian follows from imposing the constraint $\Sigma_{ab}(t,t') = - \Sigma_{ba}(t',t)$ as well as requiring $\Sigma_{ab}(t,t')$ to be purely imaginary, which is required for convergence of the path integral. The factor $2^{N/2}$ for each replica is the contribution of the free determinant $\prod_{n=0}^\infty(n+1/2)$ which, after zeta function regularisation, gives $\sqrt{2}$. Since $Z(\ii T)$ can only depend on the dimensionful variables $J$ and $T$ through the combination $JT$, its value at $J=0$ simply computes the dimension of the Hilbert space $Z(\ii T) \rvert_{J=0} = 2^{N/2}$. The normalisation chosen in \eqref{product matrix integrals} thus ensures that the spectral form factor is normalised on average.

Since the path integral \eqref{product matrix integrals} decomposes as product of decoupled matrix integrals for each mode, we can focus our attention on the contribution of a single mode:
\be \label{matrix integral mode}
Z_n = \left( \frac{N}{2\pi J^2} \right)^{2k^2} \int_{\R^{(2k)^2}} \dd \Sigma_n \exp \left\{ N \Tr \log \left(1 + \frac{\Sigma_n}{\ii \omega_n}\right) - \frac{N}{2J^2} \Tr \Sigma_n^2\right\}.
\ee
The matrix integral over each mode $\Sigma_n$ \eqref{product matrix integrals} is invariant under conjugation by a unitary matrix $\Sigma_n \to U_n \Sigma_n U_n^\dagger$ with $U_n \in \mathrm{U}(2k)$. This symmetry is simply the restriction of the symmetry \eqref{SigmaU} mentioned in the previous section to time translation invariant configurations. Using this symmetry to diagonalise $\Sigma_n$, the integral can be recast as an integral over the eigenvalues with a Jacobian factor, which computes the volume of the orbit of the diagonal matrix under the the adjoint action of $U(2k)$ \cite{mehta2004random}:
\be \label{eigenvalue integral}
Z_n = \left( \frac{N}{2\pi J^2} \right)^{2k^2} \text{vol}\left(\frac{\mathrm{U}(2k)}{\mathrm{U}(1)^{2k} \times S_{2k}}\right) \int_{\R^{2k}} \prod_a \dd \lambda_a \prod_{a<b} (\lambda_a-\lambda_b)^2 e^{-N \sum_aV(\lambda_a)}.
\ee
where
\be
V(\lambda) = -\log\left(1+\frac{\lambda}{\ii \omega_n}\right) + \frac{1}{2J^2} \lambda^2.
\ee
The Jacobian gives rise to the square of the Vandermonde determinant $\Delta(\{\lambda\}) = \prod_{a<b} |\lambda_a-\lambda_b|$. With our normalisation
\be
\text{vol} \, \mathrm{U}(x) = \frac{(2\pi)^{\frac{x(x+1)}{2}}}{G(x+1)},
\ee
where $G$ is the Barnes-G function. For large $N$, the integral \eqref{eigenvalue integral} is dominated by its saddle points and small fluctuations around them. For $k \ll N$, we can neglect the contribution of the Vandermonde determinant when determining the saddle points. The saddle points are then obtained by extremising $V(\lambda)$ which gives a quadratic equation
\be
\lambda_a^2 + \ii \omega_n \lambda_a - J^2=0.
\ee
The two possible solutions for each eigenvalue are
\be \label{lambdapm}
\lambda_\pm = \frac{-\ii \omega_n \pm \sqrt{4J^2-\omega_n^2}}{2}.
\ee
As discussed in \cite{Winer:2020mdc}, for $\omega_n < 2J$, the contour of integration for each eigenvalue may be deformed to pass through both saddle points $\lambda_\pm$, while for $\omega_n > 2J$ it is only possible to deform the contour to pass through the saddle point $\lambda_+$. This situation is quite different from that in section \ref{sec:general k} for $q > 2$, where the exchange of dominance between the two saddle points occurs at different times rather than at different frequencies.

First consider the simpler case that $\omega_n > 2J$. Adding a small fluctuation $\mu_a$ to the saddle point $\lambda_a = \lambda_+ + \mu_a$ we get
\be
\ba
Z_n =&\;e^{-2kNV(\lambda_+)} \left( \frac{N}{2\pi J^2} \right)^{2k^2} \text{vol}\left(\frac{\mathrm{U}(2k)}{\mathrm{U}(1)^{2k} \times S_{2k}}\right) \\
& \times\int_{\R^{2k}} \prod_a \dd \mu_a \prod_{a<b} (\mu_a-\mu_b)^2 e^{-\frac{N}{2J^2}\left(1+\frac{\lambda_+^2}{J^2}\right) \sum_a \mu_a^2}.
\ea
\ee
The one loop contribution is most easily computed by interpreting the integral over $\mu_a$ as the integral over the eigenvalues of a $2k \times 2k$ hermitian matrix $X$. We then have
\be \label{omega>2J}
\ba
Z_n &= e^{-2kNV(\lambda_+)} \left( \frac{N}{2\pi J^2} \right)^{2k^2} \int_{\R^{(2k)^2}} \dd X e^{-\frac{N}{2J^2}\left(1+\frac{\lambda_+^2}{J^2}\right) \Tr X^2}\\
&= e^{-2kNV(\lambda_+)} \left(1+\frac{\lambda_+^2}{J^2}\right)^{-2k^2}.
\ea
\ee
For $\omega_n < 2J$ each eigenvalue integral is given by a sum of over two saddle points. Since the integral is invariant under permutations of eigenvalues there are $2k+1$ distinct contributions
\be
Z_n = \sum_{\ell=0}^{2k} Z_n^{(\ell)},
\ee
where $Z_n^{(\ell)}$ is the contribution where $\ell$ eigenvalues are expanded around the $\lambda_-$ saddle point and the remaining $2k-\ell$ eigenvalues are expanded around the $\lambda_+$ saddle point. Expanding around the saddle points $\lambda_a = \lambda_- + \mu_a^-$ for $a=1,\dots,\ell$ and $\lambda_a = \lambda_+ + \mu_a^+$ for $a=\ell+1,\dots,2k-\ell$ we get
\be
\ba
Z_n^{(\ell)} =&\; e^{-\ell NV(\lambda_-)-(2k-\ell)NV(\lambda_+)} \left( \frac{N}{2\pi J^2} \right)^{2k^2} (\lambda_+-\lambda_-)^{2\ell(2k-\ell)}\text{vol}\left(\frac{\mathrm{U}(2k)}{\mathrm{U}(1)^{2k} \times S_{2k}}\right) \\
&\times \int_{\R^\ell} \prod_a \dd \mu_a^- \prod_{a<b} (\mu_a^--\mu_b^-)^2 e^{-\frac{N}{2J^2}\left(1+\frac{\lambda_-^2}{J^2}\right) \sum_a (\mu_a^-)^2} \\
&\times\int_{\R^{2k-\ell}} \prod_a \dd \mu_a^+ \prod_{a<b} (\mu_a^+-\mu_b^+)^2 e^{-\frac{N}{2J^2}\left(1+\frac{\lambda_+^2}{J^2}\right) \sum_a (\mu_a^+)^2}.
\ea
\ee
As before, the one loop contributions are most easily computed by interpreting the integral over $\mu_a^-$ as the integral over eigenvalues of an $\ell \times \ell$ hermitian matrix and the integral over $\mu_a^+$ as the integral over eigenvalues of a $(2k-\ell) \times (2k-\ell)$ hermitian matrix. This leads to the following result
\be
\ba \label{Zn sum over saddles}
Z_n = \sum_{\ell=0}^{2k} &\;e^{-\ell NV(\lambda_-)-(2k-\ell)NV(\lambda_+)} \left( \frac{N}{2\pi J^2} (\lambda_+-\lambda_-)^2\right)^{\ell(2k-\ell)} \text{vol}\left(\frac{\mathrm{U}(2k)}{\mathrm{U}(\ell) \times \mathrm{U}(2k-\ell)}\right) \\
&\times\left(1+\frac{\lambda_-^2}{J^2}\right)^{-\frac{\ell^2}{2}} \left(1+\frac{\lambda_+^2}{J^2}\right)^{-\frac{(2k-\ell)^2}{2}}.
\ea
\ee
This formula has a simple interpretation in terms of the original matrix integral \eqref{matrix integral mode}. Due to the $\mathrm{U}(2k)$ invariance of the matrix integral the extremums of $\Tr V(\Sigma_n)$ are highly degenerate. In general they are not saddle points, rather they are saddle manifolds. Let $\Lambda(\ell) = \text{diag}(\lambda_-,\dots,\lambda_-,\lambda_+,\dots,\lambda_+)$, where $\ell$ denotes the multiplicity of $\lambda_-$. Then, each of the terms in sum above has the interpretation as the result of the integral along the orbits of the diagonal saddle points $\Lambda(\ell)$ under the action of $\mathrm{U}(2k)$ together with the one loop contribution in the directions orthogonal to the orbit. The orbit space is the quotient $\mathrm{U}(2k)/G$ where $G$ is the subgroup of $\mathrm{U}(2k)$ which leaves the diagonal saddle point fixed. For $\Lambda(\ell)$, $G=\mathrm{U}(\ell) \times \mathrm{U}(2k-\ell)$ so the orbit space is the Grassmannian $\text{Gr}(\ell,2k) = \mathrm{U}(2k)/\mathrm{U}(\ell) \times \mathrm{U}(2k-\ell)$, the space of $\ell$-dimensional linear subspaces of $\C^{2k}$. The measure on the orbit comes from commuting $\mathfrak{g}^\bot$ with $\Lambda(\ell)$, where $\mathfrak{g}^\bot$ is the orthocomplement to the Lie algebra $\mathfrak{g}$ of $G$. Concretely, $\mathfrak{g}^\bot$ is the $2\ell(2k-\ell)$-dimensional vector space spanned by $2k \times 2k$ antihermitian matrices that have vanishing entries in the top-left $\ell \times \ell$ block and bottom-right $(2k-\ell) \times (2k-\ell)$ block. The commutator of any matrix in $\mathfrak{g}^\bot$ with $\Lambda(\ell)$ is proportional to $\lambda_+-\lambda_-$, so the volume of the orbit is proportional to $(\lambda_+-\lambda_-)^{2\ell(2k-\ell)}$. This gives rise to the last two terms in the first line of \eqref{Zn sum over saddles}.

Since $\text{Re} V(\lambda_+) = \text{Re} V(\lambda_-)$ the dominant contribution in the sum over saddles \eqref{Zn sum over saddles} is the term with $\ell=k$ as it maximises the volume of the orbit space, which scales as $N^{\ell(2k-\ell)}$. Only keeping this contribution gives
\be 
Z_n = e^{-kN[V(\lambda_-)+V(\lambda_+)]} \left( \frac{N}{2\pi J^2} (\lambda_+-\lambda_-)^2\right)^{k^2} \text{vol}\left(\frac{\mathrm{U}(2k)}{\mathrm{U}(k)^2}\right) \left(1+\frac{\lambda_-^2}{J^2}\right)^{-\frac{k^2}{2}} \left(1+\frac{\lambda_+^2}{J^2}\right)^{-\frac{k^2}{2}}.
\ee
Using the formula \eqref{lambdapm} for $\lambda_\pm$ we get
\be
Z_n = \left(\frac{J}{\omega_n}\right)^{2kN} e^{kN\left(\frac{\omega_n^2}{2J^2}-1\right)} \left(\frac{N}{2\pi} \sqrt{4-\frac{\omega_n^2}{J^2}}\right)^{k^2} \text{vol}\left(\frac{\mathrm{U}(2k)}{\mathrm{U}(k)^2}\right).
\label{omega<2J}
\ee
Notice that if we take $k \gg 1$ but $k \ll N$ we find
\be
Z_n \sim \left(\frac{J}{\omega_n}\right)^{2kN} e^{kN\left(\frac{\omega_n^2}{2J^2}-1\right)} \left(\frac{N}{k} \sqrt{4-\frac{\omega_n^2}{J^2}}\right)^{k^2},
\ee
the $N/k$ factor in the last parenthesis indicates a change of behaviour as $k$ approaches a fixed fraction of $N$.

Now we can multiply the contributions for all the single modes $Z_n$ calculated in \eqref{omega>2J} and \eqref{omega<2J} to calculate the moments of the spectral form factor.
For early times such that $JT < \frac{\pi}{2}$, all the frequencies $\omega_n > 2J$, so using \eqref{omega>2J} the spectral form factor is given by the infinite product
\be
\langle |Z(\ii T)|^{2k} \rangle = 2^{kN} \prod_{n=0}^\infty e^{-2kNV(\lambda_+)} \left(1+\frac{\lambda_+^2}{J^2}\right)^{-2k^2}.
\ee
To leading order for small $JT$ this gives
\be
\langle |Z(\ii T)|^{2k} \rangle = 2^{kN} e^{-\frac{kN}{8} (JT)^2},
\ee
displaying the characteristic decaying behaviour typical of the slope. For $k=1$ this agrees with \cite{Winer:2020mdc}.
For later times there will be a contribution from the frequencies $\omega_n$ with $\omega_n < 2J$. Using \eqref{omega>2J} and \eqref{omega<2J}, the spectral form factor is then given by the infinite product
\be
\ba
\langle |Z(\ii T)|^{2k} \rangle = 2^{kN} & \prod_{n=0}^{n_T} \left(\frac{J}{\omega_n}\right)^{2kN} e^{kN\left(\frac{\omega_n^2}{2J^2}-1\right)} \left(\frac{N}{2\pi} \sqrt{4-\frac{\omega_n^2}{J^2}}\right)^{k^2} \text{vol}\left(\frac{\mathrm{U}(2k)}{\mathrm{U}(k)^2}\right)\\
\times & \prod_{n={n_T}+1}^\infty e^{-2kNV(\lambda_+)} \left(1+\frac{\lambda_+^2}{J^2}\right)^{-2k^2},
\ea
\ee
where $n_T = \lfloor \frac{JT}{\pi}-\frac{1}{2} \rfloor$ is the number of frequencies $\omega_n$ with $\omega_n < 2J$. Notice that the number of zero modes increases by one at discrete time intervals of size $\frac{\pi}{J}$, with the first zero mode appearing at $T=\frac{\pi}{2J}$, consistent with the numerical simulation in figure \ref{fig:SFFSYK2}. For large $JT$, the frequencies are very closely spaced so we can approximate the discrete variable $\omega_n$ by a continuous variable. This gives
\be
\langle |Z(\ii T)|^{2k} \rangle = 2^{kN} \left[ \left(\frac{32N}{\pi e^3}\right)^{k^2} \text{vol}\left(\frac{\mathrm{U}(2k)}{\mathrm{U}(k)^2}\right) \right]^{\frac{JT}{\pi}}.
\label{eq:SYK2_moments_small_k}
\ee
The contribution of the classical action vanishes to leading order in $JT$ as there is no exponential in $N$ contribution (except for the $2^{kN}$), leaving only the contribution from the zero mode volume, which generates an exponential in $T$ ramp. Taking $k \gg 1$ but $k \ll N$ we find
\be \label{syk:naive extrap}
\langle |Z(\ii T)|^{2k} \rangle \sim 2^{kN} \left( \frac{16}{e^{3/2}} \frac{N}{k} \right)^{k^2 \frac{JT}{\pi}}.
\ee
This indicates a significant change in behvaiour of the moments as $k$ approaches a fixed fraction of $N$. In particular, naively extrapolating to $k \gg N$ suggests that the moments begin to decrease. We will return to this problem in section \ref{syk:sec: large k}.

In these formulas we have only considered the one loop contribution from configurations which are invariant under a diagonal time translation. The full one loop contribution can be understood analytically for late times $T$. In this limit the dominant contribution is from the soft modes, which corresponds to making the zero mode configurations depend slightly on $t+t'$. The calculation is a straightforward generalisation of the one in \cite{Winer:2020mdc}, so we defer the details to appendix \ref{syk:app:one loop}. The result is
\be
\langle |Z(\ii T)|^{2k} \rangle = 2^{kN} \left[ \left(\frac{8N}{\pi e^2 JT}\right)^{k^2} \text{vol}\left(\frac{\mathrm{U}(2k)}{\mathrm{U}(k)^2}\right) \right]^{\frac{JT}{\pi}},
\label{eq:SYK2_moment}
\ee
where the main effect is that $N$ dependence has been rescaled, $N \to N/JT$. The moments grow exponentially quickly with $k$, in sharp contrast to what we found in the $q>2$ model, where the $k$ dependence was just quadratic. Moreover, another important different features is that the moments are strongly time dependent, indicating that the fluctuation in the spectral form factor are grows exponentially with the time, which is another different feature compared to the $q>2$ cases.

Notably, equation \eqref{eq:SYK2_moment} indicates a breakdown of the saddle point approximation as $JT$ approaches a fixed fraction of $N$, which corresponds to the beginning of plateau region. While we don't have an understanding of how to study the plateau using the collective field variables we can understand what the result should be based on a long-time average argument. By doing the Gaussian integral over the fermions, the unaveraged spectral form factor is
\be
|Z(\ii T)|^2 = 2^N \prod_{i=1}^{N/2} \cos^2 \frac{T \lambda_i}{2},
\ee
where $\pm \ii \lambda_i$ are the eigenvalues of $J_{ij}$. A long-time average implies the moments are
\be
\langle |Z(\ii T)|^{2k} \rangle = \binom{2k}{k}^{N/2},
\label{eq:plateau_SYK2}
\ee
in contrast with the plateau value calculated in \eqref{eq:long_time_average_SYK_RMT} for a RMT. The agreement of this result with the numerical value is showed in figure \ref{fig:SFFSYK2} for $N=50$ and $k=2$. A comparison of the plateau value for larger values of $k$ or $N$ is complicated by the large fluctuations that affect the SYK$_2$ model,  which rapidly render the required simulation time unfeasible. As observed in \cite{Legramandi:2024kcv}, insufficient statistical sampling leads to a downward shift in the spectral form factor, particularly affecting the plateau region where fluctuations are most pronounced. This issue becomes even more severe for the moments, where fluctuations grow exponentially with $N$.

\subsection{Large $k$ from duality} \label{syk:sec: large k}

For $k \gg N$ it is no longer possible to neglect the contribution of the Vandermonde determinant in \eqref{eigenvalue integral} when determining the saddle points. However, there is a simple way to rewrite the original $2k \times 2k$ matrix integral \eqref{matrix integral mode}
\be \label{Zn recap}
Z_n = \left( \frac{N}{2\pi J^2} \right)^{2k^2} \int_{\R^{(2k)^2}} \dd \Sigma e^{-\frac{N}{2J^2} \Tr\Sigma^2} \det\left(\ii \omega_n+\Sigma\right)^N.
\ee
as an $N \times N$ matrix integral where the saddle point analysis is simple for $k \gg N$. For sake of notation we have added back the contribution from the free determinant in this expression. We will subtract it's contribution in the final expression. This duality is a special case of a duality found by Hikami and Brézin \cite{Brezin:2007wv}. We start by introducing complex Grassmann variables $\chi_i^a$ and $\bar{\chi}_i^a$ with $i=1,\dots,N$ and $a=1,\dots,2k$ to rewrite the determinant as
\be \label{det grassmann}
\det\left(\ii \omega_n+\Sigma\right)^N = \int \dd \chi \dd \bar{\chi} e^{\Tr B\left(\ii \omega_n+\Sigma\right)},
\ee
where $\dd \chi \dd \bar{\chi} = \prod_{i,a} \dd \chi_i^a \dd \bar{\chi}_i^a$ and $B$ is the $2k\times2k$ matrix
\be
B^{ab}=\sum_i \bar{\chi}_i^a \chi_i^b.
\ee
With the Grassmann integral representation of the determinant \eqref{det grassmann} the original matrix integral \eqref{Zn recap} becomes Gaussian. Integrating out $\Sigma$ gives
\be \label{grassmann integral B}
Z_n = \int \dd \chi \dd \bar{\chi} e^{\ii \omega_n \Tr B + \frac{J^2}{2N} \Tr B^2}.
\ee
By anticommuting the Grassmann variables we may write
\be
\ba
\Tr B^2 &= \sum_{a,b} \sum_{i,j} \bar{\chi}_i^a \chi_i^b \bar{\chi}_j^b \chi_j^a\\
&= -\sum_{a,b} \sum_{i,j} \bar{\chi}_i^a \chi_j^a \bar{\chi}_j^b \chi_i^b\\
&= - \Tr \widetilde{B}^2,
\ea
\ee
where $\widetilde{B}$ is the $N \times N$ matrix
\be
\widetilde{B}_{ij} = \sum_a \bar{\chi}_i^a \chi_j^a.
\ee
Noting that also $\Tr B = \Tr \widetilde{B}$, we may rewrite \eqref{grassmann integral B} as
\be \label{Zn B tilde}
Z_n = \int \dd \chi \dd \bar{\chi} e^{\ii \omega_n \Tr \widetilde{B} - \frac{J^2}{2N} \Tr \widetilde{B}^2}.
\ee
By introducing an $N \times N$ matrix $\widetilde{\Sigma}$ we may write
\be
e^{-\frac{J^2}{2N} \Tr \widetilde{B}^2} = \left(\frac{N}{2\pi J^2}\right)^{\frac{N^2}{2}} \int_{\R^{N^2}} \dd \widetilde{\Sigma} e^{-\frac{N}{2J^2} \Tr \widetilde{\Sigma}^2+\ii \Tr \widetilde{B} \widetilde{\Sigma}}.
\ee

\begin{figure}
	\centering
    \begin{tikzpicture}
\node at (0,0) {\includegraphics[width=0.6\textwidth]{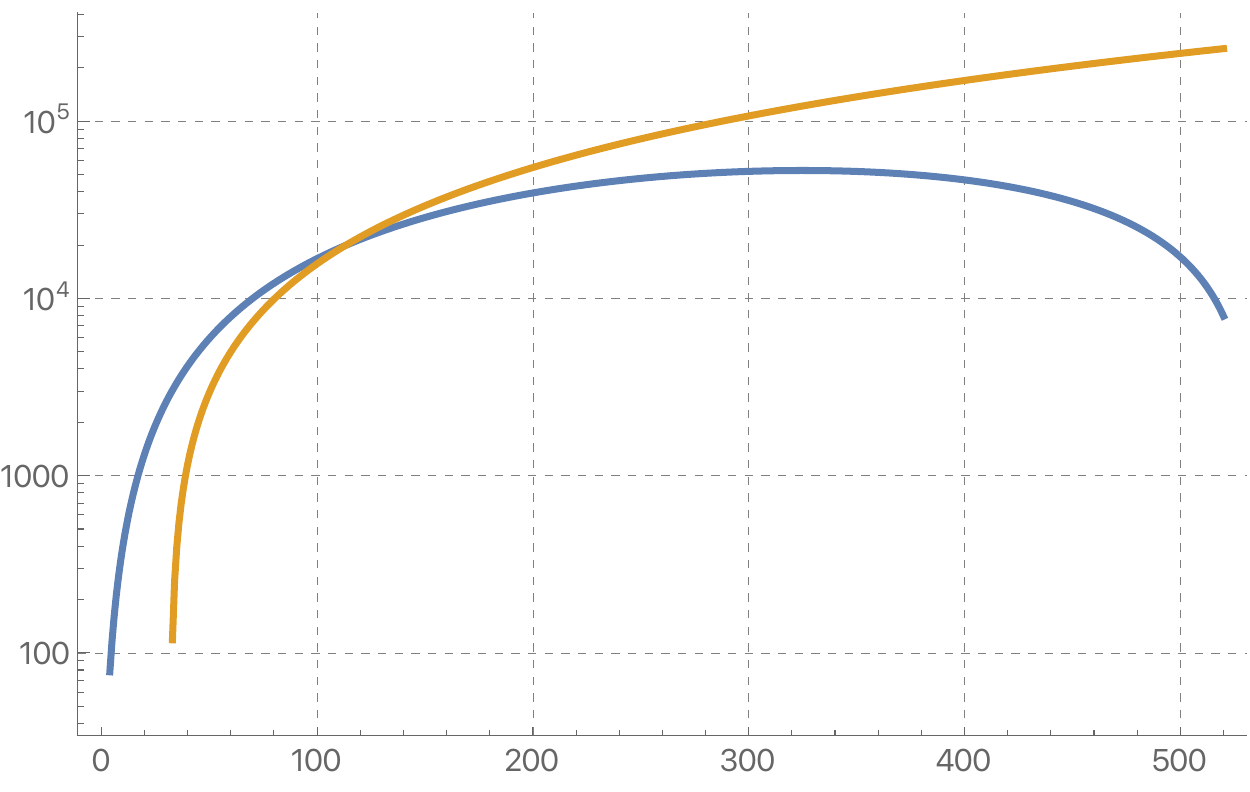}};
\node at (current bounding box.south) {{\small $k$}};
\node at (current bounding box.west) {\small \rotatebox{90}{$f(k)$}};
\end{tikzpicture}
\caption{Plot of $f(k)=\frac{\pi}{JT} \log (\langle |Z(\ii T)|^{2k}\rangle / 2^{kN} )$ for $N=150$, comparing two different expressions for the moments of the spectral form factor. The blue curve is obtained from \eqref{eq:SYK2_moments_small_k} and is valid for $k\ll N$ while the orange one, given by \eqref{eq:SYK2_moments_large_k}, is valid for $k \gg N$. Both curves are extrapolated beyond their expected range of validity. The two curves cross over at $k \sim 100$, reflecting the expectation that the formula for the low moments breaks down as $k$ approaches a fixed fraction of $N$ and the orange curve correctly captures the behaviour in the $k \gg N$ regime.}
\label{fig:SYK2_SFF}
\end{figure}

Plugging this expression into \eqref{Zn B tilde} and integrating over the Grassmann variables gives
\be
Z_n = \left(\frac{N}{2\pi J^2}\right)^{\frac{N^2}{2}} \int_{\R^{N^2}} \dd \widetilde{\Sigma} e^{-\frac{N}{2J^2} \Tr \widetilde{\Sigma}^2} \det\left[\ii(\omega_n+\widetilde{\Sigma})\right]^{2k}.
\ee
The duality exchanges the original integral over a $2k \times 2k$ hermitian matrix with a determinant insertion raised to the $N$\textsuperscript{th} power with an integral over an $N \times N$ hermitian matrix with a determinant insertion raised to the $2k$\textsuperscript{th} power.\footnote{At the level of the full path integral this duality simply undoes the $G, \Sigma$ trick, see appendix \ref{syk:app:q=2}.} Finally, subtracting the contribution from the free determinant gives
\be
Z_n = \left(\frac{N}{2\pi J^2}\right)^{\frac{N^2}{2}} \int_{\R^{N^2}} \dd \widetilde{\Sigma} e^{-\frac{N}{2J^2} \Tr \widetilde{\Sigma}^2} \det\left(1+\frac{\widetilde{\Sigma}}{\omega_n}\right)^{2k}.
\ee
The matrix integral over $\widetilde{\Sigma}$ is invariant under conjugation by a unitary matrix $\widetilde{\Sigma} \to U\widetilde{\Sigma}U^\dagger$ with $U \in \mathrm{U}(N)$. Using this symmetry to gauge fix $\widetilde{\Sigma}$ to be a diagonal matrix gives
\be
Z_n = \left( \frac{N}{2\pi J^2} \right)^{\frac{N^2}{2}} \text{vol}\left(\frac{\mathrm{U}(N)}{\mathrm{U}(1)^N \times S_N}\right) \int_{\R^N} \prod_i \dd \lambda_i \prod_{i<j} |\lambda_i-\lambda_j|^2 e^{-N \sum_i\widetilde{V}(\lambda_i)},
\ee
where
\be
\widetilde{V}(\lambda) = -\frac{2k}{N} \log\left(1+\frac{\lambda}{\omega_n}\right) + \frac{1}{2J^2} \lambda^2.
\ee
For infinite $k$ the saddle points for integral over eigenvalues are pushed to $\pm \infty$. For large but finite $k$ the quadratic term in the potential provides a small stabilisation which pulls the saddle points from $\pm \infty$ to a finite distance. The advantage of the dual description of the original matrix integral is that for $k \gg N$ we can self-consistently neglect the contribution from the Vandermonde determinant when determining the saddle points. The extremums of $\widetilde{V}(\lambda)$ are
\be
\widetilde{\lambda}_\pm = \frac{-\omega_n \pm \sqrt{\frac{8k}{N}J^2+\omega_n^2}}{2}.
\ee
The saddle point $\widetilde{\lambda}_+$ dominates for all $\omega_n$ and so
\be
Z_n = e^{-N^2 \widetilde{V}(\widetilde{\lambda}_+)} \left(1+\frac{N}{2k} \frac{\widetilde{\lambda}_+^2}{J^2}\right)^{-\frac{N^2}{2}}.
\ee
For large $JT$, the frequencies are very closely spaced so we can approximate the discrete variable $\omega_n$ by a continuous variable. This gives
\be
\langle |Z(\ii T)|^{2k} \rangle = 2^{kN} \exp\left[ \frac{4}{3} \sqrt{2kN} \left(k-\frac{3(\pi-2)}{16}N\right) \frac{JT}{\pi} \right].
\label{eq:SYK2_moments_large_k}
\ee
This indicates that the moments continue to grow as $k$ increases beyond $N$, in contrast to the naive extrapolation of \eqref{syk:naive extrap}. A comparison between
\eqref{eq:SYK2_moments_small_k} and \eqref{eq:SYK2_moments_large_k} is shown in figure \ref{fig:SYK2_SFF}. We stress that in this formula we have not accounted for the one loop determinant from the configurations which are not invariant under the diagonal time translation symmetry.

\section{Sparse SYK}
\label{sec:sparse_SYK}

In section \ref{sec:1/N}, we observed that the leading correction to the moments of the spectral form factor scales as $k^2/N^{q-2}$, indicating a change in the behaviour for the high moments, when $k \sim N^{q/2-1}$. As discussed there, the correction in equation \eqref{eq:final_result_correction} scales inversely with $N^q/q!$ which is the number of independent random parameters in the SYK Hamiltonian. In this section we try to understand in what sense this is the correct interpretation by considering a sparsified version of the SYK model where we can tune the number of random parameters. While certain signatures of quantum chaos, such as a linear ramp, are preserved in the sparse model, we expect that reducing the number of random parameters weakens its adherence to random matrix universality.

\begin{figure}
\centering
\begin{subfigure}{0.48\textwidth}
\raggedleft
\begin{tikzpicture}
\node at (0,0) {\includegraphics[width=0.95\linewidth]{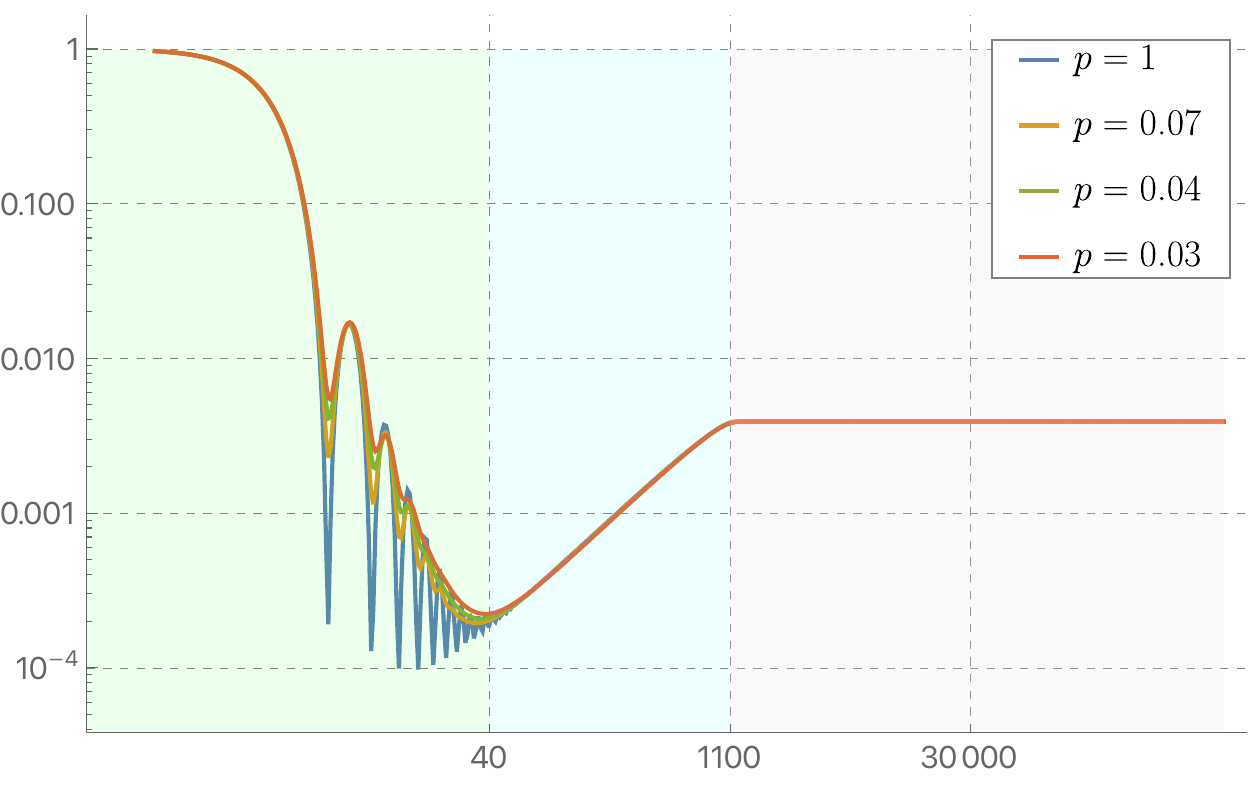}};
\node at (current bounding box.south) {{\scriptsize $JT$}};
\node at (current bounding box.west) {\scriptsize \rotatebox{90}{$\langle |Z(\ii T)|^2 \rangle$}};
\end{tikzpicture}
\end{subfigure}
% $\quad$
\begin{subfigure}{0.48\textwidth}
\raggedright
\begin{tikzpicture}
\node at (0,0) {\includegraphics[width=0.95\linewidth]{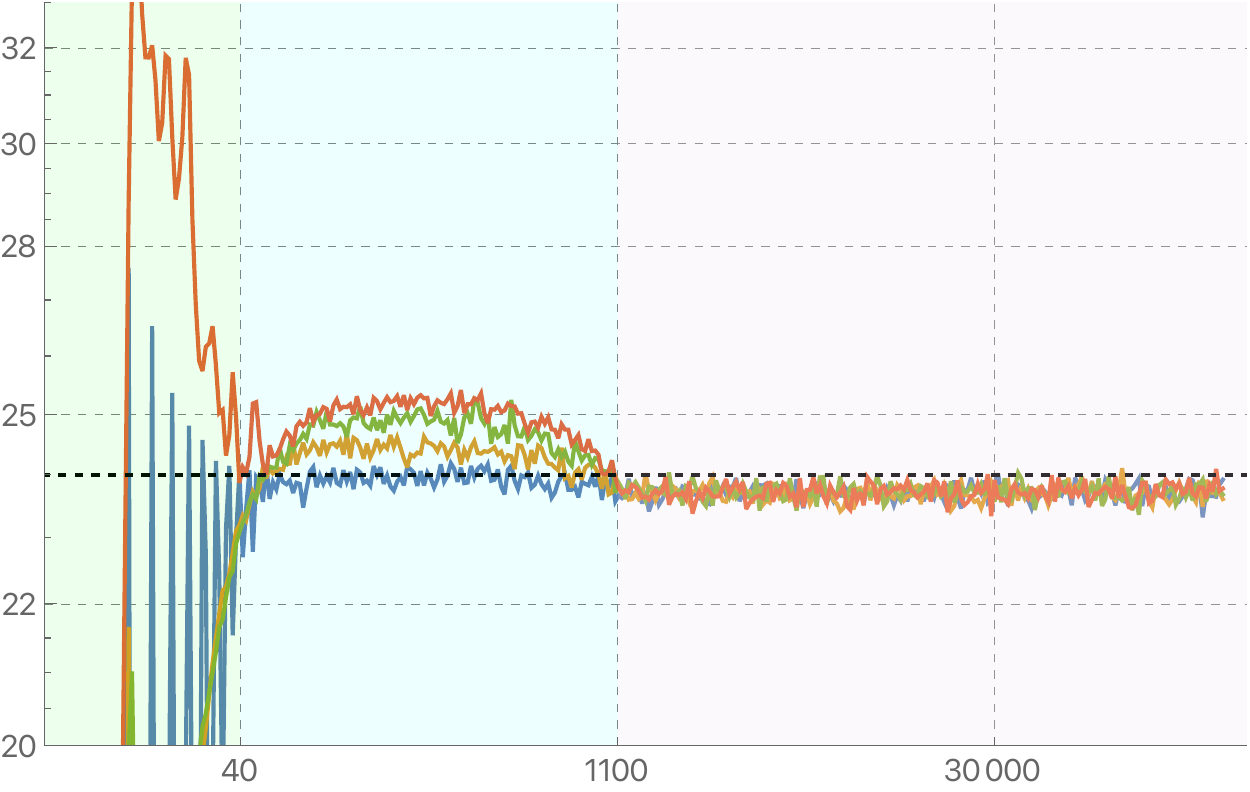}};
\node at (current bounding box.south) {{\scriptsize $JT$}};
\node at (current bounding box.west) {\scriptsize \rotatebox{90}{$\langle |Z(\ii T)|^{8} \rangle / \langle |Z(\ii T)|^2 \rangle^4$}};
\end{tikzpicture}
\end{subfigure}
%\par\medskip
\begin{subfigure}{0.48\textwidth}
\raggedright
\begin{tikzpicture}
\node at (0,0) {\includegraphics[width=0.95\linewidth]{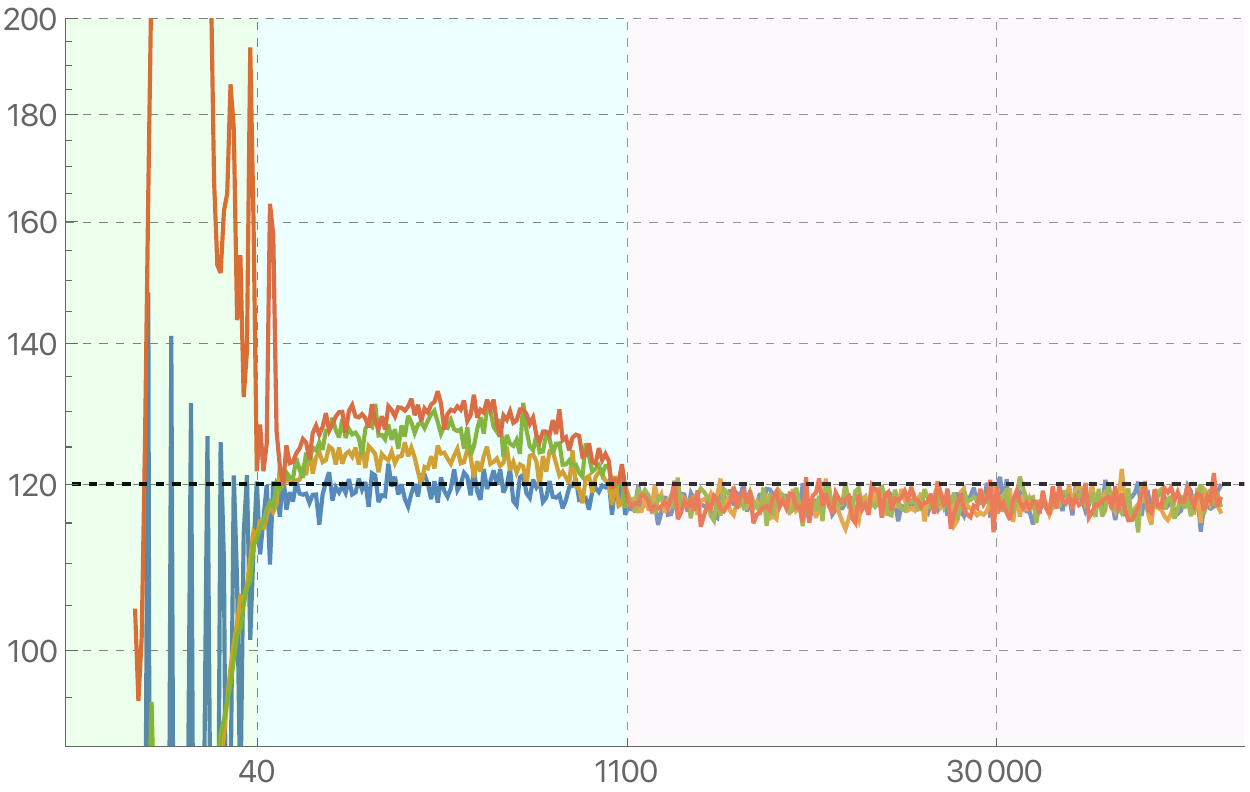}};
\node at (current bounding box.south) {{\scriptsize $JT$}};
\node at (current bounding box.west) {\scriptsize \rotatebox{90}{$\langle |Z(\ii T)|^{10} \rangle / \langle |Z(\ii T)|^2 \rangle^5$}};
\end{tikzpicture}
\end{subfigure}
%$\quad$
\begin{subfigure}{0.48\textwidth}
\raggedright
\begin{tikzpicture}
\node at (0,0) {\includegraphics[width=0.95\linewidth]{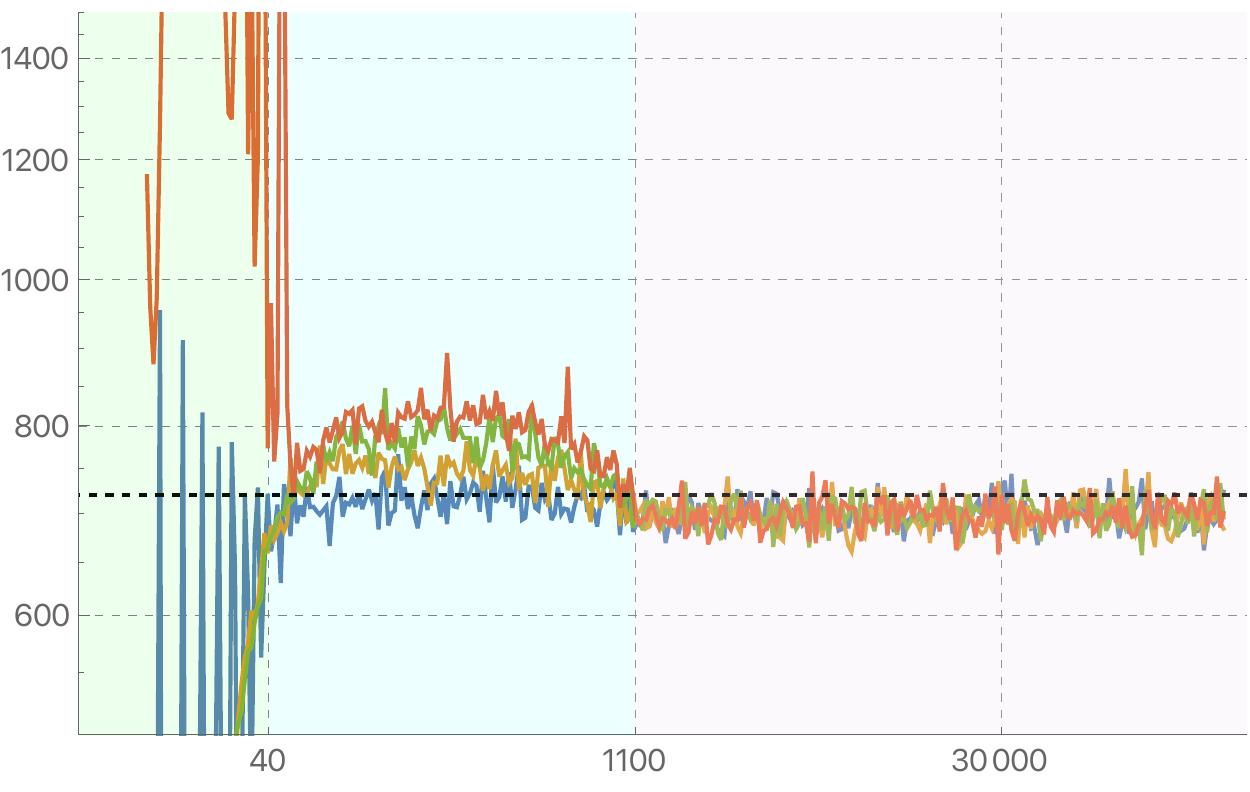}};
\node at (current bounding box.south) {{\scriptsize $JT$}};
\node at (current bounding box.west) {\scriptsize \rotatebox{90}{$\langle |Z(\ii T)|^{12} \rangle / \langle |Z(\ii T)|^2 \rangle^6$}};
\end{tikzpicture}
\end{subfigure}
\caption{The spectral form factor and its moments for the sparse SYK model with $N=18$ fermions, averaged over 1.8 million realisations. The plots correspond to $k=1, 4, 5, 6$ from left to right. Different colours represent different degrees of sparsification: $p=1$ i.e. unsparsified SYK (blue), $p=0.07$ (yellow), $p=0.04$ (green), and $p=0.03$ (red). The black dashed line marks the leading contribution of $k!$. The observed plateau lies below $k!$ due to finite $N$ effects. The sparsification probabilities have been choose in such a way that the sparse model agrees with the regular SYK model in the ramp region (cyan background). This is consistent with \cite{Orman:2024mpw}, where a transition, signaled by a growing dip, was noticed for $p \lesssim 8.7 \times 24 / N^3 \sim 0.036$ for $N=18$. However, the agreement in the ramp region is lost when considering the moments: while for regular SYK they closely approach $k!$, sparse SYK shows increasing deviations at higher sparsification. Notably, the gap between the moments and $k!$ grows significantly with $k$, making the correction more and more relevant.}
\label{fig:spars_moments}
\end{figure}

Deleting some of the terms in the SYK Hamiltonian gives rise to a model known as the sparse SYK model. It gained a lot of attention recently because it significantly reduces the computational complexity of simulations yet still displays a holographic phase in certain regimes \cite{Xu:2020shn}. The computational advantage is particularly relevant when dealing with the moments of the spectral form factor since the simulation time is enlarged by the huge statistics needed to stabilise their fluctuations. Indeed, the standard deviation associated to $|Z(iT)|^{2k} / \langle |Z(iT)|^2 \rangle^k$ is of  order
\begin{equation}
    \sqrt{(2k)!} \sim \left( \frac{2k}{e} \right)^k,
    \label{eq:fluctuations_moments_num}
\end{equation}
which is superexponential in $k$. This would be particularly problematic if the correction to the moments only becomes significant at $k \sim N^{q/2-1}$ as for regular SYK. So, from a computational perspective, sparse SYK offers a dual advantage: it reduces the simulation time while also making the corrections more prominent even for $k \ll N^{q/2-1}$.
Furthermore, studying the moments of the spectral form factor for sparse SYK allows us to examine the regime of validity for the model: the $k$\textsuperscript{th} moment of the spectral form factor can be related to $2k$-point out-of-time-order correlators\footnote{The relation becomes more precise at late time or averaging over the operator in the OTOC.} \cite{Cotler:2017jue,Saad:2019pqd}, so a substantial deviation of the moments with respect to regular SYK suggests that sparse SYK is no longer a good approximation of the model, and one should only trust it for computing lower-point functions.

The Hamiltonian for sparse SYK is
\begin{align} 
H = \ii^{q/2} \sum_{1 \leq i_1 < \dots < i_q \leq N} x_{i_1 \dots i_q} J_{i_1 \dots i_q} \psi_{i_1} \dots \psi_{i_q},
\end{align}
where $x_{i_1 \dots i_q}$ is a binary random variable that is equal to 0 with probability $1-p$ and equal to 1 with probability $p$. On average the number of nonzero independent random variables in the Hamiltonian is approximately $p N^q/q!$ for large $N$. The random couplings $J_{i_1 \dots i_q}$ have zero mean and variance
\begin{align} 
\langle J_{i_1 \dots i_q} J_{j_1 \dots j_q} \rangle = \frac{1}{p} \frac{J^2 (q-1)!}{N^{q-1}} \delta_{i_1j_1} \dots \delta_{i_qj_q}.
\end{align}
Notice that the variance of the couplings slightly different as compared with \eqref{J variance} in order to account for the fact that many of the interactions are suppressed. For the remainder of this section, we fix $q=4$. 

It has been argued that the sparse SYK model remains holographic so long as $p \gtrsim N^{-3}$ \cite{Xu:2020shn}. For smaller values, deviations from random matrix statistics cast doubt on the existence of a holographic interpretation. Numerical analysis showed that the spectral form factor exhibits no significant deviations in the dip and ramp region within the holographic phase \cite{Orman:2024mpw}. In figure \ref{fig:spars_moments} we considered the higher moments of the spectral form factor for different $p$, for regimes where the linear ramp doesn't deviate significantly from the unsparsified model.\footnote{In principle, one can also consider stronger sparsification $p< N^{-3}$, where the spectral form factor itself begins to deviate from RMT predictions. Specifically, the dip is lifted and the onset of the ramp is delayed. Since the spectral form factor is qualitatively different in this regime, no agreement with the regular SYK model can be expected for the moments of the spectral form factor, which are subject to large fluctuations until the ramp begins. After the ramp the fluctuations reduce following the analytical expectation. This is very similar to the behaviour of the $p=0.03$ case in figure \ref{fig:spars_moments}, which indeed has a $p$ very close to the transition.} Even if the spectral form factor is qualitatively the same, suggesting the system is in the holographic phase, its moments show significant deviations from the RMT expectation, which become more pronounced as the model becomes sparser.
The deviations fluctuate around a constant value for the bulk of the ramp region. We have also checked this effect for larger values of $N$, where it is more evident since the ramp region is stretches over a greater time scale.

\begin{figure}
	\centering
    \begin{tikzpicture}
\node at (0,0) {\includegraphics[width=0.6\textwidth]{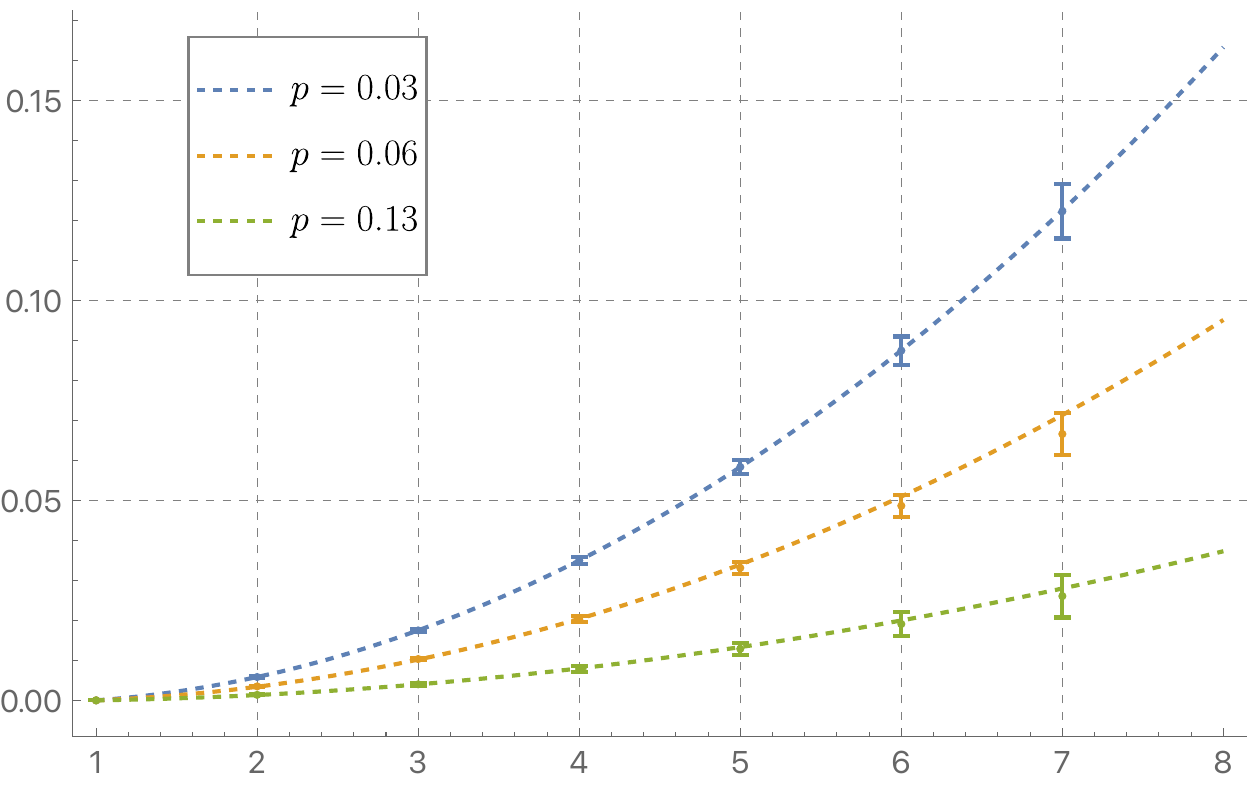}};
\node at (current bounding box.south) {{\small $k$}};
\node at (current bounding box.west) {\small \rotatebox{90}{$\overline{B}$}};
\end{tikzpicture}
	\caption{The behaviour of the time-averaged correction $\overline{B}$ defined in \eqref{eq:averaged_sparse_ramp} as a function of $k$. We considered $N=18$ sparsified SYK model with $p=0.03$ (blue), $p=0.06$ (orange) and $p=0.13$ (green) averaged over $1.8$M realizations. The error bars are calculated with respect to the time average. The dashed lines are an interpolation of the theoretical expectation $\overline{B} \propto k(k-1)$, which shows a very good agreement. As expected, $\overline{B}$ grows with as the model becomes more sparse.}
	\label{fig:B_k}
\end{figure}

Building on this qualitative picture, we aim to determine whether the corrections to the moments in the sparse SYK model, given by
\begin{equation}
    B(T) = \frac{1}{k!}\frac{\langle |Z(\ii T)|^{2k} \rangle}{\langle |Z(\ii T)|^2\rangle^k} - 1 \, ,
\end{equation}
follow the analytical result in \eqref{eq:final_result_correction} derived for the dense case. We first focus on the $k$ dependence. To simplify the analysis and avoid unnecessary complications from time dependence, we average $B(T)$ over the ramp region $[T_\text{Dip},T_{\text{Pl}}]$ defining 
\begin{equation}
    \overline{B} = \frac{1}{T_{\text{Pl}}-T_{\text{Dip}}} \int_{T_{\text{Pl}}}^{T_\text{Dip}}  B(T) \dd T .
        \label{eq:averaged_sparse_ramp}
\end{equation}
The result of this interpolation is shown in figure \ref{fig:B_k}, showing a good agreement with the analytical prediction 
\begin{equation}
    \overline{B}= \alpha k (k-1),
    \label{eq:averaged_sparse_ramp2}
\end{equation}
for different levels of sparsifications. 
A similar computation for regular SYK leads to inconclusive results, since $\overline{B}$ is significantly smaller compared to the sparse SYK case, while the size of the fluctuations is fixed by equation \eqref{eq:fluctuations_moments_num} at leading order, independent of $p$.

\begin{figure}
	\centering
    \begin{tikzpicture}
\node at (0,0) {\includegraphics[width=0.6\textwidth]{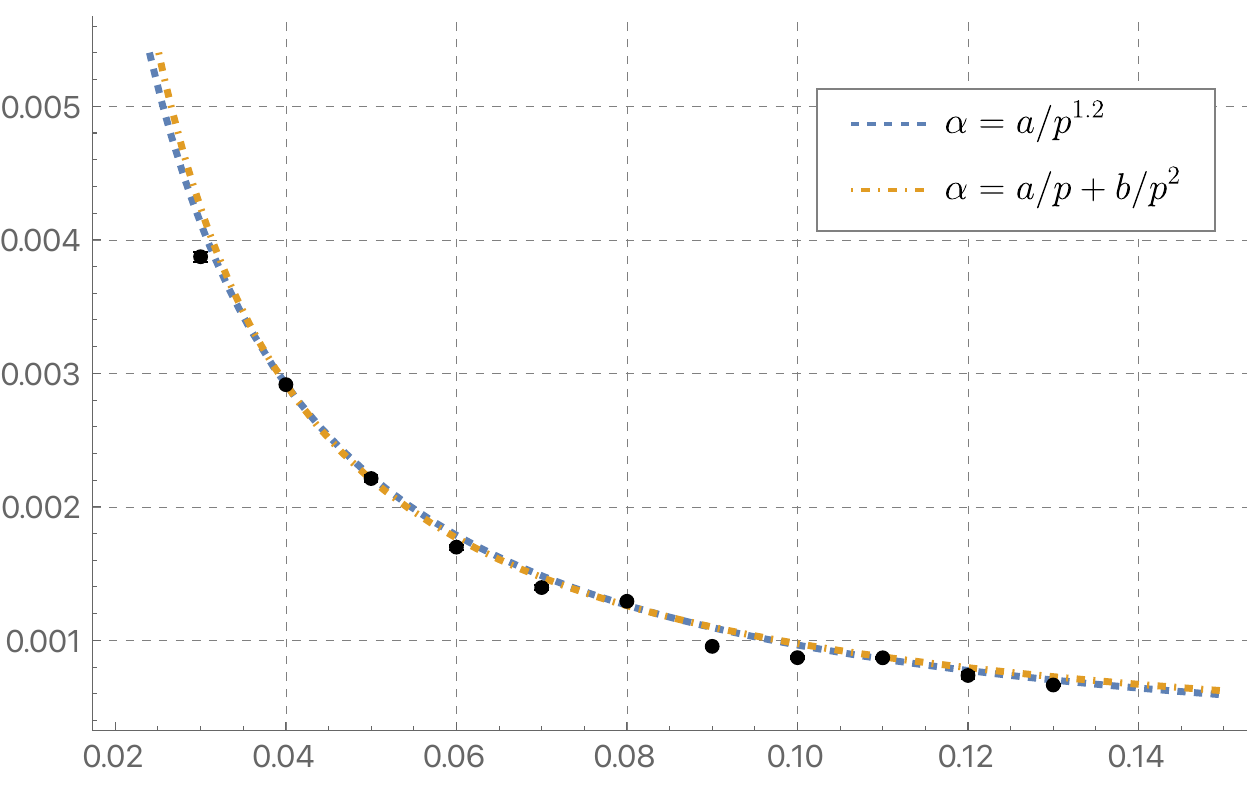}};
\node at (current bounding box.south) {{\small $p$}};
\node at (current bounding box.west) {\small \rotatebox{90}{$\alpha$}};
\end{tikzpicture}
	\caption{Dependence of $\alpha$, as defined in \eqref{eq:averaged_sparse_ramp2}, on $p$ for 1.8M samples of an $N=18$ sparsified SYK model. Each value of $\alpha$ was obtained by interpolating $\overline{B}(k)$, as shown in figure \ref{fig:B_k}. According to our expectation, the leading correction to the moments of the spectral form factor is inversely proportional to the number of independent random variables, so $\alpha \sim 1/p$. A best fit yields $\alpha \sim 1/p^{1.2}$ (blue dashed line). Additionally, we considered a fit accounting for a potential second-order correction proportional to the square of the number of independent degrees of freedom (orange dash-dotted line), which also shows good agreement with the data.}
	\label{fig:alpha}
\end{figure}

We can also analyse quantitatively the hypothesis that the size of the leading correction to the moments of the spectral form factor is proportional to the number of independent random variables. To do so, we have to determine the dependence of the coefficient $\alpha$ defined in \eqref{eq:averaged_sparse_ramp2} on the sparsification parameter $p$. Since the number of nonzero couplings is proportional to $p$, we expect 
\begin{equation}
\label{eq:alpha_expect}
\alpha \sim \frac{1}{p}.
\end{equation}
However, there can be deviations to this expectation since the subleading terms in the perturbative expansion might become quickly more relevant in sparse SYK as compared with the unsparsified model. Moreover, a meaningful estimate of $\alpha$ requires $\overline{B}$ to be large compared to the statistical fluctuations. This forces us to focus on values of $p$ which are not too far from the transition at which the onset of the dip starts to increase. We expect that this might also affect dependence of $\alpha$ on $p$, which is shown in figure \ref{fig:alpha}. The data is compatible with a power law behaviour, specifically $\alpha \sim p^{-1.2}$, which is in qualitative agreement with equation \eqref{eq:alpha_expect}. Additionally, figure \ref{fig:alpha} explores the possibility of deviations from the $1/p$ behaviour due to subleading corrections in the number of independent random variables, which is proportional to $1/p^2$. While this model fits the numerical data reasonably well, a more quantitative analysis is hindered by the lack of precise knowledge about the form of these subleading corrections.

The naive extrapolation of this analysis suggests that for the sparse SYK model we have an analogous formula as \eqref{eq:final_result_correction} but with the number of independent random variables now given by $p N^q/q!$.
If $\overline{\Delta}_E$ \eqref{bar delta E} is extensive in $N$ for any $p$, this suggests a breakdown of the perturbative series at $p \sim 1/N^2$ due to the fluctuations near the edge of the spectrum. In contrast, if we only focus on the bulk of the spectrum, i.e. away from the edge, we expect sparse SYK to resemble RMT all the way down to $p \sim 1/N^3$, as argued in \cite{Orman:2024mpw}. We leave a detailed investigation of this N dependence for future work.

We stress that the dependence of the leading corrections on the sparsification parameter $p$ is not specific to the moments of the spectral form factor. For example, a similar result has been found for the moments of the Hamiltonian in sparse SYK \cite{Garcia-Garcia:2020cdo}. In particular, it has been shown that, at leading order, the moments of the sparse SYK Hamiltonian behaves as the unsparsified case, except for corrections of the order of the number of independent interaction vertices.\footnote{This quantity is called $k N$ in the sparse SYK literature.} It would be interesting to extend that analysis to determine at which order of the moments we expect the corrections to become significant. 

\begin{figure}[!t]
\centering
\begin{subfigure}{0.48\textwidth}
\raggedleft
\begin{tikzpicture}
\node at (0,0) {\includegraphics[width=0.95\linewidth]{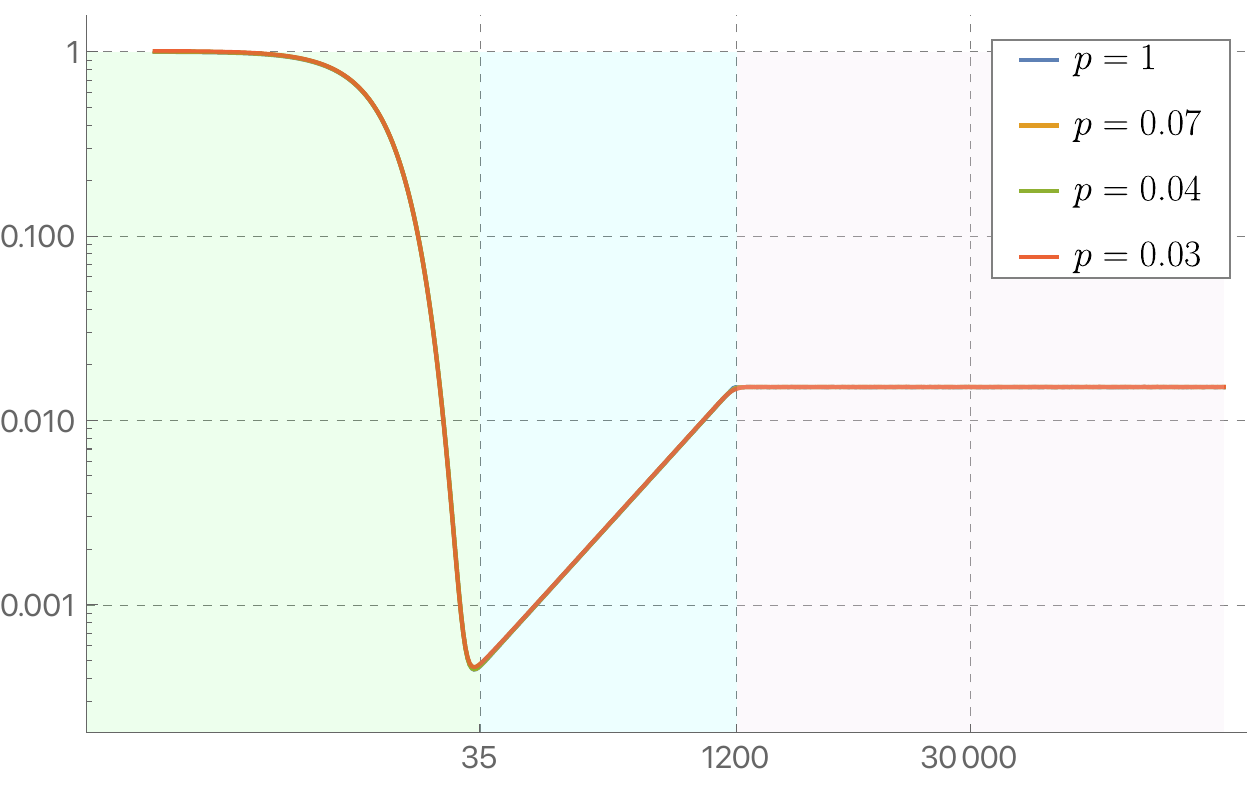}};
\node at (current bounding box.south) {{\scriptsize $JT$}};
\node at (current bounding box.west) {\scriptsize \rotatebox{90}{$\langle |Y(\ii T)|^2 \rangle$}};
\end{tikzpicture}
\end{subfigure}
% $\quad$
\begin{subfigure}{0.48\textwidth}
\raggedright
\begin{tikzpicture}
\node at (0,0) {\includegraphics[width=0.95\linewidth]{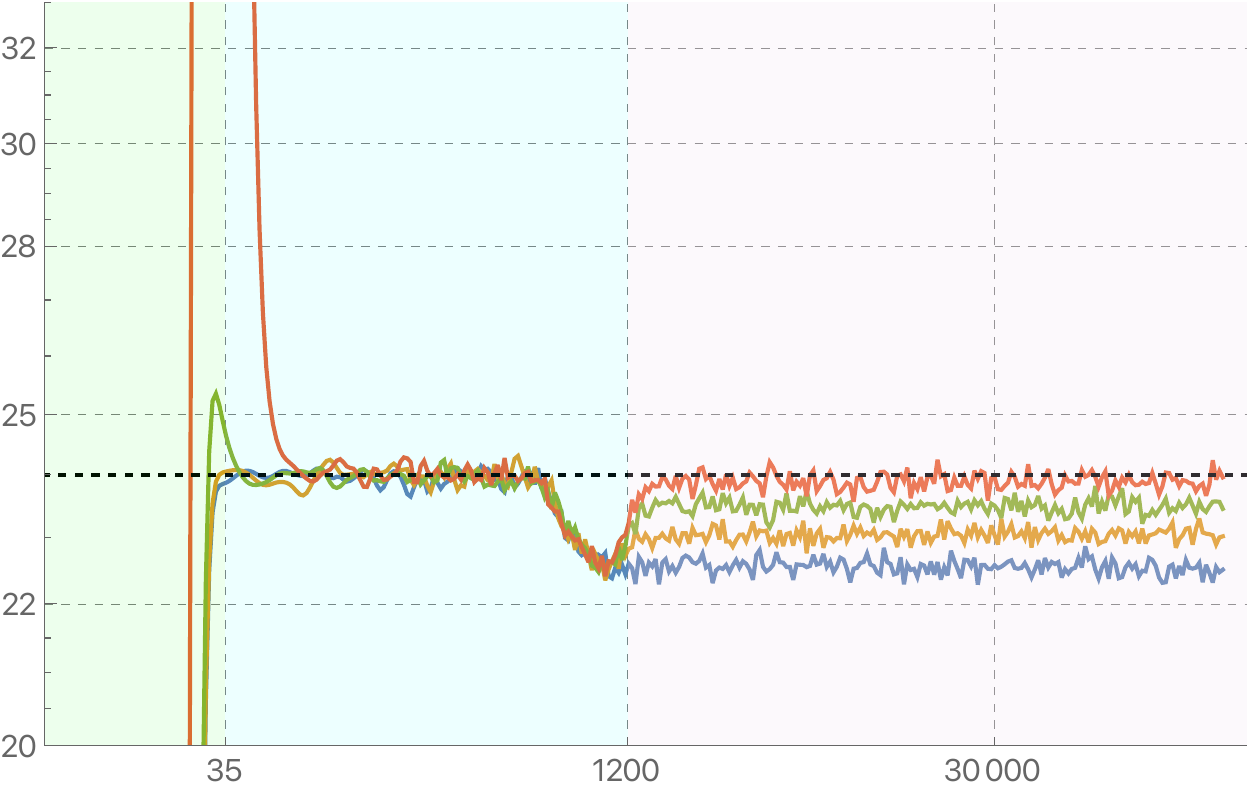}};
\node at (current bounding box.south) {{\scriptsize $JT$}};
\node at (current bounding box.west) {\scriptsize \rotatebox{90}{$\langle |Y(\ii T)|^{8} \rangle / \langle |Y(\ii T)|^2 \rangle^4$}};
\end{tikzpicture}
\end{subfigure}
%\par\medskip
\begin{subfigure}{0.48\textwidth}
\raggedright
\begin{tikzpicture}
\node at (0,0) {\includegraphics[width=0.95\linewidth]{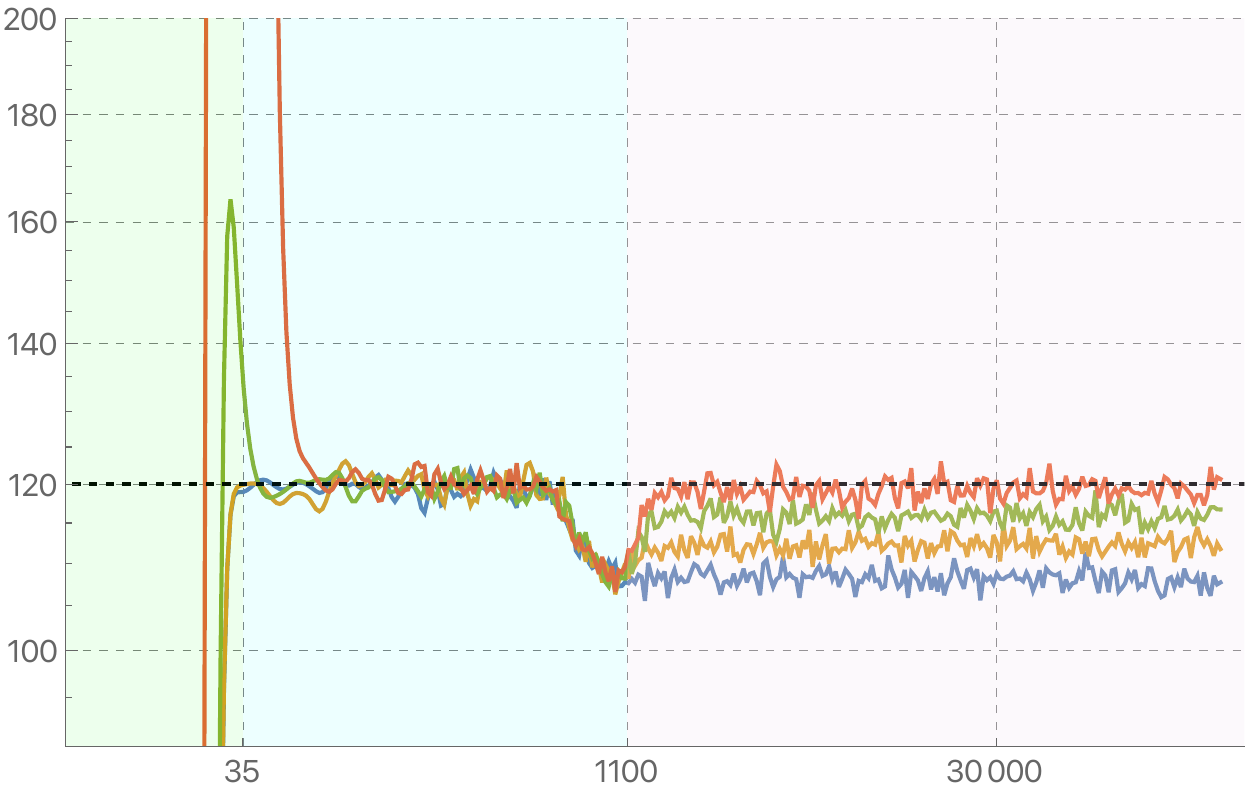}};
\node at (current bounding box.south) {{\scriptsize $JT$}};
\node at (current bounding box.west) {\scriptsize \rotatebox{90}{$\langle |Y(\ii T)|^{10} \rangle / \langle |Y(\ii T)|^2 \rangle^5$}};
\end{tikzpicture}
\end{subfigure}
%$\quad$
\begin{subfigure}{0.48\textwidth}
\raggedright
\begin{tikzpicture}
\node at (0,0) {\includegraphics[width=0.95\linewidth]{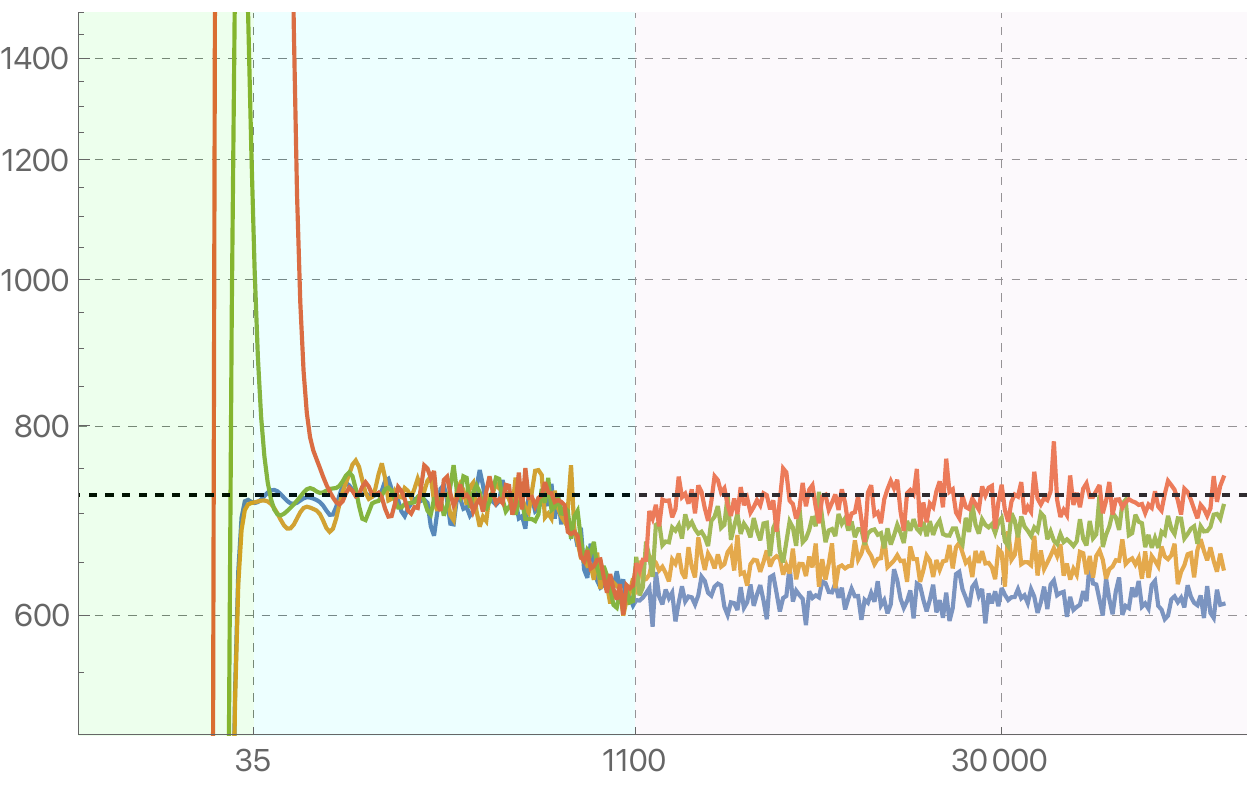}};
\node at (current bounding box.south) {{\scriptsize $JT$}};
\node at (current bounding box.west) {\scriptsize \rotatebox{90}{$\langle |Y(\ii T)|^{12} \rangle / \langle |Y(\ii T)|^2 \rangle^6$}};
\end{tikzpicture}
\end{subfigure}
\caption{The microcanonical spectral form factor $|Y(\ii T)|^2$ \eqref{eq:filtering_sparse} and its moments for the SYK model with $N=18$ fermions, averaged over 1.8 million realisations. The Gaussian filtering function is centered around zero with $\Delta = 0.1$. The plots correspond to $k=1, 4, 5, 6$ from left to right. Different colours represent varying degrees of sparsification: unsparsified SYK (blue), $p=0.07$ (yellow), $p=0.04$ (green), and $p=0.03$ (red). The black dashed line marks the leading contribution of $k!$. The filtering function reduces the effective Hilbert space dimension, affecting the plateau value in a more prominent way that what observed in figure \ref{fig:spars_moments}. In the ramp region (cyan background), in the bulk of the ramp region, the various curves fall one on top of the other, in contrast to the unfiltered spectral form factor \eqref{fig:spars_moments}. This signals that the leading correction to $k!$ behaviour is highly suppressed, as expected from the discussion in section \ref{sec:finite_beta}. Since the filtering function effectively cuts of the edges of the spectrum, this confirms the expectation that the correction to the moments of the spectral form factor during the ramp region arise from fluctuations near the edge of the spectrum.}
\label{fig:spars_moments_filtered}
\end{figure}

As a final point, we present the numerical results for the filtered or microcanonical version of the spectral form factor defined in \eqref{Yf}:
\begin{equation}
\label{eq:filtering_sparse}
|Y(T)|^2 = \Tr[f(H) e^{-\ii T H}] \Tr[f(H) e^{\ii T H}], \qquad f(H) = e^{-H^2 / 2 \Delta^2} \, .
\end{equation}
The rationale behind this choice of the filtering function $f(H)$ is that, by restricting to a sufficiently small $\Delta$, we can effectively cut off the edge of the spectrum, which, as discussed in section \ref{sec:finite_beta}, is responsible for deviations from RMT expectations during the ramp region. 
The numerical results for \(\langle |Y(\ii T)|^{2k} \rangle / \langle |Y(\ii T)|^{2} \rangle^k\) are shown in figure \ref{fig:spars_moments_filtered}, where we fixed $\Delta = 0.1$ ensuring that approximately $18\%$ of the energy levels around the middle of the spectrum contributes, as indicated by $\langle Y(0) \rangle/2^{N/2} \approx 0.18$.
It can be seen that in the bulk of the ramp region all the curves fall on on top of the other, in contrast with the unfiltered spectral form factor in figure \ref{fig:spars_moments}. 

\section{Discussion}
\label{sec:conclusion}

In this paper we analysed the moments of the spectral form factor in the SYK model and compared these results with expectations from RMT. In the large $N$ limit, the $q>2$ SYK model aligns with these expectations, however, perturbative corrections to the moments display non-universal features arising from fluctuations near the edge of the spectrum. For the high moments, the leading correction \eqref{eq:final_result_correction} scales with $k^2/N^{q-2}$ indicating a marked departure from random matrix behaviour when $k$ approaches a fixed fraction of $N^{q/2-1}$. 

The $q=2$ model is a free theory of fermions with a random mass matrix and as such is not expected to display features of many-body chaos. This is reflected in drastic changes to the spectral form factor which displays an exponential ramp with increased noise. By using a trick to rewrite the path integral, we also identified saddle points that could potentially contribute to the very high moments with $k \gg N$.

The leading correction \eqref{eq:final_result_correction} to the moments in the $q>2$ model scales inversely with the number of random parameters in the Hamiltonian. This suggests that the correction would be enhanced in a sparsified model, with fewer random parameters, leading to a faster departure from RMT expectations. In section \ref{sec:sparse_SYK} we provided numerical evidence supporting this hypothesis in the sparse SYK model, in which interaction terms are deleted with probability $1-p$.
We found notable deviations to the low order moments even for regimes where the onset of the linear ramp remains unchanged as compared with the unsparsified model.
 
This result suggests a criterion for the breakdown of random matrix universality based on the number of independent degrees of freedom and positions the moments of the spectral form factor as a more sensitive diagnostic of quantum chaos as compared to the spectral form factor itself, challenging the regime of validity of sparse models. It would be interesting to refine the numerical analysis in this paper to study the $N$ dependence of the corrections, since naively extrapolating our findings suggests that RMT behaviour may break down at a lower level of sparsification than previously predicted \cite{Xu:2020shn}, driven by fluctuations near the edge of the spectrum.

These considerations are particularly relevant in the context of experimental proposals for realising the SYK model in solid-state systems \cite{Pikulin_2017,Chew_2017,Chen_2018} as well as on quantum-simulation platforms \cite{Garcia_alvarez_2017, Babbush_2019}. Many of these implementations face scalability challenges, making the study of small $N$ effects particularly important. Furthermore, sparsification plays a critical role in these setups. Our analysis provides insights into how deviations from RMT expectations arise due to the limited number of random parameters, which is crucial for assessing the fidelity of table-top realisations.

Another intriguing direction would be to explore how the relation between the leading correction in equation \eqref{eq:final_result_correction} and the number of independent random parameters in the Hamiltonian extends to systems where the couplings are fixed.
In such cases, a running time average can be used to tame the erratic oscillations in the spectral form factor and study its moments. It would be interesting to investigate which features of the Hamiltonian drive the deviations from random matrix statistics in these scenarios. 
A possibility is that the role of the number of independent random parameters is replaced by the number of noncommuting interaction terms in the Hamiltonian, as this distinction also distinguishes dense systems like RMT and sparse systems like SYK.

Finally, we note that the techniques developed in this paper are also applicable to other disordered models, such as the SY model \cite{Sachdev1993}, as well as bosonic systems like the bosonic SYK model and spin glasses \cite{erdHos_2014,Baldwin:2019dki,Winer:2022ciz,Hanada:2023rkf}. It would be interesting to understand the behaviour of the moments of the spectral form factor in a gravitational setup with backreacting matter, as considered in \cite{Chen:2023hra}, to better understand these results from a bulk gravitational perspective.

\section*{Acknowledgments}
We would like to thank Timothy J. Hollowood and S. Prem Kumar for many helpful discussions throughout this work, and S. Prem Kumar for originally suggesting to study the moments of the spectral form factor. We also would like to thank Alex Windey for useful discussions. AL acknowledges support from Provincia Autonoma di Trento and by Q@TN, the joint lab between University of Trento. This work was supported by the Swiss State Secretariat for Education, Research and Innovation (SERI) under contract number UeMO19-5.1.
NT acknowledges support from STFC grant ST/T506473/1 and ERC-COG grant NP-QFT No. 864583 \say{Non-perturbative dynamics of quantum fields: from new deconfined
phases of matter to quantum black holes}, by the MUR-FARE2020 grant No. R20E8NR3HX \say{The Emergence of Quantum Gravity from Strong Coupling Dynamics}.

\appendix
\addtocontents{toc}{\protect\setcounter{tocdepth}{1}}

\section{Expectations from RMT} \label{app:rmt}

In this appendix we review how the behaviour of the moments of the spectral form factor during the slope and ramp regions can be understood as a perturbative effect in RMT \cite{Cotler_2017, AlSh86}. We explain this for the Wigner-Dyson and Altland-Zirnbauer ensembles and then apply these result to SYK \cite{Saad:2018bqo} after reviewing its symmetry classification \cite{Behrends_2019, You:2016ldz, Kanazawa:2017dpd}. We make little claim to originality; the purpose of this appendix is simply to provide a pedagogical review of this calculation.

\subsection{The Wigner-Dyson $\upbeta$ ensembles} \label{sec:beta_ens}

The Wigner-Dyson $\upbeta$ ensembles are characterised by a measure of the form
\be
e^{-L \frac{\upbeta}{4} \Tr H^2} \dd H,
\ee
where $H$ is either a real symmetric, complex hermitian, or quaternionic hermitian matrix, corresponding to the GOE, GUE, or GSE, respectively. The letter G indicates that these are Gaussian ensembles, while O, U, and S indicate that the measure is invariant under the adjoint action of the orthogonal, unitary, or symplectic groups, respectively. The corresponding measure for the eigenvalues of $H$ takes the following form
\be \label{eigenvalue measure beta}
\prod_{1 \leq i < j \leq L} |\lambda_i - \lambda_j|^\upbeta \prod_{i=1}^L e^{-L \frac{\upbeta}{4} \lambda_i^2} \dd \lambda_i,
\ee
with $\upbeta=1,2,$ or 4 for GOE, GUE, or GSE. For quantum systems with no antiunitary symmetry, the relevant ensemble is GUE. While for systems with time reversal symmetry $\mathsf{T}$, the relevant ensemble is GOE if $\mathsf{T}^2=1$ and GSE if $\mathsf{T}^2=-1$. For GSE, there is a two-fold degeneracy in the spectrum due to Kramer's theorem and in the measure \eqref{eigenvalue measure beta} each $\lambda_i$ really represents a pair of eigenvalues. With this convention, for GOE and GUE, $L = \dim \cH$, where $\cH$ is the Hilbert space on which $H$ acts, while for GSE, $L=\frac{1}{2} \dim \cH$.
The two‑fold degeneracy of the GSE follows immediately from the standard embedding of the quaternions into $2 \times 2$ complex matrices. A quaternionic Hermitian matrix has real quaternionic eigenvalues, which can each be represented as a real number multiplying a $2 \times 2$ identity matrix.

In the large $L$ limit, the normalised eigenvalue distribution
\be
\rho(\lambda) = \frac{1}{L}\sum_{i=1}^L \delta(\lambda-\lambda_i),
\ee
may be approximated by a continuous, non-negative function, and matrix integrals are determined by saddle points of the action
\be \label{rho action}
I = \frac{L^2 \upbeta}{4} \int \dd \lambda \rho(\lambda) \lambda^2 - \frac{L^2\upbeta}{2} \int \dd \lambda \fint \dd \lambda' \rho(\lambda) \rho(\lambda') \log |\lambda-\lambda'|,
\ee
and small fluctuations around them. The second term in the action arises from the Vandermonde determinant $\Delta(\left\{\lambda\right\})=\prod_{i<j}|\lambda_i-\lambda_j|$ in \eqref{eigenvalue measure beta} and $\fint$ indicates that the integral is regularised by taking the principal value. The saddle point is well known and is given by Wigner's semicircle distribution
\be
\label{eq:rho_semicircle}
\rho_\text{sc}(\lambda) = \frac{1}{2\pi} \sqrt{4-\lambda^2} \, \theta(4-\lambda^2).
\ee
Expanding around the saddle point $\rho(\lambda) = \rho_\text{sc}(\lambda) + \delta \rho(\lambda)$ gives a quadratic action from the second term in \eqref{rho action}
\be \label{delta rho action}
\delta I^{(2)}=-\frac{L^2\upbeta}{2} \int \dd \lambda \fint \dd \lambda' \delta\rho(\lambda) \delta\rho(\lambda') \log |\lambda-\lambda'|.
\ee
This leads to the propagator
\be \label{delta rho propagator}
\langle \delta\rho(\lambda) \delta\rho(\lambda') \rangle = - \frac{1}{\pi^2 L^2 \upbeta} \frac{1}{(\lambda-\lambda')^2},
\ee
where the dependence on the ensemble only appears through an overall coefficient. This expression can be found by decomposing $\delta \rho$ into Fourier modes
\be
\delta \rho(\lambda) = \frac{1}{2\pi} \int \dd t e^{\ii \lambda t} \delta \hat\rho(t),
\ee
where the hat denotes the Fourier transform. The quadratic action \eqref{delta rho action} then reads
\be
\delta I^{(2)}=\frac{L^2 \upbeta}{2} \int \dd t \int \dd t' \delta\hat\rho(t) \frac{\delta(t+t')}{|t-t'|} \delta\hat\rho(t'),
\ee
from which one can simply read off the propagator in Fourier space
\be \label{delta rho propagator fourier}
\langle \delta\hat\rho(t) \delta\hat\rho(t') \rangle = \frac{1}{L^2 \upbeta} \delta(t+t')|t-t'|.
\ee
To summarise, in the saddle point approximation $\rho(\lambda) = \rho_\text{sc}(\lambda) + \delta \rho(\lambda)$ is a Gaussian variable with
\be
\langle \rho(\lambda) \rangle = \rho_\text{sc}(\lambda),\qquad \langle \rho(\lambda) \rho(\lambda') \rangle - \langle \rho(\lambda) \rangle \langle \rho(\lambda') \rangle = - \frac{1}{\pi^2 L^2 \upbeta} \frac{1}{(\lambda-\lambda')^2}.
\ee
Since the real-time partition function $Z(\ii T)$ is given by the Fourier transform of the eigenvalue distribution
\be
Z(\ii T) = L \int \dd \lambda e^{-\ii T\lambda} \rho(\lambda) = L \hat{\rho}(T),
\ee
in the saddle point approxmation, $Z(\ii T)$ is also Gaussian variable with
\be \label{Z mean and var RMT}
\langle Z(\ii T) \rangle = L \hat \rho_\text{sc}(T),\qquad \langle Z(\ii T) Z(\ii T') \rangle - \langle Z(\ii T) \rangle \langle Z(\ii T') \rangle = \frac{1}{\upbeta} \delta(T+T') |T-T'|.
\ee
The connected two-point function vanishes for $T+T'\neq0$ and diverges for $T+T'=0$. The divergence is an artefact of expanding around $L=\infty$ and we regularise it as
\be
\delta(0) \to \frac{E_\text{max}-E_\text{min}}{2\pi},
\ee
where $E_\text{min}$ and $E_\text{max}$ are the lower and upper edges of the eigenvalue distribution, respectively. To summarise, in the saddle point approximation, and after regularisation, $Z(\ii T)$ is a complex Gaussian variable with mean and variance given by
\be \label{Z beta gaussian}
\langle Z(\ii T) \rangle = L \hat \rho_\text{sc}(T) = L \, \frac{J_1(2T)}{T},\qquad \langle |Z(\ii T)|^2 \rangle_c  = \frac{T}{\pi \upbeta} (E_\text{max}-E_\text{min}),
\ee
where the subscript $c$ stands for the connected contribution. For large $T$, the mean oscillates over times scales of order one with a decaying envelope proportional $T^{-3/2}$ while the variance increases linearly with time $T$. The average of the spectral form factor involves a disconnected and connected contribution:
\be \label{beta SFF}
 \langle| Z(\ii T) |^2 \rangle = |\langle Z(\ii T) \rangle|^2 + \frac{T}{\pi \upbeta} (E_\text{max}-E_\text{min}).
\ee
The first term describes the slope in the spectral form factor while the second term describes the linear ramp. The crossover between these two regions occurs at the \say{dip time}, which is approximately $t_\text{dip} \approx \sqrt{L}$. The plateau is a nonperturbative effect which is not seen in this analysis.

The behaviour of the moments can now be obtained in a straightforward way. In the following it will be convenient to work with the connected real time partition function $\delta Z(\ii T) = Z(\ii T) - \langle Z(\ii T) \rangle$. The $k$\textsuperscript{th} moment of the spectral form factor is then given by
\be
\langle |Z(\ii T)|^{2k} \rangle = \sum_{\ell,\ell'=0}^k \binom{k}{\ell} \binom{k}{\ell'} \langle Z(\ii T) \rangle^{k-\ell} \langle Z(-\ii T) \rangle^{k-\ell'} \langle \delta Z(\ii T)^\ell \delta Z(-\ii T)^{\ell'} \rangle.
\ee
Using Wick's theorem
\be
\langle \delta Z(\ii T)^\ell \delta Z(-\ii T)^{\ell'} \rangle = \delta_{\ell \ell'}\,\ell!\,\langle Z(\ii T) Z(-\ii T) \rangle_c^\ell,
\ee
where $\ell!$ counts the number of ways we can pair $\delta Z(\ii T)$ with $\delta Z(-\ii T)$. Hence
\be
\ba
\langle |Z(\ii T)|^{2k} \rangle &= \sum_{\ell=0}^k \ell! \binom{k}{\ell}^2 |\langle Z(\ii T) \rangle|^{2(k-\ell)} \,\langle |Z(\ii T)|^2 \rangle_c^\ell\\
&= |\langle Z(\ii T) \rangle|^{2k} + \dots + k! \left(\frac{T}{\pi \upbeta} (E_\text{max}-E_\text{min})\right)^k.
\ea
\ee
For large $L$, effectively only the first and last terms in the sum can dominate. For early times, the first term dominates and the spectral form factor is self-averaging, $\langle |Z(\ii T)|^{2k} \rangle \approx \langle |Z(\ii T)|^{2} \rangle^k$, as fluctuations around the mean value are suppressed. For later times, after the dip time, the spectral form factor has fluctuations of order the mean signal since $\langle |Z(\ii T)|^{2k} \rangle \approx k! \langle |Z(\ii T)|^{2} \rangle^k$, so is no longer self averaging, $\langle |Z(\ii T)|^{2k} \rangle \approx k! \langle |Z(\ii T)|^{2} \rangle^k$. This is the result presented in equation \eqref{eq:RMT_exp_intro} (see also \eqref{more:)}).

\subsection{The Altland-Zirnbauer $(\upalpha,\upbeta)$ ensembles}

The Altland-Zirnbauer $(\upalpha,\upbeta)$ ensembles are characterised by a measure for the eigenvalues of the form
\be
\prod_{1 \leq i < j \leq L} |\lambda_i^2-\lambda_j^2|^\upbeta \prod_{i=1}^L |\lambda_i|^{\upalpha} e^{-\frac{L\upbeta}{2} \lambda_i^2} \dd \lambda_i.
\ee
For the $(\upalpha,\upbeta)$ ensembles, the eigenvalues come in pairs $(\lambda,-\lambda)$, and in the above, one of the eigenvalues (which we can assume to be nonnegative) is represented. With this convention $L=\frac{1}{2}\dim \cH$. It's convenient to work with the following normalised eigenvalue distribution
\be
\rho(\lambda) = \frac{1}{2L}\sum_{i=1}^L \left[\delta(\lambda-\lambda_i)+\delta(\lambda+\lambda_i)\right].
\ee
In the large $L$ limit we work with the action
\be
I = \frac{L^2 \upbeta}{2} \int \dd \lambda \rho(\lambda) \lambda^2 - L^2\upbeta \int \dd \lambda \fint \dd \lambda' \rho(\lambda) \rho(\lambda') \log |\lambda-\lambda'|-\upalpha L \int \dd \lambda \rho(\lambda) \log|\lambda| .
\ee
Provided $\upalpha$ scales slower than $L$ (which is the case for the ensembles relevant to SYK, where $\upalpha=0$ or 2, see table \ref{RMT class q=2mod4}), the last term in the action can be neglected for the purpose of determining the saddle point. The action is then identical to the one for the $\upbeta$ ensembes \eqref{rho action}, up to an overall factor of $2$. This leads to the same expression for the saddle point solution \eqref{eq:rho_semicircle}, described by Wigner's semicircle distribution \eqref{eq:rho_semicircle}.

Next we consider fluctuations around the saddle point. To account for the constraint that $\rho(\lambda)$ is even we write
\be
\rho(\lambda) = \rho_\text{sc}(\lambda) + \delta \rho(\lambda), \qquad \delta \rho(\lambda) = \theta (\lambda) \delta\varrho(\lambda)+\theta (-\lambda) \delta\varrho(-\lambda).
\ee
The quadratic action then reads
\be
\ba
\delta I^{(2)}&=-2L^2\upbeta \int_0^\infty \dd \lambda \fint_0^\infty \dd \lambda' \delta\varrho(\lambda) \delta\varrho(\lambda') \log |\lambda^2-\lambda'^2| \\ &= -L^2\upbeta \int \dd \lambda \fint \dd \lambda' \delta\rho(\lambda) \delta\rho(\lambda') \log |\lambda-\lambda'|.
\ea
\ee
The discussion now closely mimics the one in section \ref{sec:beta_ens}. Moving to Fourier space gives rise to the propagator
\be
\langle \delta\hat\rho(t) \delta\hat\rho(t') \rangle = \frac{1}{4L^2 \upbeta} \delta(t+t')|t-t'|+ \frac{1}{4L^2 \upbeta} \delta(t-t')|t+t'|,
\ee
which reflects the fact that $\delta\hat\rho(t)$ is even, which follows from the constraint that $\delta \rho(t)$ is even. The variable $Z(\ii T)$ is now a \textit{real} Gaussian variable with mean and variance given by \eqref{Z beta gaussian}. The $k$\textsuperscript{th} moment of the spectral form factor is given by
\be
\ba
\langle |Z(\ii T)|^{2k} \rangle &= \sum_{\ell=0}^k (2\ell-1)!! \binom{2k}{2\ell} |\langle Z(\ii T) \rangle|^{2(k-\ell)} \langle |Z(\ii T)|^2 \rangle^{\ell}\\
&= |\langle Z(\ii T) \rangle|^{2k} + \dots + (2k-1)!! \left(\frac{T}{\pi \upbeta} (E_\text{max}-E_\text{min})\right)^k.
\ea
\ee

\subsection{Symmetry classification of SYK} \label{app:rmt syk}

In this section we review the symmetry classification of SYK \eqref{HSYK} following \cite{Behrends_2019} (see also \cite{Stanford:2019vob}) before applying the previous results to predict the behaviour of the moments of the spectral form factor in SYK. The analysis for $q=0$ mod 4 is essentially the same as in appendix C of \cite{Saad:2018bqo}.

SYK has a fermion parity symmetry $(-1)^\mathsf{F}$. In a basis in which this operator is block diagonal a Hamiltonian which is consistent with this symmetry is
\be
H = \begin{pmatrix}
H_1 & 0 \\
0 & H_2
\end{pmatrix}.
\ee
Classically, for $q=0$ mod 4, there is a time reversal symmetry $\mathsf{T}$ which commutes with $(-1)^\mathsf{F}$ and squares to one, $\mathsf{T}^2=1$. Naively, we might then expect a maximally random Hamiltonian to have independent GOE statistics for each of the blocks of $H_1$ and $H_2$. However, this is only true if there are no anomalies, which is only the case for $N = 0$ mod 8 \cite{You:2016ldz, Kanazawa:2017dpd}. For $N=2,6$ mod 8, $\mathsf{T}^2=1$ but $\mathsf{T}$ anticommutes with $(-1)^\mathsf{F}$ so exchanges the blocks and doesn't constrain their form. In this case, the blocks have GUE statistics, but they are not independent since they are exchanged by $\mathsf{T}$. For $N=4$ mod 8, $\mathsf{T}^2=-1$ and $\mathsf{T}$ commutes with $(-1)^\mathsf{F}$ so each of the blocks have independent GSE statistics. This is summarised in table \ref{RMT class q=0mod4}.

\begin{table}[!h]
\begin{center}
\begin{tabular}{c|c|c|c|c|c}
$N$ mod 8 & RMT class & $\upalpha$ & $\upbeta$ & degeneracy & $Z(x)$ \\ \hline\hline
& & & & \\[-10pt]
0 & $\begin{pmatrix} \text{GOE}_1 & 0\\ 0 & \text{GOE}_2 \end{pmatrix}$ & - & 1 & 1 & $Z_{\text{GOE}_1}(x)+Z_{\text{GOE}_2}(x)$ \\
& & & & \\[-10pt]\hline
& & & & \\[-10pt]
2 & $\begin{pmatrix} \text{GUE} & 0\\ 0 & \overline{\text{GUE}} \end{pmatrix}$ & - & 2 & 2 & $2Z_\text{GUE}(x)$ \\ & & & & \\[-10pt]\hline
& & & & \\[-10pt]
4 & $\begin{pmatrix} \text{GSE}_1 & 0\\ 0 & \text{GSE}_2 \end{pmatrix}$ & - & 4 & 2 & $2Z_{\text{GSE}_1}(x)+2Z_{\text{GSE}_2}(x)$ \\
& & & & \\[-10pt]\hline
& & & & \\[-10pt]
6 & $\begin{pmatrix} \text{GUE} & 0\\ 0 & \overline{\text{GUE}} \end{pmatrix}$ & - & 2 & 2 & $2Z_\text{GUE}(x)$
\end{tabular}
\caption{Symmetry classification for SYK with $q=0$ mod 4. In the second column, the subscripts on the blocks represent independent random matrices. For $N=2,6$ mod 8, the blocks are not independent but are instead exchanged by $\mathsf{T}$. The third column denotes the degeneracy of the Hamiltonian $H$. In the last column, we relate $Z(x)=\Tr[e^{-xH}]$ for the symmetry class to the standard ensembles.}
\label{RMT class q=0mod4}
\end{center}
\end{table}

For $q=2$ mod 4, $\mathsf{T}$ instead anticommutes with the Hamiltonian and so the spectrum is symmetric about zero. For $N=0$ mod 8, $\mathsf{T}^2=1$ and $\mathsf{T}$ commutes with $(-1)^\mathsf{F}$ so each of the blocks have independent BdG(D) statistics. For $N=2,6$ mod 8, $\mathsf{T}$ anticommutes with $(-1)^\mathsf{F}$ so each of the blocks have GUE statistics, but they are not independent since they are exchanged by $\mathsf{T}$. For $N=4$ mod 8, $\mathsf{T}^2=-1$ and $\mathsf{T}$ commutes with $(-1)^\mathsf{F}$ so each of the blocks have independent BdG(C) statistics. This is summarised in table \ref{RMT class q=2mod4}.

\begin{table}[!h]
\begin{center}
\begin{tabular}{c|c|c|c|c|c}
$N$ mod 8 & RMT class & $\upalpha$ & $\upbeta$ & degeneracy & $Z(x)$ \\ \hline\hline
& & & & \\[-10pt]
0 & $\begin{pmatrix} \text{BdG(D)}_1 & 0\\ 0 & \text{BdG(D)}_2 \end{pmatrix}$ & 0 & 2 & 1 & $Z_{\text{BdG(D)}_1}(x)+Z_{\text{BdG(D)}_2}(x)$ \\
& & & & \\[-10pt]\hline
& & & & \\[-10pt]
2 & $\begin{pmatrix} \text{GUE} & 0\\ 0 & -\overline{\text{GUE}} \end{pmatrix}$ & - & 2 & 1 & $Z_\text{GUE}(x)+Z_\text{GUE}(-x)$ \\ & & & & \\[-10pt]\hline
& & & & \\[-10pt]
4 & $\begin{pmatrix} \text{BdG(C)}_1 & 0\\ 0 & \text{BdG(C)}_2 \end{pmatrix}$ & 2 & 2 & 1 & $Z_{\text{BdG(C)}_1}(x)+Z_{\text{BdG(C)}_2}(x)$ \\
& & & & \\[-10pt]\hline
& & & & \\[-10pt]
6 & $\begin{pmatrix} \text{GUE} & 0\\ 0 & -\overline{\text{GUE}} \end{pmatrix}$ & - & 2 & 1 & $Z_\text{GUE}(x)+Z_\text{GUE}(-x)$
\end{tabular}
\caption{Symmetry classification for SYK with $q=2$ mod 4. For $N=2,6$ mod 8 the two blocks are not independent but instead exchanged by $\mathsf{T}$.}
\label{RMT class q=2mod4}
\end{center}
\end{table}

We now turn to the expectations for moments of the spectral form factor in SYK based on the classification in tables \ref{RMT class q=0mod4} and \ref{RMT class q=2mod4}. This is mostly an exercise in keeping track of factors of 2.

$\mathbf{q=0}$ \textbf{mod 4:} For $N=0$ mod 8, $Z(\ii T)$ $Z(\ii T)$ is a sum of two i.i.d. complex Gaussian variables, $Z(\ii T) = Z_{\text{GOE}_1}(\ii T)+Z_{\text{GOE}_2}(\ii T)$. So $Z(\ii T)$ is also a complex Gaussian variable with
\be
\langle Z(\ii T) \rangle = 2 \langle Z_\text{GOE}(\ii T) \rangle,\quad
\langle |Z(\ii T)|^2 \rangle_c = 2 \langle |Z_\text{GOE}(\ii T)|^2 \rangle_c .
\ee
For $N=2,6$ mod 8, $Z(\ii T) = 2 Z_\text{GUE}(\ii T)$, so $Z(\ii T)$ is a complex Gaussian variable with
\be
\langle Z(\ii T) \rangle = 2 \langle Z_\text{GUE}(\ii T) \rangle,\quad
\langle |Z(\ii T)|^2 \rangle_c = 4 \langle |Z_\text{GUE}(\ii T)|^2 \rangle_c.
\ee
For $N=4$ mod 8, $Z(\ii T)$ is the twice the sum of two i.i.d. complex Gaussian variables, $Z(\ii T) = 2 Z_{\text{GSE}_1}(\ii T) + 2 Z_{\text{GSE}_2}(\ii T)$. The factor of $2$ comes from the fact that eigenvalues are two-fold degenerate for GSE. So, $Z(\ii T)$ is also a complex Gaussian variable with
\be
\langle Z(\ii T) \rangle = 4 \langle Z_\text{GSE}(\ii T) \rangle,\quad
\langle |Z(\ii T)|^2 \rangle_c = 8 \langle |Z_\text{GSE}(\ii T)|^2 \rangle_c.
\ee
For any $N$, the result for $q=0$ mod 4 the connected contribution can neatly be summed up as 
\be \label{RMT q=4 var}
\langle |Z(\ii T)|^2 \rangle_c = 2 \upbeta \, \langle |Z_\upbeta(\ii T)|^2 \rangle_c = \frac{2T}{\pi}(E_\text{max}-E_\text{min}),
\ee
which is independent of $\upbeta$. During the ramp region, the moments are given by
\be
\langle |Z(\ii T)|^{2k} \rangle_c = k! \left( \frac{2T}{\pi}(E_\text{max}-E_\text{min}) \right)^k.
\ee

$\mathbf{q=2}$ \textbf{mod 4:} For $N=0,4$ mod 8, $Z(\ii T)$ is the sum of two i.i.d. real Gaussian variables $Z(\ii T) = Z_{\text{BdG}_1}(\ii T) + Z_{\text{BdG}_2}(\ii T)$. So
\be
\langle Z(\ii T) \rangle = 2 \langle Z_\text{BdG}(\ii T) \rangle,\quad \langle |Z(\ii T)|^2 \rangle_c = 2 \langle |Z_\text{BdG}(\ii T)|^2 \rangle_c.
\ee
For $N=2,6$ mod 8, $Z(\ii T) = 2 \text{Re} Z_\text{GUE}(\ii T)$. So $Z(\ii T)$ is a real Gaussian variable with
\be
\langle Z(\ii T) \rangle = 2 \text{Re} \langle Z_\text{GUE}(\ii T) \rangle,\quad \langle |Z(\ii T)|^2 \rangle_c = 2 \langle |Z_\text{GUE}(\ii T)|^2 \rangle_c.
\ee
For any $N$, the result for $q=2$ mod 4 the connected contribution can neatly be summed up as 
\be \label{RMT q=2 var}
\langle |Z(\ii T)|^2 \rangle_c = \upbeta \, \langle |Z_\upbeta(\ii T)|^2 \rangle_c = \frac{T}{\pi}(E_\text{max}-E_\text{min}),
\ee
which again is independent of $\upbeta$. During the ramp region, the moments are given by
\be
\langle |Z(\ii T)|^{2k} \rangle_c = (2k-1)!! \left( \frac{T}{\pi}(E_\text{max}-E_\text{min}) \right)^k.
\ee
The $(2k-1)!!$ in place of $k!$ stems from the fact that $Z(\ii T)$ is real Gaussian variable rather than a complex one for $q=2$ mod 4. Equations \eqref{RMT q=4 var} and \eqref{RMT q=2 var} are consistent with what is found in SYK \eqref{ziT^2}.

\section{Details on $1/N$ corrections in SYK} \label{app:SYK_perturbation}

In this appendix we provide details on the perturbative analysis in section \ref{sec:1/N}. We start by showing how the free propagators for $\delta G, \delta \Sigma$ can be computed, and then proceed to prove that \eqref{two G} is the leading contribution in the perturbative series. 

\subsection{Propagators}
\label{appendix:propagator}

As in the main text, we split the fluctuations in the collective fields into a \say{parallel} and \say{perpendicular} part. It's important to stress that not all of these components are independent because of the antisymmetry constraint $\delta G_{ab}(t_1,t_2)= -\delta G_{ba}(t_2,t_1)$ (and similarly for $\delta \Sigma$). This needs to be accounted for when determining the free propagators.

The quadratic action \eqref{quadratic action} can be separated in two contributions:
\begin{equation}
\label{eq:splitting action}
    \delta I^{(2)} = \delta I^{(2)}_\perp (\delta G_\perp,\delta \Sigma_\perp) + \delta I^{(2)}_\parallel (\delta G_\parallel,\delta \Sigma_\parallel)
\end{equation}
where
\be
\ba
\frac{1}{N} \, \delta I^{(2)}_\perp &= \frac{1}{4} \Tr[(\mathsf{G} \delta \Sigma_\perp)^2] + \frac{1}{2} \int_0^T\dd t \dd t' (\delta \Sigma_\perp)_{ab} (\delta G_\perp)_{ab} \, , \\
\frac{1}{N} \, \delta I^{(2)}_\parallel &= \frac{1}{4} \Tr[(\mathsf{G} \delta \Sigma_\parallel)^2] + \frac{1}{2} \int_0^T \dd t \dd t' \left[ (\delta \Sigma_\parallel)_{ab} (\delta G_\parallel)_{ab} - J^2 \frac{q-1}{2} s_{ab} \mathsf{G}_{ab}^{q-2} (\delta G_\parallel)_{ab}^2 \right].
\ea
\ee
Equation \eqref{eq:splitting action} follows since $\mathsf{G} = \mathsf{G}_\parallel$, which implies $(\mathsf{G} \delta \Sigma_\parallel \mathsf{G} \delta \Sigma_\perp)_\parallel=0$ and so $\Tr(\mathsf{G} \delta \Sigma_\parallel \mathsf{G} \delta \Sigma_\perp) = 0$. As a consequence of \eqref{eq:splitting action}, all the mixed parallel and perpendicular two-point functions must vanish as in \eqref{eq:2point_mixed}.

For the perturbative analysis in the next section, we will actually need just the two-point function of the perpendicular components, which can be calculated from the path integral. For example:
\begin{equation}
    \langle \delta \Sigma_\perp \delta \Sigma_\perp \rangle = \int D\delta \Sigma_\perp D\delta G_\perp e^{-I^{(2)} _\perp} \delta \Sigma_\perp \delta \Sigma_\perp = \int D\delta \Sigma_\perp e^{-\frac{N}4 \Tr[(\mathsf{G} \delta \Sigma_\parallel)^2]} \delta \Sigma_\perp \delta \Sigma_\perp \delta(\delta \Sigma_\perp) = 0 \, ,
\end{equation}
where in the second step we performed the integral over $\delta G_\perp$ which produces a delta function which sets $\delta \Sigma_\perp=0$ . The same trick of integrating $\delta G_\perp$ out first can be used to calculate all the other propagators once we rewrite
\begin{equation}
    \delta G_{cd} \exp \left\{-\frac{N}{2} \int_0^T\dd t \dd t' (\delta \Sigma_\perp)_{ab} (\delta G_\perp)_{ab} \right\} = - N \frac{D}{D\delta \Sigma_{cd}} \exp \left\{-\frac{N}{2} \int_0^T\dd t \dd t' (\delta \Sigma_\perp)_{ab} (\delta G_\perp)_{ab}\right\},
\end{equation}
where $D$ denotes the differential on the collective fields. For obtaining this result, we used the antisymmetry property $\delta G_{ab}(t_1,t_2)= -\delta G_{ba}(t_2,t_1)$ by considering independent variables to be the ones with $a<b$. Now we can calculate
\begin{equation}
\begin{split}
        \langle \delta \Sigma_\perp(t_1,t_2)_{c d} \delta G_\perp(t_1^\prime,t_2^\prime)_{c^\prime d^\prime} \rangle &=  - \frac{1}{N} \int D\delta \Sigma_\perp e^{-\frac{N}4 \Tr[(\mathsf{G} \delta \Sigma_\parallel)^2]} \delta \Sigma_\perp(t_1,t_2)_{c d}   \frac{D \, \delta(\delta \Sigma_\perp)}{D\delta \Sigma_\perp(t_1^\prime,t_2^\prime)_{c^\prime d^\prime}} \\
        &= \frac{\pi^\perp_{cd}}{N} \delta_{c c^\prime} \delta_{d d^\prime} \delta(t_1-t_1^\prime)\delta(t_2-t_2^\prime) \, ,
\end{split}
\end{equation}
where in the last step we used integration by parts and the $\pi^\perp_{cd}$ appears since the indices $cd,c^\prime d^\prime$ span the perpendicular components of the collective fields.
Lastly we have
\begin{equation}
    \langle \delta G_\perp(t_1,t_2)_{ab} \delta G_\perp(t_3,t_4)_{cd} \rangle =  \frac{1}{N^2} \int D\delta \Sigma_\perp  \frac{D^2 e^{-\frac{N}4 \Tr[(\mathsf{G} \delta \Sigma_\parallel)^2]}}{D\delta \Sigma_\perp(t_1,t_2)_{ab} D\delta \Sigma_\perp(t_3,t_4)_{cd}}  \delta(\delta \Sigma_\perp)
\end{equation}
if now we expand
\begin{equation}
\frac{1}{2} \Tr[(\mathsf{G} \delta \Sigma_\perp)^2] =  \sum_{\substack{a<b \\ c<d}} \int \prod_i \dd t_i \, \delta \Sigma_\perp(t_1,t_2)_{ab} \left[ \mathsf{G}_{ad}(t_1,t_4) \mathsf{G}_{cb}(t_3,t_2)-\mathsf{G}_{ac}(t_1,t_3) \mathsf{G}_{db}(t_4,t_2)\right] \delta \Sigma_\perp(t_3,t_4)_{cd}
\end{equation}
we get equation \eqref{eq:2point_GG}. 

For the sake of calculating the relevant terms in the perturbative expansion we don't need to calculate the propagators of the parallel fields. These is particular convenient for two reasons: first of all, it is very hard to invert the quadratic action $I^{(2)}_\parallel$ to obtain an explicit expression for the two-point functions. Secondly, the inversion of the propagator associated to $I^{(2)}_\parallel$ is subtle due to the presence of the zero modes, resulting from the saddle point spontaneously breaking the relative time translation symmetry.

\subsection{Perturbative expansion}
\label{app:perturbative_exp}

As discussed in section \ref{sec:1/N},
equation \eqref{eq:relevant_vertex_G^q} suggests that the leading term in the perturbative expansion for $\langle |Z(\ii T)|^{2k} \rangle/ \langle |Z(\ii T)|^2 \rangle^k$ comes from terms containing a $\delta G_\perp^q$ insertion.
More precisely, these terms are generated by
\begin{equation}
\label{eq:different_perturbative_terms}
    \frac{ N J^2}{2q} s_{ab} \langle   (\delta G_\perp)_{ab}^q   e^{- N \sum_{\alpha \ge 3} \delta I^{(\alpha)}}\rangle  \, ,
\end{equation}
where $\delta I^{(\alpha)}$ is the $\alpha$\ts{th} order expansion of the action $I$ in the fluctuations of the collective fields. As shown in equation \eqref{eq:1_term_expansion}, the first term in the expansion of the exponential is zero because of the Wick contraction
\be \label{selfc}
\wick{\c{(\delta G_\perp)}_{ab} \c{(\delta G_\perp)}_{ab}} =0 \, .
\ee

We are now going to prove that many of the other terms in \eqref{eq:different_perturbative_terms} also vanish. The interaction vertices appearing in the action are:
\begin{equation}
 V(\alpha)=\frac{N}{2 \alpha} \Tr(\tilde{G} \delta \Sigma)^\alpha \, , \qquad  W(\alpha)=\frac{N}{2} \iint_0^T \dd t \dd t^\prime \frac{J^2}{q} \binom{q}{\alpha}  s_{cd} \mathsf{G}^{q-\alpha}_{cd} \delta G_{cd}^\alpha  \, ,
\end{equation}
with $\alpha>2$. The most general term in the expansion \eqref{eq:different_perturbative_terms} is of the form
\begin{equation}
    \frac{ N J^2}{2q} \int_0^T \dd t \dd t^\prime  s_{ab} \langle   (\delta G_\perp)_{ab}^q  V(\alpha_1)  \dots W(\alpha_2) \dots \rangle \, .
    \label{eq:generic_term}
\end{equation}
Let's start by considering terms only containing insertions of $W$ with different $\alpha_i < q$. Then
\begin{equation}
\label{prop:no_V2}
    \frac{ N J^2}{2q} \int_0^T \dd t \dd t^\prime  s_{ab} \langle   (\delta G_\perp)_{ab}^q  W(\alpha_1)  \dots W(\alpha_n)\rangle = 0 \, .
\end{equation}
This follows for the following reason. \eqref{selfc} implies that the insertion of $\delta G_\perp$ needs to be contracted with another field to get a nonzero result. However, $W(\alpha_i < q)$ only contains $\delta G_\parallel$ insertions, due to the presence of $\mathsf{G}=\mathsf{G}_\parallel$, but this cannot give a nonzero result since $\langle \delta G_\perp \delta G_\parallel \rangle=0$. Now we can consider the complementary case, where we only have insertions of $V$. There are just two kinds of two-point functions that we need to consider:
\begin{equation}
    \langle \delta \Sigma \delta \Sigma  \rangle = \langle \delta \Sigma_\parallel \delta \Sigma_\parallel  \rangle \, , \qquad \langle \delta G_\perp \delta \Sigma  \rangle = \langle  \delta G_\perp \delta \Sigma_\perp  \rangle \, .
    \label{eq:useful_wick_V1}
\end{equation}
The expansion of the trace contained in the vertices $V$ will produce chains of terms alternating between $\delta \Sigma$ and $\mathsf{G}$ with indices which are cyclically contracted. A generic Wick contraction between two $\delta \Sigma$ reads
\begin{equation}
    \langle \dots \mathsf{G}_{a_1 b_1} \wick{\c{\delta \Sigma}_{b_1 b_2} \mathsf{G}_{b_2 a_2} \dots \mathsf{G}_{c_1 d_1} \c{\delta \Sigma}_{d_1 d_2}} \mathsf{G}_{d_2 c_2} \dots  \rangle \, .
\end{equation}
Since $\mathsf{G}=\mathsf{G}_\parallel$ and from the first equation in \eqref{eq:useful_wick_V1}, all the $a_i$ and $c_i$ must span the same diagonal block in the parallel subspace. So without loss of generality, we can write
\begin{equation}
\label{eq:contr_1}
    \langle \mathsf{G}_{a_1 b_1} \wick{\c{\delta \Sigma}_{b_1 b_2} \mathsf{G}_{b_2 a_2} \dots \mathsf{G}_{c_1 d_1} \c{\delta \Sigma}_{d_1 d_2}} \mathsf{G}_{d_2 c_2}   \rangle = \pi^\parallel_{a_1 a_2} \pi^\parallel_{c_1 c_2} \pi^\parallel_{a_1 c_1}\langle \mathsf{G}_{a_1 b_1} \wick{\c{\delta \Sigma}_{b_1 b_2} \mathsf{G}_{b_2 a_2} \dots \mathsf{G}_{c_1 d_1} \c{\delta \Sigma}_{d_1 d_2}} \mathsf{G}_{d_2 c_2} \rangle,
\end{equation}
where some of the projections are now explicit and repeated indices are not summed.
On the other hand, the generic term \eqref{eq:generic_term} always contains insertions of $\delta G_\perp$:
\begin{equation}
\label{eq:contr_2}
    \langle \wick{\c{(\delta G_\perp)}_{ab} \dots \mathsf{G}_{c_1 d_1} \c{\delta \Sigma}_{d_1 d_2}} \mathsf{G}_{d_2 c_2} \dots \rangle = \pi^\perp_{ab} \langle \dots \mathsf{G}_{c_1 a} \mathsf{G}_{b c_2} \dots \rangle
\end{equation}
It is now clear that the two Wick contractions \eqref{eq:contr_1} and \eqref{eq:contr_2} produce indices which span orthogonal spaces and give a vanishing result if they are contracted, e.g.
\begin{equation}
  \left( \pi^\perp_{ab} \mathsf{G}_{a_1 a} \mathsf{G}_{b c_2} \right)  \left(\mathsf{G}_{a_1 b_1} \mathsf{G}_{b_2 a_2} \langle (\delta \Sigma_\parallel)_{b_1 b_2}  (\delta \Sigma_\parallel)_{d_1 d_2} \rangle \mathsf{G}_{c_1 d_1} \mathsf{G}_{d_2 c_2} \right) =0 \, .
\end{equation}
This result would not change if in the second bracket we have a longer string of contracted indices involving $\mathsf{G}$ and $\langle \delta \Sigma \delta \Sigma \rangle$. The Wick contractions $\langle \delta \Sigma \delta \Sigma \rangle$ and the $\mathsf{G}$ are therefore creating a net of indices which all span the same block in the parallel subspace, and would give a vanishing result if they encounter the indices $ab$ produced by \eqref{eq:contr_2}. Excluding the case where all the fields in some insertion of $V$ contract among themselves (which doesn't produce an interesting result since they are not contracting with the insertion of $\delta G_\perp^q$), we necessarily have that at a certain point the net of indices which span a block in the parallel subspace must contract with $ab$, due to the fact that all the fields must be Wick contracted and the presence of the trace in $V$. This situation will therefore lead to a vanishing result.
This argument does not exclude the possibility of having a nonzero contribution if we consider that all the $\delta \Sigma$ are contracted with $\delta G_\perp$, which is the case, for example, if there is a single insertion of $V(q)$. However:
\begin{equation}
    \frac{ N J^2}{2q} \int_0^T \dd t \dd t^\prime  s_{ab} \langle   (\delta G_\perp)_{ab}^q  V(q) \rangle = \frac{ N J^2}{2q} \int_0^T \dd t \dd t^\prime  s_{ab} \pi^\perp_{ab} \mathsf{G}_{ab}^q = 0 \, .
\end{equation}
We would have got the same result by considering two insertions of $V(q/2)$ or any other combination containing $q$ fields $\delta \Sigma$.
What we have therefore proved is that if there are only insertions of $V$, the result vanishes,
\begin{equation}
\label{eq:V1_not_alone}
    \frac{ N J^2}{2q} \int_0^T \dd t \dd t^\prime  s_{ab} \langle   (\delta G_\perp)_{ab}^q  V(\alpha_1)  \dots V(\alpha_n)\rangle = 0 \, .
\end{equation}

So far, we argued that many of the terms in \eqref{eq:different_perturbative_terms} vanish. We now consider power counting argument with respect to $N$ that will allow us to identify the leading term in the expansion. 
The form of the action and of the propagator implies that each insertion (vertex) contributes with a factor of $N$ while each contraction (propagator) gives rise to a factor $1/N$. The generic term in the perturbative expansion is of the form \eqref{eq:generic_term}. 
In proving \eqref{prop:no_V2} we showed that any $W(\alpha < q)$ gives a vanishing result if it contract with $\delta G_\perp^q$. To get a nonzero result it must contract with something else. Therefore an insertion of $W(\alpha < q)$ gives a total contribution of $N^{1-\alpha}$ without reducing the number of $\delta G_\perp$ insertions. Since $\alpha \ge 3$, this means that an insertion of $W(\alpha)$ with $\alpha < q$ will lead always give a subleading contribution. 
Equation \eqref{eq:V1_not_alone}, on the other hand, implies that the most generic leading contribution must be of the form:
\begin{equation}
    \frac{ N J^2}{2q} \int_0^T \dd t \dd t^\prime  s_{ab} \langle   (\delta G_\perp)_{ab}^q  W(q)^m V(\alpha_1) \dots V(\alpha_n) \rangle \, ,
\end{equation}
which is of the order
\begin{equation}
    N^{1+m+n} / N^{\frac{1}{2}(q(m+1)+\sum_i \alpha_i)} \, .
\end{equation}
Since $q>2$ and $m>0$, it is maximised by setting $m=1$. Moreover, since $\alpha_i \ge 3$, $\frac{1}{2} \sum_i \alpha_i > \frac{3n}{2}$, so the leading contribution corresponds to $n=0$. This proves that \eqref{two G} is the leading contribution in the perturbative expansion.

\section{More on $q=2$}
\label{app:SYK_q=2}

In this section we give a detailed treatment of some calculations mentioned in section \ref{syk:sec:q=2} of the main text.

\subsection{One loop determinant} \label{syk:app:one loop}

In section \ref{syk:sec:q=2} we only considered the one loop contribution from configurations which are invariant under a diagonal time translation. In this section we consider the full one loop contribution. Expanding in fluctuations around the saddle points, $\Sigma = \mathsf{\Sigma}+\delta \Sigma$, the quadratic action is
\be
\ba
\delta I = \frac{N}{4 J^2} \sum_{a,b} \int \dd t_1 \dots \dd t_4 \delta \Sigma_{ab}(t_1,t_2) K_{ab}(t_1, \dots,t_4)\delta \Sigma_{ba}(t_4,t_3),
\ea
\ee
where
\be
K_{ab}(t_1,t_2,t_3,t_4) = K_{ab}(t_1-t_3,t_2-t_4) = \delta(t_1-t_3) \delta(t_2-t_4)+ J^{-2} \mathsf{\Sigma}_{aa}(t_3-t_1) \mathsf{\Sigma}_{bb}(t_2-t_4).
\ee
In Fourier space, this reads
\be
\delta I = \frac{N}{4J^2 T^2} \sum_{a,b} \sum_{\omega_1,\omega_2} |\delta \Sigma_{ab}(\omega_1,\omega_2)|^2 K_{ab}(-\omega_2,-\omega_1),
\ee
where
\be
K_{ab}(-\omega_2,-\omega_1) = 1+J^{-2} \mathsf{\Sigma}_{aa}(\omega_1) \mathsf{\Sigma}_{bb}(-\omega_2).
\ee
In order to regularise the one loop determinant, we impose that the determinant goes to $1$ as $J \to 0$. In this limit, $ \mathsf{\Sigma} \sim J^{2} / \omega$ which leads to the following normalisation
\be
Z_\text{one loop}=\frac{\det \left(\frac{N}{4 J^2 T^2} K \right)^{-1/2}}{\det \left(\frac{N}{4 J^2 T^2} \right)^{-1/2}} = \exp\left[- \frac{1}{2} \sum_{a,b} \sum_{\omega_1,\omega_2} \log K_{ab}(\omega_2,\omega_1) \right] .
\ee
The kernel $K_{ab}(\omega_2,-\omega_1)$ is zero in the range $-2J<\omega_1 = \omega_2 < 2J$ with $a,b$ spanning the off diagonal blocks. These are the zero modes, whose contribution has be accounted for in section \ref{sec-sp-q2}. In the section \ref{sec-sp-q2} only configurations which are under the diagonal time translation symmetry were considered. The dominant contribution to the one loop determinant comes from soft modes which correspond to moving a little bit away from the zero modes, specifically $\omega_2 = \omega$, $\omega_1 = \omega + \epsilon$ with $\epsilon = 2 \pi m / T$, where $m$ is an integer and $|\epsilon|\ll 2J$. Notice that this only makes sense at late times. On the off diagonal blocks we have
\be
K_{ab}(\omega,-\omega-\epsilon) = \ii \frac{\epsilon }{\sqrt{4 J^2-\omega^2}},
\ee
so we can rewrite the one loop contribution as
\be
\log Z_\text{one loop} = \frac12 \cdot 4 \cdot k^2 \sum_{m=1}^M \sum_{0<\omega < 2J} \log \left( \frac{2 \pi m}{T} \frac{ 1 }{\sqrt{4 J^2-\omega^2}} \right).
\ee
where the factor of $4$ comes from having restricted the sums on $m,\omega$ to just the positive values, while the sum over $a,b$ gave the factor $k^2$. $M$ is a positive integer counting the number of soft modes, $M \ll JT$. The presence of $M$ is quite unpleasing since there is not a natural value for it. We can proceed as in \cite{Winer:2020mdc} by using the analytical properties of the kernel $\tilde{K}$ and its behaviour at infinity. In particular, we have
\be
\int \dd \omega_1 \dd \omega_2 \log K_{ab}(\omega_1,\omega_2) = 0.
\ee
The idea now is to subtract this integral to regulate the sum as
\be
\log Z_\text{one loop} = 2 k^2 \sum_{0<\omega<2J} \left(\sum_{m=1}^M- \int_0^{M+\frac{1}{2}} \dd m\right) \log \left( \frac{2 \pi m}{T} \frac{ 1 }{\sqrt{4 J^2-\omega^2}} \right).
\ee
Using Stirling's approximation, we have
\be
\ba
\left(\sum_{m=1}^M- \int_0^{M+\frac{1}{2}} \dd m\right) \log (ma) &= \log (M! a^M) - \left( M + \frac12 \right) \left( \log \left(a\left( M + \frac12 \right) \right) -1 \right)\\
&= - \frac12 \log \frac{a}{2 \pi},
\ea
\ee
so we can write
\be
\log Z_\text{one loop} = \sum_{0<\omega<2J}k^2 \log \left(T \sqrt{4 J^2-\omega^2} \right).
\ee
The next step is to perform the sum over $\omega \sim 2 \pi n/T$
\be
\ba
\log Z_\text{one loop} =& k^2 \frac{1}{2}\sum_{n< JT/\pi} \log \left[(2\pi)^2 \left(\frac{J^2 T^2}{\pi^2}- n^2\right)\right] = k^2\log \left[(2\pi)^{\frac{2 J T}{\pi}} \left(\frac{2J T}{\pi}\right)!\right] \\
=& - k^2\frac{JT}{\pi} \log \left(\frac{e}{4 JT}\right).
\ea
\ee
Notice that for $k=1$ this result matches \cite[eq. (31)]{Winer:2020mdc}. The ramp region, corrected with the contribution of the soft modes therefore reads:
\be
\langle |Z(\ii T)|^{2k} \rangle = 2^{kN} \left[ \left(\frac{8N}{\pi e^2 JT}\right)^{k^2} \text{vol}\left(\frac{\mathrm{U}(2k)}{\mathrm{U}(k)^2}\right) \right]^{\frac{JT}{\pi}}.
\ee

\subsection{The moments as a matrix integral} \label{syk:app:q=2}

A similar trick to the one used in section \ref{syk:sec: large k} can be applied to the full path integral \eqref{moments collective fields q=2}
\begin{align}
\langle |Z(\ii T)|^{2k} \rangle = \int D \Sigma e^{-\frac{N}{4J^2} \Tr \Sigma^2} \det(\partial_t-\Sigma)^{\frac{N}{2}}.
\end{align}
The analogue of the matrix $B$ is
\begin{align}
G_{ab}(t,t') = \frac{1}{N} \sum_i \psi_i^a(t)\psi_i^b(t'),
\end{align}
while the analogue of $\widetilde{B}$ is the antisymmetric matrix
\begin{align}
\widetilde{G}_{ij} = \sum_a \int_0^T \dd t \psi^a_i(t) \psi^a_j(t).
\end{align}
Going through the steps basically undoes the $G,\Sigma$ trick and gives back
\begin{align}
\langle |Z(\ii T)|^{2k} \rangle = \left(\frac{N}{2\pi J^2}\right)^{\frac{N(N-1)}{4}} \prod_{i<j} \int \dd J_{ij} e^{-\frac{N}{2J^2} J_{ij}^2} \int D\psi \exp\left\{-\frac{1}{2} \int_0^T \dd t \left(\psi^a_i \partial_t \psi^a_i - J_{ij} \psi^a_i \psi^a_j\right)\right\}.
\end{align}
In this expression there is an implicit sum over both $i$ and $a$ in the action. Of course, since the $q=2$ model is quadratic in the fermions we could have integrated them out in the first step. This is simplest if we work in Fourier space. After decomposing the fermions into Fourier modes
\begin{align}
\psi_i^a(t) = \frac{1}{\sqrt{T}}\sum_{n\in\Z} \psi_i^a(\omega_n) e^{-\ii \omega_n t},\qquad \omega_n = \frac{2\pi(n+\frac{1}{2})}{T}.
\end{align}
we get an infinite product of decoupled Grassmann integrals for each mode. This gives
\begin{align} \label{cosine}
\prod_{n=0}^\infty \det(\ii \omega_n \delta_{ij} + J_{ij})^{2k} = 2^{kN} \prod_{i=1}^{N/2} \cos^{2k} \frac{T \lambda_i}{2}.
\end{align}
In this expression $2^{kN}$ is the contribution of the free determinant. By an orthogonal transformation, the antisymmetric matrix $J_{ij}$ can be brought into block diagonal form with blocks
\begin{align}
\begin{pmatrix}
0 & \lambda_i\\
-\lambda_i & 0
\end{pmatrix},\qquad \lambda_i \geq 0.
\end{align}
Expressing the determinant in terms of the $\lambda_i$, $i=1,\dots,N/2$, and doing the infinite product over the modes give the second term in \eqref{cosine}. Writing the matrix integral over $J_{ij}$ as an integral over the $\lambda_i$ gives
\begin{align}
\langle |Z(\ii T)|^{2k} \rangle = 2^{kN} \left(\frac{N}{2\pi J^2}\right)^{\frac{N(N-1)}{4}} \text{vol}\left(\frac{O(N)}{SO(2)^{\frac{N}{2}} \times S_{\frac{N}{2}}}\right) \prod_i \int_0^\infty \dd \lambda_i \prod_{i<j} (\lambda_i^2-\lambda_j^2)^2 e^{-N\sum_iV(\lambda_i)}.
\end{align}
where
\begin{align} \label{cosine potential}
V(\lambda) = \frac{1}{2J^2} \lambda^2-\frac{k}{N}\log \cos^2 \frac{T \lambda}{2}.
\end{align}
This is the same approach used in \cite{Liao:2020lac} for the complex SYK case, which leads to an alternative treatment of the spectral form factor compared to the collective fields action.

\printbibliography

\end{document}